\def\editmode{0}  

\def\bibfilenames{bibman_refs}

\def\singlenarrowcol{0}
\if\singlenarrowcol0
    \documentclass[conference, 10pt]{IEEEtran}
    \IEEEoverridecommandlockouts
\else
    \documentclass[onecolumn]{IEEEtran}
    \usepackage[pass]{geometry}
    \newgeometry{textwidth=252pt, textheight=672pt}
    
\fi
\usepackage{include_v7b}
\usepackage{flushend}

\newcommand{\figpath}[1]{fig/#1}

\usepackage{import}

\def\journal{1}
\usepackage{environ}

\if\journal0    
    \NewEnviron{journalonly}{%    
    }
    \NewEnviron{conferenceonly}{%    
    \BODY
    }
\else
    \if\journal1
        \NewEnviron{journalonly}{%
        \BODY
        }   
        \NewEnviron{conferenceonly}{%            
        }  
    \else
        \if\journal2
        \newenvironment{journalonly}{\color{darkyellow}}{}
        \newenvironment{conferenceonly}{\color{bluegreen}}{}
        \fi
    \fi
\fi

\def\extended{1} % Extended version (e.g. proofs)
\if\extended0    
    \NewEnviron{extendedonly}{%    
    }
    \NewEnviron{nonextendedonly}{%    
    \BODY
    }
\else
    \if\extended1
        \NewEnviron{extendedonly}{%
        \BODY
        }   
        \NewEnviron{nonextendedonly}{%            
        }  
    \else
        \if\extended2
        \newenvironment{extendedonly}{\color{CadetBlue}}{}
        \newenvironment{nonextendedonly}{\color{Cerulean}}{}
        \fi
    \fi
\fi

\def\intermediatesteps{0}
\if\intermediatesteps1
    \NewEnviron{intermed}{%    
        \color{purple}
        \BODY
    }
    \newcommand{\jumpline}{\\}
    \newcommand{\alignchar}{&}
    \newcommand{\jlac}{\\&}
\else
    \NewEnviron{intermed}{%    
    }
    \newcommand{\jumpline}{}
    \newcommand{\alignchar}{}
    \newcommand{\jlac}{}
\fi

\renewcommand{\define}{\triangleq}
\renewcommand{\expected}{\mathbb{E}}
\newcommand{\spanc}{\overline{\spanv}}
\newcommand{\auxfun}{{\hc{g}}}
\newcommand{\auxset}{{\hc{B}}}
\newcommand{\measure}{{\hc{\mu}}}
\newcommand{\fourier}{\mathcal{F}}
\newcommand{\eigmax}{\hc{\mathcal{\lambda}}_{\text{max}}}
\newcommand{\bw}{{\hc{B}}} % bandwidth

\newcommand{\aconst}{\hc{a}}
\newcommand{\auxvec}{{\hc{\bm v}}} % auxiliary vector
\newcommand{\auxmat}{{\hc{\bm A}}} % auxiliary matrix
\newcommand{\auxmatent}{{\hc{A}}} % auxiliary matrix entry
\newcommand{\auxvar}{{\hc{x}}} % auxiliary variable

\newcommandoa{\mapregion}{\hc{\mathcal{R}}}{\hc{\mathcal{R}}^{(#1)}}

\newcommandoa{\srcregion}{\hc{\mathcal{V}}}{\hc{\mathcal{V}}^{(#1)}}
\newcommand{\dimregion}{{\hc{d}}}
\newcommand{\loc}{{\hc{\bm r}}}
\newcommand{\locx}{{\hc{r}}_{x}}
\newcommand{\normlocx}{{\hc{\check r}}_{x}} % normalized locx

\newcommand{\locxu}{\stackrel{\lrcorner}{\hc{r}}_x} % locx upper
\newcommand{\locxl}{\stackrel{\llcorner}{\hc{r}}_x} % locx lower

\newcommand{\locy}{{\hc{r}}_{y}}
\newcommand{\locz}{{\hc{r}}_{z}}

\newcommand{\locfull}{{\hc{\underline{\bm r}}}}
\newcommandoa{\srclocfull}{{\hc{\underline{\acute{\bm r}}}}}{{\hc{\underline{\acute{\bm r}}}}_{#1}} % source location
\newcommand{\auxsrclocfull}{{\hc{\underline{\check{\bm r}}}}}
\newcommand{\auxsrclocx}{{\hc{\check r}}_x}
\newcommand{\auxsrclocy}{{\hc{\check r}}_y}
\newcommand{\auxsrclocz}{{\hc{\check r}}_z}

\newcommand{\locp}{{\hc{\bm r}}'} % loc prime
\newcommand{\locxsp}[1]{{\hc{r}}_{#1}} % locx sampling point
\newcommandoa{\locxspset}{{\hc{\tilde{\mathcal{R}}}}}{{\hc{\tilde{\mathcal{R}}}_{#1}}} % locx sampling point set
\newcommand{\pmapsv}[1]{{\hc{\pmap}}_{#1}} % pmap sampling value
\newcommand{\pmapdelta}[1]{{\hc{\Delta\pmap}}_{#1}} % pmap delta
\newcommand{\pa}[1]{|{\hc{\Delta\pmap}}_{#1}|} % pmap delta abs
\newcommand{\relvar}[1]{\hc{\xi}_{#1}} % relative variation
\newcommand{\relvaropt}[1]{\hc{\xi}_{#1}^*} % optimal relative variation
\newcommand{\locxdelta}[1]{{\hc{\Delta r}}_{#1}} % location delta
\newcommand{\locxmp}[1]{{\hc{\tilde r}}_{#1}} % location mean point
\newcommand{\spind}{{\hc{n}}} % sampling point index
\newcommand{\spnum}{{\hc{N}}} % number of sampling points
\newcommand{\spindset}{{\hc{\mathcal{N}}}} %  index set of sampling points

\newcommand{\ubound}[1]{{\hc{f}}^{(u)}_{#1}} % upper bound
\newcommand{\lbound}[1]{{\hc{f}}^{(l)}_{#1}} % lower bound
\newcommand{\bound}[1]{{\hc{f}}_{#1}} % bound
\newcommand{\cbound}[1]{\hc{\tilde f}_{#1}} % combined bound

\newcommand{\wnx}{{\hc{k}}_{x}} % wavenumber x

\newcommand{\wavelen}{{\hc{\lambda}}} %
\newcommandoa{\txpow}{{\hc{P}_\text{Tx}}}{{\hc{P}^{(#1)}_\text{Tx}}} %
\newcommand{\txantgain}{{\hc{G}_\text{Tx}}} %
\newcommand{\rxantgain}{{\hc{G}_\text{Rx}}} % 
\newcommand{\mindist}{{\hc{\eta}}_{\text{min}}} % min distance to apply Friis' equation
\newcommand{\minsrcregdist}{{\hc{d}_{\text{min}}}} % min distance to srcregion

\newcommandoa{\srcloc}{{\hc{\acute{\bm r}}}}{{\hc{\acute{\bm r}}}_{#1}} % source location
\newcommandoa{\srclocx}{{\hc{\acute r}}_x}{{\hc{\acute{r}}}_{x,#1}} % source location x-coord
\newcommandoa{\srclocy}{{\hc{\acute r}}_y}{{\hc{\acute r}}_{y,#1}} % source location y-coord
\newcommandoa{\srclocz}{{\hc{\acute r}}_z}{{\hc{\acute r}}_{z,#1}} % source location z-coord
\newcommandoa{\rdists}{{\hc{\beta}}^2}{{\hc{\beta}}^2_{#1}} % squared distance to the map region
\newcommandoa{\rdist}{{\hc{\beta}}}{{\hc{\beta}}_{#1}} %  distance to the map region
\newcommand{\auxrdist}{{\hc{b}}}
\newcommandoa{\normrdist}{{\hc{\check \beta}}}{{\hc{\check \beta}}_{#1}} %  normalized distance to the map region
\newcommand{\rdistmin}{{\hc{\beta}}_{\text{min}}} %  minimum distance to the map region
\newcommand{\rdistmax}{{\hc{\beta}}_{\text{max}}} %  maximum distance to the map region
\newcommandoa{\rdistsset}{{\hc{\mathcal{B}}}}

\newcommand{\srcnum}{{\hc{S}}} % number of sources
\newcommand{\srcind}{{\hc{s}}} % index of source
\newcommand{\srcindp}{{\hc{s'}}} % index of source prime
\newcommandoa{\srccoef}{{\hc{\alpha}}}{{\hc{\alpha}}_{#1}} % source coefficient

\newcommand{\prox}{\hc{\rho}} % Proximity factor

\newcommand{\tol}{{\hc{\epsilon}}} % for all epsilon, there exists...

\newcommand{\etfield}{{\hc{\bm e}}} % E field in time
\newcommand{\pmap}{{\hc{\gamma}}} % power map
\newcommand{\pmapval}{{\hc{g}}} % power map value
\newcommand{\pmapvec}{{\hc{\bm \pmap}}} % 
\newcommand{\pmapestvec}{{\hc{\hbm \pmap}}} % 
\newcommand{\pmapoptval}{{\hc{g}^*}} % power map opt val
\newcommand{\normpmap}{{\hc{\check \gamma}}} % normalized arg power map
\newcommand{\pmapest}{{\hc{\hat \gamma}}} % power map estimate
\newcommand{\pmapft}{{\hc{\Gamma}}} % power map Fourier transform
\newcommand{\auxpmap}{{\hc{\bar\gamma}}} % auxiliary power map
\newcommand{\auxpmapt}{{\hc{\check\gamma}}} % auxiliary power map two
\newcommand{\pmapset}{{\hc{\mathcal{G}}}} % power map set
\newcommandoa{\fspmapset}{{\hc{\pmapset_\text{FS}}}}{{\hc{\pmapset_\text{FS}^{(#1)}}}} % free-space power map set [restriction dim]
\newcommandoa{\auxfspmapset}{{\hc{\check\pmapset_\text{FS}}}}{{\hc{\check\pmapset_\text{FS}^{(#1)}}}} % free-space power map set [restriction dim]
\newcommand{\fspmaprdset}[1]{{\hc{\pmapset_\text{FS}^{(#1,\rdists)}}}} % free-space power map rdists set [restriction dim]

\newcommand{\hplaneset}{{\hc{\mathcal{H}}}} % horizontal plane
\newcommand{\lineset}{{\hc{\mathcal{L}}}} % line

\newcommand{\deriub}{{\hc{m}}} % derivative upper bound

\newcommand{\meas}{\pmap} % measurement
\newcommand{\noise}{{\hc{\zeta}}} % measurement noise

\newcommand{\kernel}{{\hc{\kappa}}} % reproducing kernel

\def\timenotation{0}

\if\timenotation1

    \renewcommand{\time}{\hc{t}}
    \newcommand{\sampind}{\hc{n}}
    \newcommand{\sampint}{\hc{T}}
    \newcommand{\sig}{\hc{x}} % signal
    \newcommand{\sigft}{\hc{X}}
    \newcommand{\sigest}{\hc{\hat x}}
    \newcommand{\sigestft}{\hc{\hat X}}
    \newcommand{\shiftsig}{\hc{y}\sigshiftnot{\sigshift}} % shifted signal
    \newcommand{\shiftsigft}{\hc{Y}\sigshiftnot{\sigshift}} % shifted signal FT
    \newcommand{\sigshift}{{\hc{\tau}}} % signal shift
    \newcommand{\sigshiftf}{{\hc{\theta}}} % signal shift fourier dual variable
    \newcommand{\shiftsigest}{\hc{\hat y}\sigshiftnot{\sigshift}} % shifted signal estimate
    \newcommand{\highband}{{\hc{\mathcal{B}}}}

    \newcommand{\auxseq}{\hc{z}} % aux sequence
    \newcommand{\auxseqft}{\hc{Z}} % aux sequence FT
    \newcommand{\freq}{{\hc{\omega}}}

\else

    \renewcommand{\time}{\locx}
    \newcommand{\sampind}{\spind}
    \newcommand{\sampint}{\hc{\locxdelta{}}}
    \newcommand{\sig}{\pmap} % signal
    \newcommand{\sigft}{\pmapft}
    \newcommand{\sigest}{\pmapest}
    \newcommand{\sigestft}{\hat{\pmapft}}
    \newcommand{\sigshift}{{\hc{\nu}}} % signal shift
    \newcommand{\sigshiftf}{{\hc{\theta}}} % signal shift fourier dual variable

    \newcommand{\shiftsig}{\hc{\phi}\sigshiftnot{\sigshift}} % shifted signal
    \newcommand{\shiftsigft}{\hc{\Phi}\sigshiftnot{\sigshift}} % shifted signal FT
    \newcommand{\shiftsigest}{\hc{\hat \phi}\sigshiftnot{\sigshift}} % shifted signal estimate
    \newcommand{\shiftsigestcs}[1]{\hc{\hat \phi}\sigshiftnot{#1}} % shifted signal estimate custom shift
    \newcommand{\highband}{{\hc{\mathcal{B}}}}

    \newcommand{\auxseq}{\hc{\psi}} % aux sequence
    \newcommand{\auxseqft}{\hc{\Psi}} % aux sequence FT
    \newcommand{\freq}{\wnx}

\fi

\newcommand{\err}{\hc{E}}
\newcommand{\errp}{\hc{E}'} % error prime
\newcommand{\avgerr}{\hc{\bar E}}
\newcommand{\avgerrp}{\hc{\bar E}'}
\newcommand{\sigshiftnot}[1]{^{(#1)}} % signal shift notation
\newcommand{\compind}{{\hc{m}}}% component index
\newcommand{\rpulse}{\hc{\Pi}} % rectangular pulse

\newcommand{\boundfun}{\hc{c}} % bound function

\newcommand{\radius}{\hc{R}} % bound function

\newcommand{\pathlossexp}{{\hc{\upsilon}}} % path loss exponent

\newcommand{\changeo}[1]{#1}
\newenvironment{changes}{}{}

\allowdisplaybreaks

\begin{document}

%%%%%%%%%%%%%%%%%%%%%%%%%%%%%%%%%%%%%%%%%%%%%%%%%%%%%%%%%%%%%%%%%%%%%
\title{Theoretical Analysis of the
    \\ Radio Map Estimation Problem\thanks{This research has been funded in part by the Research Council of Norway under IKTPLUSS grant 311994.}}

\author{
    \IEEEauthorblockN{Daniel Romero, Tien Ngoc Ha, Raju Shrestha}
    \IEEEauthorblockA{\textit{
            Dept. of ICT},
        \textit{University of Agder}\\
        Grimstad, Norway \\
        \{daniel.romero,tien.n.ha,raju.shrestha\}@uia.no}
    \and
    \IEEEauthorblockN{Massimo Franceschetti}
    \IEEEauthorblockA{%\textit{
        % Department of ECE  },
        \textit{University of California, San Diego}\\
        San Diego, USA \\
        mfranceschetti@ucsd.edu}

}

%\author{\today\thanks{Thanks to XYZ agency for funding.}}

\maketitle
%%%%%%%%%%%%%%%%%%%%%%%%%%%%%%%%%%%%%%%%%%%%%%%%%%%%%%%%%%%%%%%%%%%%%

\begin{abstract}
    Radio maps provide radio frequency metrics, such as the received signal
    strength, at every location of a geographic area. These maps, which are
    estimated  using a set of  measurements collected at multiple positions, find a
    wide range of applications in wireless communications, including the prediction
    of coverage holes, network planning, resource allocation, and path planning for
    mobile robots. Although a vast number of estimators have been proposed, the
    theoretical understanding of the radio map estimation (RME) problem has not been
    addressed. The present work aims at filling this gap along two directions.
    First, the complexity of the set of radio map functions is quantified by means
    of lower and upper bounds on their spatial variability, which offers valuable
    insight into the required spatial distribution of measurements and the
    estimators that can be used. Second, the reconstruction error for power maps in
    free space is upper bounded for three conventional spatial interpolators. The
    proximity coefficient, which is a decreasing function of the distance from the
    transmitters to the mapped region, is proposed to quantify the complexity of the
    RME problem. Numerical experiments assess the tightness of the obtained bounds
    and the validity of the main takeaways in complex environments.

\end{abstract}


% \begin{journalonly}%
%     hello jour
% \end{journalonly}%

% \begin{conferenceonly}%
%     hello conf
% \end{conferenceonly}%

\section{Introduction}

% \begin{align}
%     line & 1 \\
%     \begin{intermed}
%         line
%     \end{intermed}
%     \alignchar
%     \begin{intermed}
%         2
%     \end{intermed}
%     \jumpline
%     line & 3
% \end{align}

\label{sec:intro}
\begin{bullets}

    \blt[radio maps]%
    \begin{bullets}%
        \blt[def]Radio maps, also known as radio environment maps, provide a radio frequency (RF) metric of interest across a geographical region~\cite{romero2022cartography}.
        \blt[power maps]For example, in \emph{power maps}, which constitute a prominent example of radio maps, the  metric of interest is the power that a sensor would measure when placed at each location. An example of a power map constructed with real data is shown in Fig.~\ref{fig:truemap}.
        \blt[other examples]Other examples of RF metrics include the received power spectral density (PSD), outage probability, and channel gain.

        \blt[applications]Radio maps are of interest in a large number of applications such as cellular communications, device-to-device communications, network planning, frequency planning, robot path planning, dynamic spectrum access, aerial traffic management in unmanned aerial systems, fingerprinting localization, and so on; see e.g.~\cite{abouzeid2013predictive,subramani2011practical,cai2011ran,zalonis2012femtocell,romero2022cartography} and references therein.
        \begin{bullets}%
            \blt A recently popular application of power maps is to determine how the coverage of a cellular or broadcast network can be improved by deploying new base stations or relays, either terrestrial or aerial \cite{romero2022aerial,viet2022introduction,viet2023roadmaps}.
        \end{bullets}

    \end{bullets}% 

    \blt\cmt{RME SOA}
    \begin{bullets}%
        \blt[RME def]In radio map estimation (RME),  a radio map is constructed using  measurements collected across the area of interest.
        \blt[estimators] Many estimators have been proposed in the literature, mostly based on some form of interpolation or regression. By far, power maps are the radio maps that garnered most interest.
        \begin{bullets}%
            \blt[parametric]
            \blt[kernel] One of the  simplest kinds of estimators relies on kernel-based learning (see~\cite{romero2015cartography} and references therein), which overcome the limitations of (the simpler) parametric estimators~\cite[Sec. ``Linear Parametric RME"]{romero2022cartography}.
            \blt[kriging] Other popular estimators are based on Kriging~\cite{alayafeki2008cartography,agarwal2018spectrum,shrestha2022surveying},
            \blt[sparsity]sparsity-based inference~\cite{bazerque2010sparsity,bazerque2011splines,jayawickrama2013compressive},
            \blt[Matrix completion]matrix completion~\cite{khalfi2018airmap,schaufele2019tensor},
            \blt[dictionary learning] dictionary learning~\cite{kim2013dictionary}, and graphical models~\cite{ha2024location}.
            \blt[deep learning]The most recent trend capitalizes on deep neural networks; see e.g.~\cite{krijestorac2020deeplearning,levie2019radiounet,han2020power,teganya2020rme}.
        \end{bullets}%
        \blt[disclaimer] Note that the aforementioned list of works is not exhaustive due to space limitations. For a more comprehensive list of references, see~\cite{romero2022cartography}.

    \end{bullets}% 

    \blt[Limitations]
    \begin{bullets}
        \blt[RME literature \ra lack of theoretical analysis]Despite the large
        volume of research in this area, the vast majority of works adhere to a
        common profile: they propose an estimator and validate it with synthetic
        data generated using a statistical propagation model or with ray-tracing software. A small number of works utilize also real data~\cite{shrestha2023empirical,xiang2005hidden,hu2013efficient,yang2016compressive,niu2018recnet}.
        However, no \emph{theoretical} analysis on the  fundamental aspects of the RME problem as well as on the  performance of estimation algorithms has been carried out.
        \blt[most related]Indeed, the most related work in this context is two-fold.
        \begin{bullets}%
            \blt[kriging]On the one hand, the estimation error of some schemes can be derived if the field of interest  adheres to a certain model~\cite{shrestha2022surveying,krijestorac2020deeplearning}. However, these models are generic, not necessarily accurate for radio maps.
            \blt[wave th. of info.]On the other hand, \changeo{the \emph{wave theory of information} (WTI)} studied the problem of reconstructing the electromagnetic field across space and time using arrays of synchronized sensors~\cite{franceschetti2018}. Nonetheless, this problem is fundamentally different from RME, where sensors are not typically synchronized, the metrics of interest involve temporal averages of the electromagnetic field, and
            the targeted spatial resolution is much lower.
        \end{bullets}%

    \end{bullets}

    \begin{figure}[t]
        \centering
        \includegraphics[width=\columnwidth]{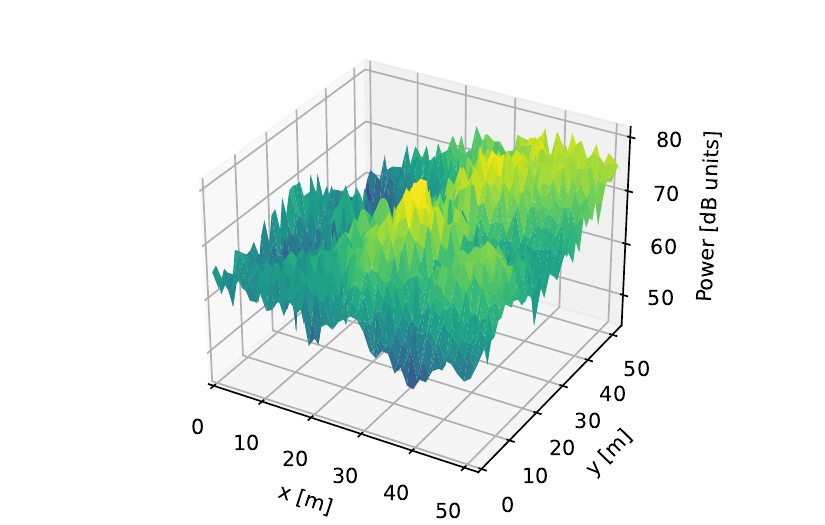}
        \caption{Example of power map where a spatially dense set of measurements was collected using an unmanned aerial vehicle~\cite{shrestha2023empirical}.}
        \label{fig:truemap}
    \end{figure}

    \blt[contributions]This paper\footnote{A conference version of this paper was submitted to the IEEE Vehicular Technology Conference, Spring 2024. Relative to that paper, the present one considers also 2D maps, contains  \Cref{prop:approxdiff}, \Cref{prop:density}, \Cref{prop:densitycircle}, the analysis of the reconstruction using sinc interpolation, the proofs of all results, and further discussions and numerical experiments.
    } takes a step to address this gap by means of a quantitative theoretical analysis of the RME problem.
    \begin{bullets}%
        \blt[spatial variability]In particular, the difficulty of the RME problem is first assessed by analyzing the spatial variability of power maps. \begin{changes}An important   finding in this context is  that the spatial variations of power maps in free space are relatively slow. Most of their energy is concentrated at low spatial frequencies,  which motivates estimators based on this property.\end{changes}

        \blt[estimator performance]Second, the estimation performance of zeroth-order, first-order, and sinc interpolators is quantified in terms of $L^1$, $L^2$, and $L^\infty$ error metrics. Many of these bounds turn out to be proportional to a quantity referred to as the \emph{proximity coefficient}, which is directly proportional to the transmitted power and inversely proportional to the cube of the distance from the transmitters to the mapped region. As a result, the analysis reveals that  a larger spatial density of measurements is required when the sources are closer to the mapped region. Error analysis of the sinc interpolator yields bounds with a faster decay rate than zeroth- and first-order interpolators, but the latter are seen to be preferable in practice, where the  number of samples is finite.

        \blt[non-free space]%
        \begin{changes}%
            Finally, although the aforementioned results assume free-space
            propagation, their generalization to more involved propagation
            phenomena is briefly addressed, both theoretically and by means of a
            numerical experiment.
        \end{changes}
    \end{bullets}

    \blt[paper structure]The rest of the paper is structured as follows.  Sec.~\ref{sec:problem} formulates the RME problem \begin{changes}and introduces useful notation\end{changes}. Sec.~\ref{sec:variability} analyzes the spatial variability of power maps. Sec.~\ref{sec:reconstruction} derives error bounds for the considered interpolators. Finally, Sec.~\ref{sec:experiments} presents numerical experiments and
    Sec.~\ref{sec:conclusions} concludes the paper. The proofs can be found in the appendices.
    \begin{nonextendedonly}
        Extended versions of the proofs can be found in \cite{romero2023theoretical}.~
    \end{nonextendedonly}%
    \begin{changes}%
        A note on the  generalizability of the results here to non-free space environments is given in Appendix~\ref{sec:plexponent}.
    \end{changes}

    \blt[notation]\emph{Notation:}
    \begin{bullets}
        \blt[eqdef]Symbol $\define$ indicates equality by definition.
        \blt[closure]If $\mathcal{A}$ is a set in a metric space, $\bar{\mathcal{A}}$ denotes its closure.
        \blt[span]If $\mathcal{A}$ is a set in a vector space, $\spanv(\mathcal{A})$ denotes the set of all linear combinations of finitely many elements of $\mathcal{A}$.
        \blt[rfield]$\mathbb{N}$ is the set of natural numbers, $\mathbb{Z}$ the set of integers, and  $\rfield$ is the field of real numbers.
        \blt[vector/matrix] Boldface lowercase (uppercase) letters denote column vectors (matrices).
        \blt[]Vertical concatenation is represented with a semicolon, e.g. $[\bm a;\bm b]$.
        \blt[functions]A function $f$ is represented by a letter, whereas the result of evaluating such a function at a point $x$ is denoted as $f(x)$.
    \end{bullets}
\end{bullets}

\section{Estimation of Radio Maps}
\label{sec:problem}

\begin{bullets}

    \blt[overview]
    \begin{changes}
        This section introduces power maps and formulates the problem of
        estimating them. Subsequently, useful notation is presented for power maps
        in free space.
    \end{changes}

    \begin{bullets}%
        \blt\cmt{Power maps in general}%
        \begin{bullets}%
            \blt\cmt{Area}            Let $\mapregion\subset \rfield^3$ comprise the Cartesian coordinates of all points in the geographic area of interest.
            \blt\cmt{Tx.} A set\footnote{
                \begin{changes}
                    \label{footnote:sources}
                    There may be other sources in the region so long as the measurement process can separate out their aggregate contribution. This can be achieved e.g. by means of spreading codes or pilot sequences. This allows the construction of a wide variety of maps, including signal maps, interference maps,  noise maps, and signal-to-interference-plus-noise-ratio (SINR) maps. This is useful e.g. to determine the coverage of a network; see~\cite{romero2022cartography} for a list of applications.
                \end{changes}
            } of $\srcnum$ sources (also referred to as transmitters) in a region
            $\srcregion\subset\rfield^3$ produce an aggregate electric field
            $\etfield(\locfull, t)\in \rfield^3$ at every point $\locfull\in\mapregion$, where
            $t$ denotes time. The underscore notation $\locfull$ will  represent full location vectors in $\rfield^3$, whereas the notation $\loc$ will be used later when introducing restrictions.

            \blt\cmt{Power map} Neglecting for simplicity polarization
            effects and modeling $\etfield(\locfull, t)$ as \begin{changes}an ergodic\end{changes} wide-sense
            stationary random process over $t$ for all $\locfull$, the power
            of the signal received by a sensor with an isotropic antenna at
            $\locfull\in\mapregion$ does not depend on $t$ so it can be represented by a function
            $\pmap:\mapregion\rightarrow \rfield_+$. Function $\pmap$, which
            therefore indicates how  power spreads across space, is
            a special case of a {radio map} termed \emph{power map} and depends on the
            transmitted signals, the transmitter locations, and the propagation environment.
        \end{bullets}

        \blt\cmt{RME}The  problem is to estimate a power map given a set of  measurements in $\mapregion$.
        \begin{bullets}%
            \blt\cmt{Measurements} Specifically, let
            $\meas_1,\ldots,\meas_\spnum$ denote the power measured at a set of locations
            $\locxspset\define\{\locfull_1,\ldots, $ $\locfull_{\spnum}\}\subset\mapregion$.
            For the ensuing analysis, it is not relevant whether  \begin{changes} sensors are static, which implies that each one measures at a single spatial location, or mobile, which means that they can   measure at multiple spatial locations.\end{changes}\footnote{
                \begin{changes}
                    Since $\etfield(\locfull, t)$ is modeled as an ergodic wide-sense
                    stationary random process, the power is a constant, i.e.,  it does not depend on time. Thus, theoretically  there is  not a maximum allowed difference between the times at which the measurements must be collected.
                    In practice, $\etfield(\locfull, t)$ is non-stationary and one can think of power as a function of time. Thus, one needs to specify a time scale under which the power  does not significantly change. All the $\spnum$ measurements must therefore be collected within this time scale.
                \end{changes}
            }

            \begin{bullets}

                \blt[power measurement] Due to the finite observation time spent by a sensor at $\locfull_\spind$
                to measure the received power,  $\meas_\spind$ does not generally equal
                $\pmap(\locfull_\spind)$. Instead, certain measurement error
                must be expected. This is oftentimes expressed  as
                $\meas_\spind=\pmap(\locfull_\spind)+\noise_\spind$, where
                $\noise_\spind$ is the measurement error.

                % \blt[location] The measurement location $\locfull_\spind$ itself
                % is also typically unknown and must be measured (or estimated)
                % by the sensor. This is usually accomplished by global
                % navigation satellite systems (GNSSs) such as GPS. This yields
                % a location measurement $\locfullmeas_\spind$ of
                % $\locfull_\spind$.

            \end{bullets}

            \blt\cmt{Statement}The power map estimation problem can be formulated as,
            \begin{bullets}%
                \blt\cmt{Given} given $\{(\locfull_\spind,
                    \meas_\spind)\}_{\spind=1}^{\spnum}$,
                \blt\cmt{Requested}
                estimate the function $\pmap$ or, equivalently, the values
                $\pmap(\locfull)$ for all $\locfull\in\mapregion$.  The map
                estimate will be denoted as $\pmapest$.
            \end{bullets}%
            \blt[common assumptions]In this formulation, no information is given about the propagation environment, the positions of the sources, the transmitted power, the radiation pattern of the transmit antennas, and so on.
            % \begin{journalonly}%                     
            %     \begin{bullets}
            %         \blt[no loc. error]
            %         %
            %         \blt[no synchronization among sensors]
            %         %
            %         \blt[the tx. power and source location is unknown]
            %         %
            %         \blt[the geometry of the environment is unknown, else, ray
            %             tracing.]
            %         %
            %     \end{bullets}%
            % \end{journalonly}%
            %
            \blt[existing approaches]This is why most estimators in the
            literature are based on interpolation algorithms rather than on electromagnetic propagation models. A detailed taxonomy of these estimators along with relevant references can be found in \cite{romero2022cartography}.

            %% \begin{enumerate}
            %%   \item \cmt{F1}
            %% \end{enumerate}

        \end{bullets}
    \end{bullets}

\end{bullets}

\subsection{Power Maps in Free Space}
\label{sec:powermapsinfreespace}

\begin{bullets}
    \blt[overview]
    \begin{changes}
        Since many of the results in this paper focus on free-space propagation, this section introduces useful notation  for this class of maps. The case of general propagation effects will be addressed when discussing some general results and it will be
        the focus of  Sec.~\ref{sec:raytracing}, Appendix~\ref{sec:plexponent}, and future work.
    \end{changes}

    % \blt[assumptions]
    % \begin{bullets}
    %   \blt[synchronization]
    %   % \begin{bullets}
    %   % \blt[Synchronized Sensors]
    %   % \begin{bullets}

    % % \blt[sensor synchronization] \ra one can estimate the E field and
    % % then square. 
    % % \blt[WTI \ra  dof ]
    % % \end{bullets}
    % % \blt[Unsynchronized Sensors]
    % %   \end{bullets}

    % % \begin{align}
    % % \label{eq:euler}
    % % e^{j\pi}=-1
    % % \end{align}

    % \end{bullets}

    % \blt[power map set] $\pmapset$ is the set of all physically possible
    % power maps up to the simplifying assumptions pointed out earlier. 

    \blt[free-space model]
    \begin{bullets}
        \blt[Friis]
        Recall that Friis' propagation law
        establishes that the power that a terminal at $\locfull$ receives from a transmitter
        at $\srclocfull$ when propagation takes place in free space is given by
        \begin{align}
            \label{eq:friisfull}
            \pmap(\locfull) = \txpow \txantgain \rxantgain \left[\frac{ \wavelen}{ 4 \pi
                    \|\locfull - \srclocfull\|  }\right]^2,
        \end{align}
        where
        \begin{bullets}%
            \blt $\wavelen$ is the wavelength,
            \blt $\txpow$ is the transmitted power,
            \blt $\txantgain$ is the antenna gain of the transmitter, and
            \blt $\rxantgain$ is the antenna gain of the receiver.
        \end{bullets}%
        \blt[simplified Friis] Suppose for simplicity that both terminals use
        isotropic antennas, i.e. $\txantgain=\rxantgain=1$. Upon letting
        $\srccoef\define \txpow( \wavelen/4\pi)^2$, expression \eqref{eq:friisfull}
        reduces to
        \begin{align}
            \label{eq:friis}
            \pmap(\locfull) = \frac{ \srccoef}{
                \|\locfull - \srclocfull\|^2  }.
        \end{align}
        \blt[Singularity] Observe that, as per \eqref{eq:friis}, $\pmap(\locfull)
            \rightarrow +\infty$ as $\locfull \rightarrow \srclocfull$, which is
        not physically possible. The reason for this disagreement between
        \eqref{eq:friis} and the physical reality is that \eqref{eq:friis} is an
        approximation valid only in the far field, i.e., when $\|\locfull -
            \srclocfull\|$ is significantly larger than $\wavelen$. Thus, it will be
        required throughout that $\|\locfull -
            \srclocfull\|\geq \mindist$, where $\mindist$ is a constant sufficiently
        larger than $\wavelen$.

    \end{bullets}

    % \newcommand{\becc}[2]{
    %   \let\po#1
    %   \let\pt#2
    %   uii\po\pt owwl
    % }

    % \newcommand{\dcom}[2]{
    %   \newcommand{\mycommand}[2]{
    %     \expandafter c1:##1,c2:##2,po:#1,pt:#2
    %     }
    % }

    % \dcom{1}{2} 

    % \mycommand{3}{4}
    \blt[free-space power map set]In the presence of multiple sources that transmit uncorrelated\footnote{This assumption excludes setups with coordinated multipoint or with multiantenna transmitters that use space-time coding or beamforming. \begin{changes}Recall also
            Footnote \ref{footnote:sources}.\end{changes}} signals, the individual contributions of each one to the total received power add up and, therefore, the set of all possible power maps is  given by
    \begin{align}
        \nonumber
        \fspmapset = \bigg\{\pmap:\mapregion \rightarrow \rfield_+~|~
        \pmap(\locfull) = \sum_{\srcind=1}^{\srcnum}\frac{\srccoef[\srcind]}{
        \|\locfull - \srclocfull[\srcind]\|^2  }, & \\~\srclocfull[\srcind] \in \srcregion,~\srccoef[\srcind]\geq 0,~\srcnum\in\mathbb{N}
        \bigg\},
        \label{eq:3Dfsppowermapset}
    \end{align}
    where
    \begin{bullets}%
        \blt $\srcnum$ denotes the number of sources, and
        \blt $\srccoef[\srcind]$
        \blt and $\srclocfull[\srcind]$ are respectively the $\srccoef$ coefficient
        and location of the $\srcind$-th source.
        Due to the minimum distance assumption introduced earlier,
        $\srcregion$ must be such that
        \begin{align}
            \label{eq:minsrcregdist}
            \minsrcregdist(\locfull)\define\inf\{\|\locfull-\srclocfull \|~|~
            \srclocfull\in\srcregion \}\geq \mindist\forall \locfull\in\mapregion.
        \end{align}
        %    \blt $\deneps$ is a small positive constant. 

    \end{bullets}
    %Clearly, $\fspmapset\subset\pmapset$. 

    \subsection{1D and 2D Restrictions}
    \label{sec:restrictions}
    \blt[restrictions]
    \begin{bullets}
        \blt[motivation] The maps in \eqref{eq:3Dfsppowermapset} are functions
        of three spatial coordinates. However, most works in the literature consider restrictions of such maps to two or one spatial dimensions. This is because the case of two spatial dimensions is of interest when users are on the ground, whereas the case of one spatial dimension is relevant e.g. when one wishes to construct a map along a road or railway. The case of three spatial dimensions is still rare in the literature, but it has already been successfully applied to deploy aerial base stations and aerial relays~\cite{viet2023roadmaps,romero2022aerial}.

        \blt[2D] \textbf{2D Restriction.}           To consider the restriction of power maps to two spatial
        dimensions, focus without loss of generality (w.l.o.g.) on the values that $\pmap$ takes on the
        horizontal plane
        $\hplaneset\define\{[\locx;\locy;\locz]\in\rfield^3~|~\locz=0\}$. To
        this end, the domain $\mapregion$ where $\pmap$ is defined must be a
        subset of $\hplaneset$, i.e., $\mapregion =
            \{[\locx;\locy;\locz]~|~[\locx;\locy]\in\mapregion[2],~\locz=0\}$ for
        some $\mapregion[2]\subset \rfield^2$. Since each point in
        $\mapregion$ can be identified by its x and y coordinates, which are
        collected in  $\mapregion[2]$, the sought restriction of
        $\pmap$ will be defined on $\mapregion[2]$.

        Before presenting the expression for this restriction,  some notation is introduced. Upon letting
        $\locfull=[\locx;\locy;0]$ and
        $\srclocfull=[\srclocx;\srclocy;\srclocz]$, equation \eqref{eq:friis} becomes
        \begin{align}
            \pmap(\locfull) =\frac{ \srccoef}{
                \|\locfull - \srclocfull\|^2  }
            =\frac{ \srccoef}{
                \|\loc - \srcloc\|^2 + \rdists  } \define \pmap(\loc),
        \end{align}
        where
        \begin{bullets}
            \blt $\loc\define[\locx;\locy]$ and
            \blt $\srcloc\define[\srclocx;\srclocy]$ respectively contain the
            horizontal coordinates of the evaluation and source locations with respect
            to $\hplaneset$, whereas
            \blt $\rdists\define \srclocz^2$ is the squared distance from the source
            location to $\hplaneset$.
        \end{bullets}
        Thus, although the points where $\pmap$ will be evaluated are on
        $\hplaneset$, the source locations are not required to be on
        $\hplaneset$.

        \begin{figure}[t]
            \centering
            \includegraphics[width=\columnwidth]{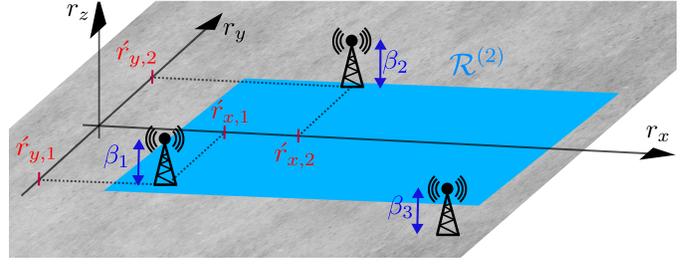}
            \caption{Visual depiction of the setup for estimating a power map in two spatial dimensions. This is the most common setup in the literature.
            }
            \label{fig:setup2D}
        \end{figure}

        With this notation, restricting the maps in
        \eqref{eq:3Dfsppowermapset} to $\hplaneset$ yields
        \begin{align}
            \label{eq:mapset2d}
            \nonumber
            \fspmapset[2] & = \bigg\{\pmap:\mapregion[2] \rightarrow \rfield_+~|~
            \pmap(\loc) = \sum_{\srcind=1}^{\srcnum}\frac{\srccoef[\srcind]}{
            \|\loc - \srcloc[\srcind]\|^2 + \rdists[\srcind]  },                  \\&
            ~\srcloc[\srcind] \in \srcregion[2],~\rdists[\srcind]\in\rdistsset(\srcloc[\srcind]),~\srccoef[\srcind]\geq 0,~\srcnum\in\mathbb{N}
            \bigg\}   ,
        \end{align}
        where
        \begin{bullets}
            \blt $\srcregion[2]$ is the set where the horizontal coordinates of
            the sources are allowed to be (which results from the projection of $\srcregion$ onto
            $\hplaneset$)
            \blt and $\rdistsset(\srcloc)$ contains the allowed values of
            $\rdists$ for each vector of source horizontal coordinates $\srcloc$, that is,
            $\rdistsset(\srcloc)\define
                \{\rdists ~|~ \exists \auxsrclocfull=[\auxsrclocx;\auxsrclocy;\auxsrclocz]\in
                \srcregion~:~\srcloc=[\auxsrclocx;\auxsrclocy]~\text{and}~\auxsrclocz^2=\rdists\}$.
        \end{bullets}
        Fig.~\ref{fig:setup2D} illustrates the main symbols used in the 2D restriction.

        \blt[1D]\textbf{1D Restriction.} Since it is the most  insightful case, \begin{changes}most results in this paper focus on \end{changes} radio maps in a single spatial dimension, i.e., when the functions in \eqref{eq:3Dfsppowermapset} are restricted to a line. \changeo{For the same reason, this approach has also been adopted in the WTI~\cite[Ch. 8]{franceschetti2018}.} RME on a line was considered e.g. in~\cite{romero2015onlinesemiparametric}.

        Consider w.l.o.g. the line
        $\lineset\define\{[\locx;\locy;\locz]\in\rfield^3~|~\locy,\locz=0\}$ and suppose that $\mapregion$ is a subset of $\lineset$. Thus, one can write
        $\mapregion =
            \{[\locx;\locy;\locz]~|~[\locx]\in\mapregion[1],~\locy,\locz=0\}$ for
        some $\mapregion[1]\subset \rfield^1$, where $\rfield^1$ is the set of
        all vectors with one real entry. Vector notation is sometimes used for
        scalars to simplify the statement of some of the upcoming  results.

        When $\locfull=[\locx;0;0]$ and
        $\srclocfull=[\srclocx;\srclocy;\srclocz]$, \eqref{eq:friis} becomes
        \begin{align}
            \pmap(\locfull) =\frac{ \srccoef}{
                \|\locfull - \srclocfull\|^2  }
            =\frac{ \srccoef}{
                \|\loc - \srcloc\|^2 + \rdists  } \define \pmap(\loc),
        \end{align}
        where
        \begin{bullets}
            \blt $\loc\define[\locx]$ and
            \blt $\srcloc\define[\srclocx]$ are the longitudinal
            coordinates of the evaluation and source locations, whereas
            \blt $\rdists\define \srclocy^2 + \srclocz^2$ is the squared distance
            from the source location to the mapped line.
        \end{bullets}%
        Thus,  restricting  the maps in \eqref{eq:3Dfsppowermapset} to
        $\lineset$  yields
        \begin{align}
            \nonumber
            \fspmapset[1] & = \bigg\{  \pmap:~\mapregion[1] \rightarrow \rfield_+~|~
            \pmap(\loc) = \sum_{\srcind=1}^{\srcnum}\frac{\srccoef[\srcind]}{
            \|\loc - \srcloc[\srcind]\|^2 + \rdists[\srcind]  },                     \\&
            ~\srcloc[\srcind] \in \srcregion[1],~\rdists[\srcind]\in\rdistsset(\srcloc[\srcind]),~\srccoef[\srcind]\geq 0,~\srcnum\in\mathbb{N}
            \bigg\},
            \label{eq:mapset1d}
        \end{align}
        where, similarly to the 2D case,
        \begin{bullets}%
            \blt $\srcregion[1]$ results from the orthogonal projection of $\srcregion$ onto
            $\lineset$
            \blt and $\rdistsset(\srcloc)$ is the set of allowed values for $\rdists$ when the x-coordinate of the source location is $\srclocx$, that is,
            $\rdistsset(\srcloc)\define \{\rdists ~|~ \exists
                \auxsrclocfull=[\auxsrclocx;\auxsrclocy;\auxsrclocz]\in
                \srcregion~:~\srcloc=[\auxsrclocx]~\text{and}~\auxsrclocy^2 + \auxsrclocz^2=\rdists\}$.
        \end{bullets}
        Fig.~\ref{fig:setup} illustrates the geometric meaning of the main symbols in \eqref{eq:mapset1d} while Fig.~\ref{fig:mapexample} shows an example of a power map in $\fspmapset[1]$.

    \end{bullets}

    \begin{figure}[t]
        \centering
        \includegraphics[width=\columnwidth]{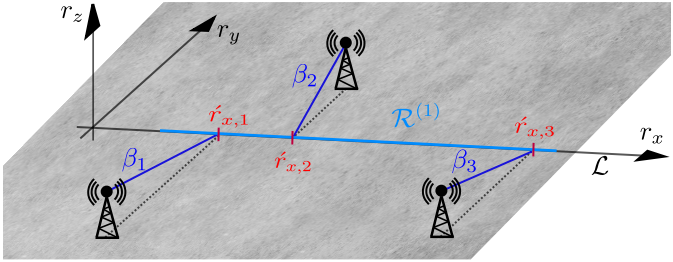}
        \caption{Visual depiction of the setup for estimating a power map in one spatial dimension. This is of interest e.g. when a map must be estimated along a road.
        }
        \label{fig:setup}
    \end{figure}

    \begin{figure}[t]
        \centering
        \includegraphics[width=\columnwidth]{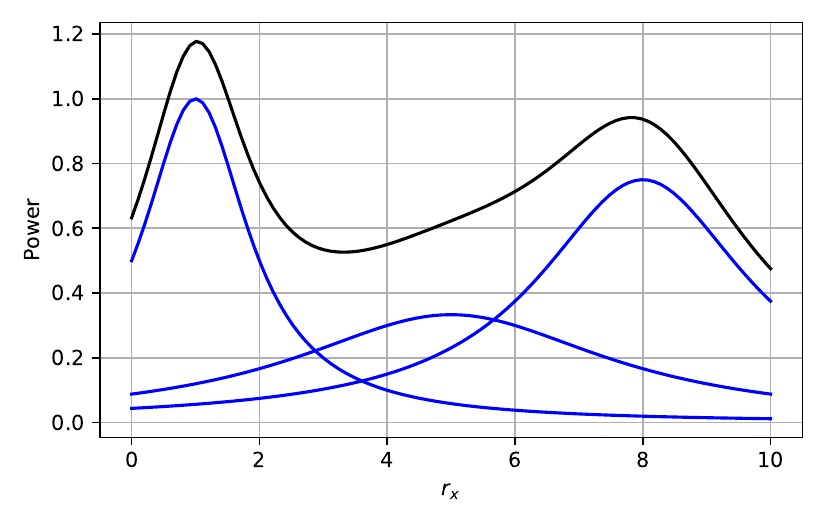}
        \caption{The black curve shows an example of a power map in  $\fspmapset[1]$ where $\srcnum=3$, $[\srclocx[1],\srclocx[2],\srclocx[3]]=[1,5,8]$,  $[\rdist[1],\rdist[2],\rdist[3]]=[1,3,2]$, and $[\srccoef[1],\srccoef[2],\srccoef[3]]=[1,3,3]$. The blue lines correspond to the contribution of each source. The maximum of each one is at the corresponding value of $\srclocx[\srcind]$.
        Although source $\srcind=1$ has the lowest power, it is closer to $\lineset$ than the other sources and this results in the largest contribution to $\pmap$ and its derivative~$\pmap'$. }
        \label{fig:mapexample}
    \end{figure}

\end{bullets}

\section{Spatial Variability of Radio Maps}
\label{sec:variability}

\begin{bullets}%

    \blt[sec. overview]
    % This section characterizes the variability of $\pmap$ across
    % space. The results presented here are of interest on their own and are
    % also used to derive error bounds for radio map estimators in
    % Sec.~\ref{sec:reconstruction}.
    \begin{changes}
        Having formalized the classes of maps under study,
        the rest of this section will analyze the variability of the functions
        in $\fspmapset[1]$ and $\fspmapset[2]$. Specifically,
        Sec.~\ref{sec:highvariability} and Sec.~\ref{sec:lowvariability} will
        respectively present high- and low-variability results. Subsequently,
        Sec.~\ref{sec:reconstruction} builds upon these results to derive
        performance bounds for three interpolation algorithms.
    \end{changes}
\end{bullets}%

\subsection{High-variability Result}
\label{sec:highvariability}

\begin{bullets}
    \blt[overview]This section establishes that power maps constitute a
    considerably rich class of functions.  It will follow that, under general
    conditions, power maps cannot be estimated exactly with a finite
    number of measurements, even in the absence of noise. In
    other words, a certain error must be expected. Importantly, these
    observations are not confined to free-space propagation: they hold in
    the presence of arbitrary propagation phenomena such as reflection,
    refraction, and diffraction.

    \blt[map differences] These conclusions follow from the next
    result, which establishes that any continuous function can be
    approximated up to arbitrary accuracy as the difference between two
    power maps in free space.

    \begin{mytheorem}
        \label{prop:approxdiff}
        Let $\mapregion[\dimregion]$ be a compact subset of
        $\rfield^\dimregion$, where $\dimregion$ is 1 or 2. Then, there exists
        $\srcregion\subset\rfield^3$ such that the following condition holds:
        for every continuous function $\pmap:\mapregion[\dimregion] \rightarrow \rfield$
        and every $\tol>0$,
        % If
        % \begin{bullets}
        %   \blt $\srcregion$ is compact
        %   \blt $\mapregion\subset \srcregion$
        %   \blt $\tol>0$
        %   \blt $\pmap:\srcregion\rightarrow \rfield$ is an arbitrary
        %   continuous function (not necessarily a power map)
        % \end{bullets}
        % Then
        \begin{align}
            \label{eq:approxdiffcond}
            \exists \pmap_+,\pmap_- \in \fspmapset[\dimregion]~:~\sup_{\loc \in \mapregion[\dimregion]}|\pmap(\loc)
            - (\pmap_+(\loc) - \pmap_-(\loc))|<\tol.
        \end{align}
        The set $\srcregion$ can be chosen to be any set that satisfies
        \begin{enumerate}[label=(C\arabic*)]
            \item \label{eq:condsubset} $\mapregion[\dimregion]\subset\srcregion[\dimregion]$ and
            \item \label{eq:condrdists} there exists
                  $\rdists>0~:~\rdists\in\rdistsset(\srcloc)~\forall\srcloc\in
                      \mapregion[\dimregion]$.
        \end{enumerate}

    \end{mytheorem}

    \begin{IEEEproof}
        See Appendix~\ref{sec:approxdiff}.
    \end{IEEEproof}

    \blt[observations] Several
    observations are relevant.
    \begin{bullets}%
        \blt[arbitrary function] First, function $\pmap$ need not be a power map -- it is an arbitrary
        continuous function which can even take negative values.
        \blt[conditions] Second, conditions \ref{eq:condsubset} and \ref{eq:condrdists} just
        require sufficient flexibility to find suitable source locations. One simple example where both conditions hold is to set $\srcregion$ so that the sources are allowed to be anywhere above a minimum positive height.
        \blt[infinite dim] Third, \Cref{prop:approxdiff} establishes that
        $\spanv\fspmapset[\dimregion]$ is dense in the space of continuous
        functions defined on any given compact subset of $\rfield^\dimregion$. As a result,
        $\fspmapset[\dimregion]$ is clearly infinite dimensional. Thus, \emph{one
            should not expect to be able to reconstruct a power map exactly with a
            finite number of measurements, even if those measurements are noiseless.}
        \blt[arbitrary maps] Finally, let $\pmapset$ denote the set of physically possible power maps,
        that is, the set of power maps consistent with Maxwell's equations. Up
        to the simplifying assumptions in Friis' transmission equation, it
        holds that $\fspmapset\subset\pmapset$. Thus, the family of functions
        $\pmapset$ is at least as rich as $\fspmapset$ and, consequently, the
        above conclusions carry over to arbitrary power maps, not just
        free-space maps.

    \end{bullets}

    \blt[negative] In view of \Cref{prop:approxdiff}, one may think that
    there is no hope that power maps can be satisfactorily estimated when
    the set of measurement locations is finite or even
    countable. Fortunately, a closer look at \Cref{prop:approxdiff}
    reveals that the variability of the functions in
    $\fspmapset[\dimregion]$ may not be as large as it may seem.
    \begin{bullets}
        \blt[cone]First and foremost, \Cref{prop:approxdiff} uses the
        \emph{difference} $ \pmap_+-\pmap_-$ rather than a single map to
        approximate $\pmap$. This is because of the requirement on
        non-negativity of the $\srccoef[\srcind]$'s in
        \eqref{eq:3Dfsppowermapset}. In other words, if
        $\fspmapset[\dimregion]$ were a subspace rather than just a convex
        cone, it would follow from \Cref{prop:approxdiff} that it is possible
        to find two power maps that, given any arbitrary discrete set $\locxspset$
        of measurement locations, (i) they take the same values at $\locxspset$
        and (ii) they differ arbitrarily at any given point of  $\mapregion[\dimregion]-\locxspset$. This would imply
        that the error of any reconstruction algorithm that relies on measurements at $\locxspset$, even in the absence of noise, would be unbounded. However, this is fortunately not the case \begin{changes}and is extensively discussed in the next section\end{changes}.

        \blt[energy/number of sources] Second, \Cref{prop:approxdiff} does not
        constrain the transmitted power or the number of sources in $ \pmap_+$
        and $\pmap_-$. This means that some of these
        quantities may arbitrarily increase as $\tol\rightarrow 0$. Thus, in the
        presence of a constraint on the transmitted power or number of
        sources, the variability of power maps may be much more limited than
        it may seem at first glance from \Cref{prop:approxdiff}.

    \end{bullets}

\end{bullets}

%\import{fig}{figure.pdf_tex}

\subsection{Low-variability Results}
\label{sec:lowvariability}

\begin{bullets}

    \blt[Overview]    This section provides upper bounds on the variability of
    power maps. To facilitate the intuitive understanding of the
    fundamental phenomena to be studied, the focus will be on the case
    $\dimregion=1$, in which case any $\pmap\in\fspmapset[\dimregion]$ can
    be written as
    \begin{align}
        \label{eq:friis1D}
        \pmap(\loc) = \pmap(\locx) = \sum_{\srcind=1}^{\srcnum}\frac{\srccoef[\srcind]}{
            (\locx - \srclocx[\srcind])^2 + \rdists[\srcind]   },
    \end{align}
    where $\srclocx[\srcind]$ and $\rdist[\srcind]$ are such that
    \begin{bullets}
        \blt $\srcloc[\srcind] =[ \srclocx[\srcind]; \srclocy[\srcind]; \srclocz[\srcind]]\in \srcregion$
        \blt and $\rdist[\srcind] =\sqrt{ \srclocy[\srcind]^2 + \srclocz[\srcind]^2}$.
    \end{bullets}

    \subsubsection{Spatial Change Rate of Power Maps}

    \blt[bounded derivatives]
    \begin{bullets}%
        \blt[Statement]

        The first  result  upper bounds the first derivative of power maps.
        \begin{mylemma}
            \label{prop:boundedderivative}
            Let
            $\mapregion[1]$ be open and let
            $\pmap \in \fspmapset[1]$. Then,
            \begin{align}
                \label{eq:boundedderivative}
                |\pmap'(\locx)|\leq
                \frac{3^{3/2}}{8} \sum_{\srcind=1}^{\srcnum}
                \frac{\srccoef[\srcind]
                }{ \rdist[\srcind]^3  }.
            \end{align}
        \end{mylemma}

        \begin{IEEEproof}
            See Appendix~\ref{proof:boundedderivative}.
        \end{IEEEproof}

        \blt[tightness]The bound in \Cref{prop:boundedderivative} is
        tight. It can be seen that equality is attained for a specific arrangement
        where all the sources lie on a plane that is perpendicular to $\lineset$.

        \blt[rewrite]To facilitate the  interpretation of \eqref{eq:boundedderivative}, recall  that  $\srccoef[\srcind]$ can be expressed as $\srccoef[\srcind]\define \txpow[\srcind]( \wavelen/4\pi)^2$, where $\txpow[\srcind]$ is the transmitted power of the $\srcind$-th source. Thus, \eqref{eq:boundedderivative} can be written~as
        \begin{align}
            \label{eq:boundedderivative2}
            |\pmap'(\locx)|\leq \frac{3^{3/2}}{128\pi^2}   \wavelen^2\sum_{\srcind=1}^{\srcnum}
            \frac{\txpow[\srcind]
            }{ \rdist[\srcind]^3  }.
        \end{align}
        \blt[observations] %Several observations are in order.
        \begin{bullets}%
            % \blt[wavelength]
            % First,  the rate at which power changes
            % over space increases with the wavelength, which is consistent with the fact that  low-frequencies tend to spread more evenly across space.
            %
            \blt[distance and power]\begin{changes}Observe that  this
                rate decreases cubically with the distance $\rdist[\srcind]$ from the sources to
                $\lineset$ while it increases linearly with the transmitted power. Thus, the influence of the distance to the sources is much more significant: reducing $\rdist[\srcind]$ by a factor of 2 has the same effect as increasing $\txpow[\srcind]$ by a factor of 8.\end{changes}
            \blt[sensor locations] Also, the fact that the derivative of $\pmap$ in \eqref{eq:friis1D} decreases to zero as $\locx$ becomes arbitrarily farther away from $\srclocx$ implies that the largest variability occurs in the vicinity of sources. By the above considerations, this variability is largest near the sources that lie close to $\lineset$. This suggests that  \emph{radio map estimators will generally benefit from collecting a larger number of measurements in those parts of  $\lineset$ that are near the sources.} \changeo{Interestingly, this is fully consistent with the WTI, which predicts that a larger spatial density of sensors is required near the sources~\cite[Secs. 8.5.2 and 8.6]{franceschetti2018}.}
        \end{bullets}% 

        \blt[re-rewrite]        It is also interesting to express \eqref{eq:boundedderivative2} after
        normalization by $\wavelen$. In particular, consider the normalized distances  $\normlocx \define
            \locx/\wavelen$ and
        $\normrdist[\srcind]\define \rdist[\srcind]/\wavelen$.
        The radio map expressed in terms of $\normlocx$ becomes $\normpmap(\normlocx)\define \pmap(\wavelen\locx)$ and its derivative satisfies
        $\normpmap'(\normlocx)\define d \normpmap(\normlocx)/d \normlocx = (d
            \pmap(\wavelen\locx)/d \locx) (d \locx / d\normlocx ) = \wavelen
            \pmap'(\wavelen\locx)$. Thus, it follows from \eqref{eq:boundedderivative2} that
        \begin{align}
            \label{eq:boundedderivativenorm}
            |\normpmap'(\normlocx)|\leq
            \frac{3^{3/2}}{128\pi^2}\sum_{\srcind=1}^{\srcnum}
            \frac{\txpow[\srcind]
            }{  \normrdist[\srcind]^3  }.
        \end{align}
        As expected from electromagnetic theory, this expression no longer depends on $\wavelen$. Thus, the variability of a power map in the scale of the wavelength is just dependent on the distance of the sources to $\lineset$ \emph{in units of} $\wavelen$. The RME problem is invariant to scaling  both $\wavelen$ and all distances  by the same factor. This means, for instance, that if one decreases $\wavelen$ and wishes to attain the same estimation performance, the distance between measurements needs to be decreased by the same factor. Conversely, for a given set of measurement locations, the estimation performance will be worse the shorter $\wavelen$ is.

    \end{bullets}%  

    \begin{changes}
        \blt[variability bounds]
        \begin{bullets}%
            \blt[theorem]The next result provides a different view on the variability of radio maps in $\fspmapset[1]$. Unlike \Cref{prop:boundedderivative}, which depends on the parameters of each source, the next result provides bounds on the values that a radio map can take at one point given the value that it takes at another point:
            \begin{mytheorem}
                \label{prop:varbounds}
                Let
                $\pmap \in \fspmapset[1]$.                If $[\locx],
                    [\locx+\locxdelta{}]\in \mapregion[1]
                $, then
                \begin{align}
                    \label{eq:varbounds}
                    \pmap(\locx)
                    \frac{\boundfun(\locxdelta{})-1}{\boundfun(\locxdelta{})+1}\leq
                    \pmap(\locx+\locxdelta{})
                    \leq
                    \pmap(\locx)
                    \frac{\boundfun(\locxdelta{})+1}{\boundfun(\locxdelta{})-1},
                \end{align}
                where
                \begin{align}
                    \boundfun(\locxdelta{})\define \sqrt{1+4\left[\frac{\mindist}{\locxdelta{}}\right]^2}.
                \end{align}
                Furthermore, if $\srcregion[1]=\rfield^1$ and
                $\mindist^2\in\rdistsset([\srclocx])~\forall\srclocx$,
                the bounds in \eqref{eq:varbounds} are tight, which means that, given $\locx$, $\locxdelta{}$, and $\pmap(\locx)$, there exists $\pmap\in \fspmapset[1]$ that satisfies either bound in \eqref{eq:varbounds} with equality.

            \end{mytheorem}
            \begin{IEEEproof}
                See Appendix~\ref{sec:varbounds}.
            \end{IEEEproof}

            \blt[Figure]Fig.~\ref{fig:varibounds} illustrates the bounds in \eqref{eq:varbounds} for an example of power map when $\locx=0$. The areas above the upper bound and below the lower bound are forbidden regions.

            \blt[Observations]
            \begin{bullets}%        
                \blt[dependence]When seen as functions of $\locxdelta{}$ for fixed $\pmap(\locx)$, the only parameter governing the bounds in \eqref{eq:varbounds} is $\mindist$. Thus, when it comes to the relative change
                $\pmap(\locx+\locxdelta{})/\pmap(\locx)$, the main factor determining the maximum variability of $\pmap$ is the minimum distance between the sources and the mapped region.
                \blt[total variability]Furthermore, since the lower bound increases with $\mindist$ whereas the upper bound decreases with $\mindist$,
                the maximum variability of $\pmap$ is largest when $\mindist$ is smallest.

            \end{bullets}%      

            \begin{figure}[t]
                \centering
                \includegraphics[width=\columnwidth]{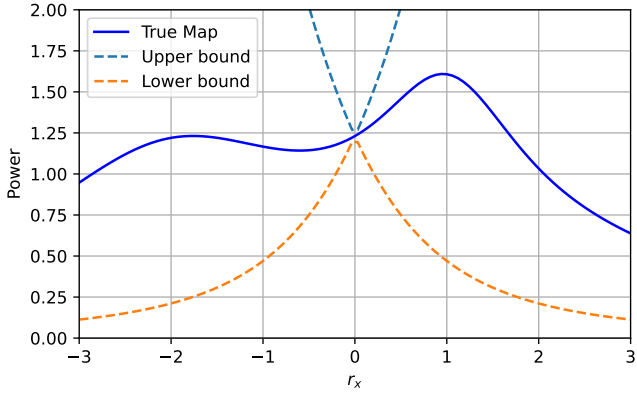}
                \caption{
                    \begin{changes}
                        Illustration of the bounds on the variability of a radio map provided by \Cref{prop:varbounds}. Given the value of $\pmap$ at a point $\locx$ ($\locx=0$ in the figure), the values that $\pmap$ can take at any other point are restricted by \eqref{eq:varbounds}. Thus, the areas above the upper bound and below the lower bound are forbidden regions.
                    \end{changes}
                }
                \label{fig:varibounds}
            \end{figure}

        \end{bullets}%

        \blt[density in $C_+$]
        \begin{bullets}%
            \blt[R1]

            Recall that \Cref{prop:approxdiff} established that the set of \emph{differences} of power maps is dense in the space of continuous functions. It was mentioned that it would be highly problematic if this applied also to the set of power maps themselves. Fortunately, the following follows from \Cref{prop:varbounds}:
            \begin{mycorollary}
                \label{prop:density}
                $\fspmapset[1]$ is not dense in the space of positive continuous functions defined on $\mapregion[1]$ and with the uniform metric.
            \end{mycorollary}
            \begin{IEEEproof}
                Trivial.
                % Choose $\locx$            
                % and $\locxdelta{}$ such that $\locx,\locx + \locxdelta{}\in\mapregion[1]$. Then construct a continuous function violating the bounds in \eqref{eq:varbounds}. 
            \end{IEEEproof}
            In other words, there are continuous functions for which a power map cannot be found that is arbitrarily close to that function.

            \blt[circle] A different proof technique can be used to establish a similar result for the case where the map is defined on a circle.
            \begin{mycorollary}
                \label{prop:densitycircle}
                Let $\mapregion[2]=\{[\locx,\locy]~|~\locx^2 + \locy^2 = \radius \}$, where $\radius>0$. Then $\fspmapset[2]$ is not dense in the space of positive continuous functions defined on $\mapregion[2]$.
            \end{mycorollary}
            \begin{IEEEproof}
                W.l.o.g. assume that $\radius=1$.
                It follows from \eqref{eq:minsrcregdist} that the series in \cite[eq. (0.8)]{hayman1990bases} contains only a finite number of terms, which implies that the resulting series cannot diverge. Hence, it follows from \cite[Th. 2]{hayman1990bases} that $\fspmapset[2]$ is not dense in the space of positive continuous functions defined on $\mapregion[2]$.
            \end{IEEEproof}
        \end{bullets}%

    \end{changes}

    \subsubsection{Spatial Bandwidth of Power Maps}

    \blt[approx. lowpass] The rest of the section establishes that radio
    maps in free space are approximately lowpass in terms of spatial frequency. This is not only relevant for purely theoretical reasons, but it is also important to motivate the usage of estimators that rely on this property. Such estimators would go along the lines of  what is discussed in \cite[Ch. 8]{franceschetti2018} about the spatial bandwidth of the electromagnetic field itself.

    \begin{bullets}
        \blt[Map FT]Consider the Fourier transform of $\pmap$:
        \begin{align}
            \pmapft(\wnx) \define \int_{-\infty}^{\infty}\pmap(\locx)e^{-j\wnx \locx}d\locx,
        \end{align}
        where $\wnx$ is the spatial frequency.
        \blt[FT]% \begin{journalonly}%
        %     Although the bounds in \eqref{eq:initialboundpmapft} and
        %     \eqref{eq:initialboundpmapfthp} suggest that the magnitude of the
        %     Fourier transform and the high-pass energy decrease inversely
        %     proportionally to the spatial frequency, a more laborious calculation
        %     reveals that the decrease is at least exponential:
        % \end{journalonly}% 
        %
        The following result\footnote{
            It is considerably easier to  establish that $|\pmapft(\wnx)|$ decreases at least as
            $\mathcal{O}(1/|\wnx|)$ as $\wnx\rightarrow \infty$ just by relying on the identity $ j \wnx \pmapft(\wnx) =
                \int_{-\infty}^{\infty}\pmap'(\locx)e^{-j\wnx \locx}d\locx$. \Cref{prop:ftbounds} is more involved to prove but it yields a much stronger result.
        } characterizes the frequency content of~$\pmap$:
        \begin{mytheorem}
            \label{prop:ftbounds}
            Let $\rdistmin\define \min_{\srcind} \rdist[\srcind]$, $\rdistmax\define \max_{\srcind} \rdist[\srcind]$, and $\bw>0$. The following holds:
            \begin{salign}
                \label{eq:pmapftbound}
                |\pmapft(\wnx)|
                &\leq\left[\frac{\pi}{\rdistmin}\sum_{\srcind=1}^{\srcnum}\srccoef[\srcind]
                    \right]e^{-\rdistmin|\wnx|}\\
                \label{eq:hpenergyub}
                \int_{\bw}^\infty |\pmapft(\wnx)|^2 d\wnx
                &\leq
                \frac{\pi^2\srcnum\sum_{\srcind=1}^{\srcnum}\srccoef[\srcind]^2}{2\rdistmin^3}
                e^{-2\rdistmin\bw}\\
                \label{eq:energylowerbound}
                \int_{0}^\infty |\pmapft(\wnx)|^2 d\wnx
                &\geq\frac{\pi^2}{2}
                \sum_{\srcind=1}^{\srcnum}
                \frac{\srccoef[\srcind]^2}{\rdist[\srcind]^3}
                \geq
                \frac{\pi^2\sum_{\srcind=1}^{\srcnum}\srccoef[\srcind]^2}{2\rdistmax^3}.
            \end{salign}
        \end{mytheorem}

        \begin{IEEEproof}
            See Appendix~\ref{sec:ftbounds}.
        \end{IEEEproof}
        Expression \eqref{eq:pmapftbound} establishes that $\pmapft$ cannot be high-pass. More precisely, one can combine  \eqref{eq:hpenergyub} and \eqref{eq:energylowerbound} to quantify the fraction of  energy of $\pmapft$  at high frequencies:
        \begin{salign}
            %\label{eq:hpenergyub}
            \frac{\int_{\bw}^\infty |\pmapft(\wnx)|^2 d\wnx}
            {
                \int_{0}^\infty |\pmapft(\wnx)|^2 d\wnx
            }
            &\leq
            \srcnum\left[\frac{\rdistmax}{\rdistmin}\right]^3
            e^{-2\rdistmin\bw}.
            %\label{eq:energylowerbound}            
        \end{salign}
        This shows that the energy of $\pmapft$ is concentrated at low frequencies. Furthermore, this concentration becomes exponentially more pronounced as
        $\bw$ increases. Besides, by increasing
        $\rdistmin$,  the concentration of the energy of $\pmapft$ at \emph{low} frequencies rapidly grows. \changeo{Finally, it is also worth pointing out that the WTI also uses the relation between the counterparts of $\rdistmin$ and $\rdistmax$ therein to quantify the complexity of the field through a notion of spatial bandwidth~\cite[Eq. (8.75)]{franceschetti2018}.}

    \end{bullets}
\end{bullets}

\section{Reconstruction Error Bounds}
\label{sec:reconstruction}
\begin{table*}[!t]
    \renewcommand{\arraystretch}{1.3}
    \caption{Upper bounds on the reconstruction error}
    \label{tab:bounds}
    \centering
    \begin{changes}
        \begin{tabular}{|c||c|c|c|}
            \hline
                                                                                                             & Zeroth-order interpolation & First-order interpolation & Sinc
            interpolation                                                                                                                                                    \\ \hline
            \multirow{2}{*}{Interpolator}                                                                    &
            \multirow{2}{*}{\begin{tabular}[c]{@{}c@{}}$\pmapest(\locx)\define
                                    \pmapsv{\spind},$ \\
                                    $\text{where}~\spind=\argmin_{\spind'}|\locx
                                        -\locxsp{\spind'}|$\end{tabular}}                       &
            \multirow{2}{*}{$\pmapest(\locx)\define \frac{
                        \pmapdelta{\spind}
                    }{
                        \locxdelta{\spind}
                    }
            (\locx - \locxsp{\spind})+\pmapsv{\spind}$}                                                      &
            \multirow{2}{*}{$\pmapest\sigshiftnot{\sigshift}(\time)\define
                    \sum_{\sampind=-\infty}^{\infty}\sig\left(\locxsp{\spind}\sigshiftnot{\sigshift}\right)\sinc\left(\frac{\time-\locxsp{\spind}\sigshiftnot{\sigshift}}{\sampint}\right)$}
            \\&&&\\\hline\hline

            \multirow{2}{*}{$L^1$}                                                                           &
            \multirow{2}{*}{$\frac{3\sqrt{3}}{32}
                    \prox
            \sum_{\spind=1}^{\spnum-1} \locxdelta{\spind}^2$}                                                &
            \multirow{2}{*}{$\frac{27\sqrt{3}}{256}  \prox \sum_{\spind=1}^{\spnum-1} \locxdelta{\spind}^2$} &
            \multirow{2}{*}{}                                                                                                                                                \\ &&&\\\hline

            \multirow{2}{*}{$L^2$}                                                                           &
            \multirow{2}{*}{$\frac{3}{16} ~\prox~
            \sqrt{\sum_{\spind=1}^{\spnum-1} \locxdelta{\spind}^3}$}                                         &
            \multirow{2}{*}{$\sqrt{\frac{144\sqrt{2} - 117}{2048 }}\prox \sqrt{
            \sum_{\spind=1}^{\spnum-1}\locxdelta{\spind}^3}$}                                                &
            \multirow{2}{*}{$\frac{\pi\srcnum\sum_{\srcind=1}^{\srcnum}\srccoef[\srcind]^2}{\rdistmin^3}
            e^{-2\pi\rdistmin/\sampint}$}                                                                                                                                    \\ &&&\\\hline

            \multirow{2}{*}{$L^\infty$}                                                                      &
            \multirow{2}{*}{$\frac{3^{3/2}}{16} \prox
            \max_{\spind}\locxdelta{\spind}$}                                                                &
            \multirow{2}{*}{$\frac{3^{3/2}}{16} \prox
            \max_{\spind}\locxdelta{\spind}$}                                                                &
            \multirow{2}{*}{}                                                                                                                                                \\ &&&\\\hline
        \end{tabular}
    \end{changes}
\end{table*}

\begin{bullets}
    \blt[Overview]This section analyzes the reconstruction performance of three simple radio map estimators.
    The analysis for more sophisticated algorithms will be addressed by future publications. \begin{changes}The obtained bounds are summarized in Table~\ref{tab:bounds}. \end{changes}

    \blt[assumptions]The reconstruction error has multiple components. One is due to the specific variability of radio maps, which was quantified in Sec.~\ref{sec:variability}. Another  is due to measurement noise and occurs in any interpolation problem. %In practice, there are further sources of error, such as the uncertainty in the measurement locations.
    \begin{bullets}%
        \blt[no noise]To focus on the first  of these components, it will be assumed that $\noise_\spind=0$ for all~$\spind$.
        %\blt[exact location] $\locfullmeas_\spind=\locfull_\spind$
    \end{bullets}

    \blt[problem]%
    \begin{bullets}%
        \blt[original]Recall that the RME problem formulation from Sec.~\ref{sec:problem} is to
        estimate $\pmap$ given  $\{(\locfull_\spind,
            \meas_\spind)\}_{\spind=1}^{\spnum}$, where $\locfull_\spind\in\mapregion~\forall\spind$.
        \blt[restriction]Focusing on the 1D  restriction introduced in Sec.~\ref{sec:restrictions}, one can rewrite this formulation as estimating $\pmap$ given  $\{(\locxsp{\spind},
            \meas_\spind)\}_{\spind=1}^{\spnum}$, where $\locxsp{\spind}\in\mapregion[1]$ is the x-coordinate of the $\spind$-th measurement location.
        \blt[general index set]However, to simplify some expressions, it is convenient to also allow a countable set of measurements. Thus, the problem will be reformulated as estimating $\pmap$ given  $\{(\locxsp{\spind},
            \meas_\spind)\}_{\spind\in\spindset}$, where $\locxsp{\spind}\in\mapregion[1]~\forall\spind$ and $\spindset\subset\mathbb{Z}$ is a (possibly infinite) countable set of indices. Obviously, the previous formulation is recovered by setting $\spindset=\{1,\ldots,\spnum\}$.
        \blt[sorting]Besides, it will be assumed  w.l.o.g. that $\locxsp{\spind}<\locxsp{\spind+1}$ for all $\spind$.
    \end{bullets}%

    % $\locfull\in\mapregion$.
    % \begin{journalonly}%

    %     The function $\pmap$ is observed at points
    %     $\locxspset\define\{\locxsp{\spind}\}_{\spind\in\spindset}$, where $\spindset\subset
    %         \mathbb{Z}$ contains the indices of those points and where
    %     $\locxsp{\spind}<\locxsp{\spind+1}$ for all $\spind$.
    % \end{journalonly}%
    % The problem is to
    % reconstruct $\pmap$ given
    % $\{(\locxsp{\spind},\pmapsv{\spind})\}_{\spind=1}^\spnum$, where
    % $\pmapsv{\spind}= \pmap(\locxsp{\spind})$.
    % For notational simplicity, it will be further assumed that  the measurement locations
    % $\locxspset\define\{\locxsp{\spind}\}_{\spind=1}^{\spnum}$ are sorted so that
    % $\locxsp{\spind}<\locxsp{\spind+1}$ for all $\spind$.

    \blt[metrics]The performance metrics to be investigated are the  conventional $L^1$ and $L^2$ norms used in Lebesgue spaces as well as the $L^\infty$ norm used in spaces of continuous bounded functions:
    \begin{salign}[eq:errmetrics]
        \|\pmap -\pmapest\|_1 &\define
        \int_{\mapregion[1]}|\pmap(\locx)-\pmapest(\locx)|d\locx\\
        \|\pmap -\pmapest\|_2^2 &\define
        \int_{\mapregion[1]}|\pmap(\locx)-\pmapest(\locx)|^2d\locx\\
        \|\pmap -\pmapest\|_\infty &\define
        \sup_{\locx\in{\mapregion[1]}}|\pmap(\locx)-\pmapest(\locx)|.
    \end{salign}
    %In this case, the integrals can be thought of as Riemann integrals since  both $\pmap$ and $\pmapest$ are continuous.

    \blt[proximity]Many of the bounds will be seen to be increasing
    functions of the following quantity, which will be referred to as the
    \emph{proximity coefficient}:
    \begin{align}
        \label{eq:proxcoeff}
        \prox \define \sum_{\srcind=1}^{\srcnum}
        \frac{\srccoef[\srcind]
        }{ \rdist[\srcind]^3  } = \left(\frac{\wavelen}{4\pi}\right)^2
        \sum_{\srcind=1}^{\srcnum}
        \frac{\txpow[\srcind]
        }{ \rdist[\srcind]^3  }.
    \end{align}
    In view of this weighted sum of the terms $1/\rdist[\srcind]^3$, one will conclude that  a poor estimation performance is expected if relatively
    strong sources are near the mapped region. This agrees with  the findings in Sec.~\ref{sec:variability}.

\end{bullets}

\subsection{Zeroth-Order Interpolation}
\label{sec:nninterp}

\begin{bullets}

    \blt[problem] Suppose that  $\spindset=\{1,\ldots,\spnum\}$ and\footnote{Strictly speaking, $\mapregion[1]$ does not contain $\locxsp{\spnum}$, which violates the assumptions in Sec.~\ref{sec:problem}. This has clearly no impact, it just simplifies the exposition.
    } $\mapregion[1]=[\locxsp{1},\locxsp{\spnum})$.
    \blt[interpolator] The zeroth-order interpolator considered here is the \emph{nearest-neighbor} estimator.  For each $\locx$, this estimator produces
    \begin{align}
        \label{eq:nnest}
        \pmapest(\locx)\define
        \pmapsv{\spind},~\text{where}~\spind=\argmin_{\spind'}|\locx -\locxsp{\spind'}|.
    \end{align}

    \blt[performance]

    \begin{mytheorem}
        \label{prop:nninterperr}
        Let $\pmapest$ be given by \eqref{eq:nnest} and let $\locxdelta{\spind}\define \locxsp{\spind+1} - \locxsp{\spind}$. Then:
        \begin{salign}[eq:nninterperrsb]
            \label{eq:nninterpl1errsb}
            \|\pmap - \pmapest\|_1 &\leq
            \begin{intermed}
                \frac{\deriub}{4} \sum_{\spind=1}^{\spnum-1} \locxdelta{\spind}^2
                =
                \frac{1}{4}\frac{3^{3/2}}{8} \sum_{\srcind=1}^{\srcnum}
                \frac{\srccoef[\srcind]
                }{ \rdist[\srcind]^3  }
                \sum_{\spind=1}^{\spnum-1} \locxdelta{\spind}^2
                =
                \frac{3\sqrt{3}}{32} \sum_{\srcind=1}^{\srcnum}
                \frac{\srccoef[\srcind]
                }{ \rdist[\srcind]^3  }
                \sum_{\spind=1}^{\spnum-1} \locxdelta{\spind}^2
                =
            \end{intermed}
            \frac{3\sqrt{3}}{32}
            \prox
            \sum_{\spind=1}^{\spnum-1} \locxdelta{\spind}^2
            \jumpline
            \begin{intermed}
                \|\pmap -\pmapest\|_2^2
            \end{intermed}
            \alignchar
            \begin{intermed}
                \leq
                \frac{\deriub^2}{12} \sum_{\spind=1}^{\spnum-1} \locxdelta{\spind}^3
                = \frac{1}{12} \bigg(\frac{3^{3/2}}{8}\bigg)^2 \bigg[\sum_{\srcind=1}^{\srcnum}
                    \frac{\srccoef[\srcind]
                    }{ \rdist[\srcind]^3  } \bigg]^2
                \sum_{\spind=1}^{\spnum-1} \locxdelta{\spind}^3
                \nonumber
                =
            \end{intermed}
            \jlac
            \begin{intermed}
                \frac{1}{12} \frac{3^{3}}{8^2} \bigg[\sum_{\srcind=1}^{\srcnum}
                    \frac{\srccoef[\srcind]
                    }{ \rdist[\srcind]^3  } \bigg]^2
                \sum_{\spind=1}^{\spnum-1} \locxdelta{\spind}^3
            \end{intermed}
            \jlac
            \begin{intermed}
                =
                \frac{9}{256} \bigg[\sum_{\srcind=1}^{\srcnum}
                    \frac{\srccoef[\srcind]
                    }{ \rdist[\srcind]^3  } \bigg]^2
                \sum_{\spind=1}^{\spnum-1} \locxdelta{\spind}^3
                =
                \frac{9}{256} \prox^2
                \sum_{\spind=1}^{\spnum-1} \locxdelta{\spind}^3
            \end{intermed}
            \\
            \label{eq:nninterpl2errsb}
            \|\pmap -\pmapest\|_2 &\leq
            \frac{3}{16} ~\prox~
            \sqrt{\sum_{\spind=1}^{\spnum-1} \locxdelta{\spind}^3}
            \\
            \label{eq:nninterplinferrsb}
            \|\pmap -\pmapest\|_\infty &\leq
            \begin{intermed}%
                \frac{
                    \deriub
                }{
                    2
                }\max_{\spind}\locxdelta{\spind}
                =
            \end{intermed}%
            \jlac
            \begin{intermed}
                \frac{3^{3/2}}{16} \sum_{\srcind=1}^{\srcnum}
                \frac{\srccoef[\srcind]
                }{ \rdist[\srcind]^3  }
                \max_{\spind}\locxdelta{\spind}
                =
            \end{intermed}
            \frac{3^{3/2}}{16} \prox
            \max_{\spind}\locxdelta{\spind}.
        \end{salign}
    \end{mytheorem}
    \begin{IEEEproof}
        See Appendix~\ref{sec:nninterperr}.
    \end{IEEEproof}

    \blt[Observations]
    \begin{bullets}%

        \blt[sampling spacing]First, observe that the error becomes 0 if  $\locxdelta{\spind}\rightarrow 0~\forall \spind$. This is expected since  $\pmap$ is continuous.
        \blt[coefficient] Second,  the bounds for all these metrics depend on the quantities
        defining the map (wavelength, transmit power, and source position) only through the proximity coefficient $\prox$, which therefore
        condenses the impact of these magnitudes effectively.

        %the three metrics depend on the coefficient $\sum_{\srcind=1}^{\srcnum}
        %     ({\srccoef[\srcind]
        %     }/{ \rdist[\srcind]^3  })
        %     =( \wavelen/4\pi)^2 \sum_{\srcind=1}^{\srcnum}( \txpow[\srcind]       /{ \rdist[\srcind]^3  })
        % $
        %
        \blt[relative error]Applying Parseval's theorem to \eqref{eq:energylowerbound} yields
        \begin{salign}
            \|\pmap\|_2^2 &\define
            \int_{-\infty}^\infty |\pmap(\locx)|^2 d\locx
            \begin{intermed}%
                =
                \frac{1}{2\pi}
                \int_{-\infty}^\infty |\pmapft(\wnx)|^2 d\wnx
            \end{intermed}
            \jlac
            =\frac{1}{\pi}
            \int_{0}^\infty |\pmapft(\wnx)|^2 d\wnx\geq\frac{\pi}{2}
            \sum_{\srcind=1}^{\srcnum}
            \frac{\srccoef[\srcind]^2}{\rdist[\srcind]^3}.
        \end{salign}
        The relative error can therefore be upper bounded as
        \begin{align}
            \label{eq:relerr}
            \frac
            {
                \|\pmap  -
                \pmapest\|_2^2
            }{
                \|\pmap\|_2^2
            } & \leq
            \begin{intermed}%
                \frac{9}{256
                }~\frac{
                    \left[ \sum_{\srcind=1}^{\srcnum}
                        \frac{\srccoef[\srcind]
                        }{ \rdist[\srcind]^3  }
                        \right]^2
                }{
                    \frac{\pi}{2}
                    \sum_{\srcind=1}^{\srcnum}
                    \frac{\srccoef[\srcind]^2}{\rdist[\srcind]^3}
                }\sum_{\spind=1}^{\spnum-1}\locxdelta{\spind}^3
            \end{intermed}%
            \jlac
            \begin{intermed}
                =
            \end{intermed}
            \frac{9}{128\pi
            }~\frac{
                \left[ \sum_{\srcind=1}^{\srcnum}
                    \frac{\srccoef[\srcind]
                    }{ \rdist[\srcind]^3  }
                    \right]^2
            }{
                \sum_{\srcind=1}^{\srcnum}
                \frac{\srccoef[\srcind]^2}{\rdist[\srcind]^3}
            }\sum_{\spind=1}^{\spnum-1}\locxdelta{\spind}^3.
        \end{align}
        Interestingly, if $\srccoef[\srcind]=\srccoef~\forall \srcind$ and $\rdist[\srcind]=\rdist~\forall \srcind$, then the relative error bound becomes
        \begin{align}
            \label{eq:relerrsimp}
            \frac
            {
                \|\pmap  -
                \pmapest\|_2^2
            }{
                \|\pmap\|_2^2
            }\leq
            \begin{intermed}
                \frac{9}{128\pi
                }~\frac{
                    \left[ \srcnum
                        \frac{\srccoef
                        }{ \rdist^3  }
                        \right]^2
                }{
                    \srcnum
                    \frac{\srccoef^2}{\rdist^3}
                }\sum_{\spind=1}^{\spnum-1}\locxdelta{\spind}^3
                =
            \end{intermed}
            \srcnum\frac{9}{128\pi
            }\sum_{\spind=1}^{\spnum-1}\left[\frac{
                    \locxdelta{\spind}
                }{
                    \rdist}\right]^3.
        \end{align}
        This again suggests that, the closer the sources are to $\lineset$, the smaller the sample spacing $\locxdelta{\spind}$ necessary for a target relative error. It is also remarkable that \eqref{eq:relerrsimp} does not depend on the transmitted power  in this simple scenario. In fact, $\srccoef[\srcind]$ (equivalently $\txpow[\srcind]$) can be thought of as a factor in \eqref{eq:relerr} that weights the effect of each $\rdist[\srcind]$ on the error.

    \end{bullets}

\end{bullets}

\subsection{First-order Interpolation}
\label{sec:lininterp}

\begin{bullets}

    \blt[problem] As in Sec.~\ref{sec:nninterp}, let
    $\spindset=\{1,\ldots,\spnum\}$ and $\mapregion[1]=[\locxsp{1},\locxsp{\spnum})$.
    \blt[interpolator] The considered first-order interpolator is the  linear interpolator returning a function  on $\mapregion[1]$ that takes the values
    \begin{align}
        \label{eq:linest}
        \pmapest(\locx)\define \frac{
            \pmapdelta{\spind}
        }{
            \locxdelta{\spind}
        }
        (\locx - \locxsp{\spind})+\pmapsv{\spind},
    \end{align}
    where
    \begin{bullets}%
        \blt $\pmapdelta{\spind}\define\pmap(\locxsp{\spind+1})-\pmap(\locxsp{\spind})$,
        \blt $\locxdelta{\spind}\define \locxsp{\spind+1} - \locxsp{\spind}$,
        \blt and $\spind$ is the only integer
        such that $\locx\in[\locxsp{\spind},\locxsp{\spind+1})$.
    \end{bullets}%

    \blt[therorem]

    \begin{mytheorem}
        \label{prop:lininterperr}
        The estimator  $\pmapest$ defined in \eqref{eq:linest} satisfies:
        \begin{salign}[eq:lininterperrsb]
            \label{eq:lininterpl1errsb}
            \|\pmap  -
            \pmapest\|_1 &\leq
            \begin{intermed}
                \frac{9}{32} \deriub\sum_{\spind=1}^{\spnum-1} \locxdelta{\spind}^2
                =
                \frac{9}{32}  \frac{3^{3/2}}{8} \sum_{\srcind=1}^{\srcnum}
                \frac{\srccoef[\srcind]
                }{ \rdist[\srcind]^3  }\sum_{\spind=1}^{\spnum-1} \locxdelta{\spind}^2
                =
            \end{intermed}
            \jlac
            \begin{intermed}
                \frac{27\sqrt{3}}{256}   \sum_{\srcind=1}^{\srcnum}
                \frac{\srccoef[\srcind]
                }{ \rdist[\srcind]^3  }\sum_{\spind=1}^{\spnum-1} \locxdelta{\spind}^2
                =
            \end{intermed}
            \frac{27\sqrt{3}}{256}  \prox \sum_{\spind=1}^{\spnum-1} \locxdelta{\spind}^2
            \jumpline
            \begin{intermed}
                \|\pmap  -
                \pmapest\|_2^2
            \end{intermed}
            \alignchar
            \begin{intermed}%
                \leq
                \frac{16\sqrt{2}-13}{96
                }\deriub^2
                \sum_{\spind=1}^{\spnum-1}\locxdelta{\spind}^3
            \end{intermed}%
            \jlac
            \begin{intermed}
                =
                \frac{144\sqrt{2} - 117}{2048
                }\left[ \sum_{\srcind=1}^{\srcnum}
                    \frac{\srccoef[\srcind]
                    }{ \rdist[\srcind]^3  }
                    \right]^2
                \sum_{\spind=1}^{\spnum-1}\locxdelta{\spind}^3
                =
            \end{intermed}
            \jlac
            \begin{intermed}
                \frac{144\sqrt{2} - 117}{2048
                }\prox^2
                \sum_{\spind=1}^{\spnum-1}\locxdelta{\spind}^3
            \end{intermed}
            \\
            \label{eq:lininterpl2errsb}
            \|\pmap  -
            \pmapest\|_2&\leq \sqrt{\frac{144\sqrt{2} - 117}{2048
                }}\prox \sqrt{
                \sum_{\spind=1}^{\spnum-1}\locxdelta{\spind}^3
            }
            \\
            \label{eq:lininterplinferrsb}
            \|\pmap  -
            \pmapest\|_\infty
            &\leq
            \begin{intermed}%
                \frac{
                    \deriub
                }{
                    2
                }\max_{\spind}\locxdelta{\spind}
                =
                \frac{3^{3/2}}{16} \sum_{\srcind=1}^{\srcnum}
                \frac{\srccoef[\srcind]
                }{ \rdist[\srcind]^3  }
                \max_{\spind}\locxdelta{\spind}
                =
            \end{intermed}
            \jlac
            \frac{3^{3/2}}{16} \prox
            \max_{\spind}\locxdelta{\spind}.
        \end{salign}

    \end{mytheorem}

    \begin{IEEEproof}
        See Appendix~\ref{sec:lininterperr}.
    \end{IEEEproof}

    \blt[observations]
    \begin{bullets}%
        \blt[overall same] Observe that the bounds in \Cref{prop:lininterperr} are the same as in \Cref{prop:nninterperr} except for  multiplicative factors. Therefore, similar observations to those in Sec.~\ref{sec:nninterp} apply here.
        \blt[constants]However, contrary to what was expected, the constants in
        \Cref{prop:lininterperr} are in fact larger than the ones in \Cref{prop:nninterperr}. This is because the latter bounds are tighter than the former since
        the worst cases implicitly considered in the proof of \Cref{prop:lininterperr} are  more extreme. Notwithstanding, a more tedious  derivation\footnote{The idea would be to enforce continuity and the derivative bound at the midpoint of each interval $[\locxsp{\spind},\locxsp{\spind+1}]$. Then, one can maximize the worst-case error with respect to the value that $\pmap$ takes at this point. Unfortunately, the derivation becomes cumbersome due to the large number of cases that must be considered.} is expected to result in  upper bounds for first-order interpolation that are lower than those for zeroth-order interpolation.

    \end{bullets}%

\end{bullets}

\subsection{Sinc Interpolation}
\label{sec:sincinterp}

\begin{bullets}

    \blt[motivation]The sinc interpolator gives rise to an \emph{exact} reconstruction of a bandlimited signal given a set of uniformly-spaced samples that satisfy the Nyquist criterion.
    Using such a sinc interpolator for reconstructing  $\pmap$ is therefore motivated by
    \Cref{prop:ftbounds}, which establishes that $\pmap$ is approximately bandlimited. This interpolator also plays a central role  in the WTI~\cite{franceschetti2018}.

    \blt[samples]Suppose that $\pmap$ is observed at a set of uniformly-spaced locations
    $\locxspset[\sigshift]\define\{\locxsp{\spind}\sigshiftnot{\sigshift},~{\spind\in\mathbb{Z}}\}$,
    where $\locxsp{\spind}\sigshiftnot{\sigshift}\define\spind\sampint + \sigshift$ are the sampling instants corresponding to offset
    $\sigshift\in\mathbb{R}$  and $\sampint>0$ is the spatial sampling interval.
    Consequently,
    $\spindset=\mathbb Z$ and let $\mapregion[1]=\rfield$.
    In practice, $\locxspset[\sigshift]$ may correspond to a scenario where a vehicle moves along $\lineset$  and collects measurements at  regular intervals.

    \blt[interpolator]The sinc interpolator is defined as
    \begin{align}
        \label{eq:sincinterpdef}
        \pmapest\sigshiftnot{\sigshift}(\time)\define
        \sum_{\sampind=-\infty}^{\infty}\sig\left(\locxsp{\spind}\sigshiftnot{\sigshift}\right)\sinc\left(\frac{\time-\locxsp{\spind}\sigshiftnot{\sigshift}}{\sampint}\right),
    \end{align}
    where $\sig(\locxsp{\spind}\sigshiftnot{\sigshift})$ are the measurements. Consider the following:

    \begin{mytheorem}
        \label{prop:sincinterperr}
        \begin{changes}
            Let $\sig$ be a function with Fourier transform $\sigft$. If one lets
        \end{changes}
        \begin{bullets}
            \blt
            \begin{align}
                \label{eq:erroffsetsinc}
                \err(\sigshift)
                \define \|\sig -\pmapest\sigshiftnot{\sigshift}\|^2
                \define & \int_{-\infty}^{\infty}|
                \sig(\time)-\pmapest\sigshiftnot{\sigshift}(\time)|^2d\time,
            \end{align}
            then
            \begin{align}
                \label{eq:sincinterpavgerr}
                \avgerr \define\frac{1}{\sampint}\int_{0}^{\sampint}  \err(\sigshift) d\sigshift = \frac{2}{\pi}\int_{\pi/\sampint}^\infty|\sigft(\freq)|^2d\freq.
            \end{align}

        \end{bullets}
    \end{mytheorem}

    \begin{IEEEproof}
        See Appendix~\ref{sec:sincinterperr}.
    \end{IEEEproof}

    \blt[Remarks]
    \begin{bullets}
        \blt[average]This theorem establishes that the average error across all offsets $\sigshift$ is proportional to the energy of $\pmap$ outside  $[-\pi/\sampint,$ $\pi/\sampint]$. This is
        therefore an \emph{aliasing error}, since there is no physical way of spatially low-pass filtering $\pmap$ before acquiring the measurements, as would be  performed by an analog-to-digital converter (ADC) in the time domain. It can also be interpreted as the expected error when the offset is uniformly distributed in $[0,\sampint)$, which captures the fact that the measurement locations do not generally depend on the coordinate system or $\pmap$ itself.

        \blt[exact] Observe also that  \eqref{eq:sincinterpavgerr} is an equality, i.e., it is not a bound, \begin{changes} and that it applies to arbitrary power maps, not necessarily in free space. \end{changes}
        \blt[bound] Substituting \eqref{eq:hpenergyub} with $\bw=\pi/\sampint$
        in \eqref{eq:sincinterpavgerr} yields the bound
        \begin{align}
            \label{eq:sincinterpavgerrmappars}
            \avgerr \leq
            \begin{intermed}
                \frac{2}{\pi}
                \frac{\pi^2\srcnum\sum_{\srcind=1}^{\srcnum}\srccoef[\srcind]^2}{2\rdistmin^3}
                e^{-2\pi\rdistmin/\sampint}
                =
            \end{intermed}
            \frac{\pi\srcnum\sum_{\srcind=1}^{\srcnum}\srccoef[\srcind]^2}{\rdistmin^3}
            e^{-2\pi\rdistmin/\sampint}.
        \end{align}
        This bound decreases much faster than the bounds in \Cref{prop:nninterperr} and \Cref{prop:lininterperr} as $\rdistmin\rightarrow\infty$ or $\sampint \rightarrow 0$ upon setting $\locxdelta{\spind}=\sampint~\forall \spind$. Furthermore, the error in  \eqref{eq:sincinterpavgerrmappars} is the total error in $\rfield$, whereas the bounds in \Cref{prop:nninterperr} and \Cref{prop:lininterperr} apply only to a bounded interval $[\locxsp{1},\locxsp{\spnum})$. In fact, the latter bounds diverge as one considers a longer support (just let $\spnum\rightarrow \infty$ with constant $\sampint$). Thus, the performance guarantees for the sinc interpolator are much stronger.
        A more detailed comparison between these bounds is provided in Sec.~\ref{sec:experiments}.

    \end{bullets}
\end{bullets}

\begin{changes}
    \begin{myremark}
        \label{rem:onesource}
        The case of a single transmitter is a relevant special case of the general problem formulated in Sec.~\ref{sec:problem}, where the number of sources is arbitrary. Some of the results in this paper can be readily specialized to this case by setting  $\srcnum=1$. This is the case of \Cref{prop:boundedderivative} and the bounds in Sec.~\ref{sec:reconstruction}. \Cref{prop:varbounds} remains unaltered if one sets $\srcnum=1$. In contrast,  other results, such as \Cref{prop:approxdiff}, \Cref{prop:density}, and \Cref{prop:densitycircle}, inherently require an arbitrary number of transmitters, so they do not apply to the case $\srcnum=1$. Further  considerations regarding the case $\srcnum=1$ can be found in~\cite{romero2022cartography}.
    \end{myremark}
\end{changes}

%replacing $\spind\sampint + \sigshift$ with $\locxsp{\spind}\sigshiftnot{\sigshift}$

\section{Numerical Experiments}
\label{sec:experiments}
\begin{changes}
    \begin{bullets}
        \blt[Overview] This section provides experiments that empirically corroborate
        the theoretical findings of the paper.
    \end{bullets}
\end{changes}

\subsection{Tightness of the Reconstruction Error Bounds}
\begin{bullets}
    \blt[overview]
    \begin{changes}This section verifies and assesses the tightness of the bounds
        in Sec.~\ref{sec:reconstruction}.
    \end{changes}
    \blt[data] To this end, $\pmap \in \fspmapset[1]$ is
    generated by placing 3 transmitters at a
    distance $\rdist$ from $\lineset$.
    Let $\lambda=1$ to express all lengths in terms of the wavelength and consider $\{ (\srclocx[\srcind], \rdist[\srcind], \srccoef[\srcind])\}_{\srcind=1}^{\srcnum}=$ $\{(1000,\rdist,(4 \pi)^2),(5000,\rdist,(4 \pi)^2),~~~$ $(8000,$ $\rdist,(4 \pi)^2)\}$.  The  $\spnum=11$ measurement locations are
    $\locxsp{\spind}=(\spind-1) \sampint,~\spind=1,\ldots,\spnum$, where $\sampint=1000$.
    \blt[estimates]
    Using these
    measurements, each algorithm  returns an interpolated function
    $\pmapest$,
    \blt[metrics]
    which is evaluated at 1000 uniformly spaced points in the interval
    $[\locxsp{1},\locxsp{\spnum}]$ to approximate the error metrics in \eqref{eq:errmetrics}. \begin{changes}The values of these parameters are set to capture typical cases in cellular communications.\end{changes}

    \blt[experiments] Figs.~\ref{fig:nninterp} and~\ref{fig:linearinterp} depict these  metrics for zeroth- and first-order interpolation along with  their upper bounds in \eqref{eq:nninterperrsb} and \eqref{eq:lininterperrsb}, respectively. Observe that the decay rates of the bounds accurately match the decay rate
    of the corresponding error metrics.
    Second, the error decreases more
    slowly than exponential, which would manifest itself as a straight
    line.
    Also, the bounds are considerably tight: observe for example that the upper bounds
    for the $L^2$ error are lower than the $L^1$ error. As anticipated, the bounds are tighter for zeroth-order interpolation than for first-order interpolation. However, the error  for the latter is lower than for the former. Thus, first-order interpolation is preferable in terms of performance.

    \blt[Sinc interpolation] The third experiment investigates the error
    of sinc interpolation. Since the upper bound in
    \eqref{eq:sincinterpavgerrmappars} pertains to the average of the
    $L^2$ error across sampling offsets $\sigshift$
    (cf. \eqref{eq:sincinterpavgerr}), the error is approximated for 20
    different offsets uniformly spaced in $[0,\sampint)$ and then
    averaged. It is observed in Fig.~\ref{fig:sincinterp} that, for a
    sufficiently small $\rdist$, both the error metrics and the bound
    decrease at the same rate, which furthermore is seen to be
    \emph{exponential}. Thus, the decrease rate of sinc interpolation is
    much faster than for zeroth and first-order interpolation. There is,
    however, an important caveat: as described in
    Sec.~\ref{sec:sincinterp}, the bound in
    \eqref{eq:sincinterpavgerrmappars} is applicable when the sampling
    grid spans the entire real line, which  will be abbreviated as
    $\spnum=\infty$. However, in practice and in a simulation, the number
    $\spnum$ of sampling locations $\locxsp{\spind}$ is finite and,
    therefore, confined to an interval with finite length. Thus, for the
    upper bound to hold, it is necessary that the interpolation error with
    finite $\spnum$ is sufficiently close to the theoretical interpolation
    error when $\spnum=\infty$. For this to hold, the energy of $\pmap$
    must be sufficiently concentrated on the observed interval. Otherwise,
    the omitted terms in \eqref{eq:sincinterpdef}, i.e. those
    corresponding to unobserved values of
    $\sig(\locxsp{\spind}\sigshiftnot{\sigshift})$, have a significant
    impact on the interval where the error metric is being
    approximated. Fig.~\ref{fig:sincinterp} shows the sharp transition
    between both regimes when $\rdist$ increases. For sufficiently small
    $\rdist$, function $\pmap$ is concentrated in the observation
    interval. Remarkably, the transition occurs at a rather small value of
    $\rdist$: just note the difference between the scale of
    $\srclocx[\srcind]$ and the scale of the  $\rdist$ at which the
    transition occurs. Thus, although the sinc interpolator is very promising from a theoretical perspective, the finite length of the sampling interval may render it impractical.

    \begin{figure}[t]
        \centering
        \includegraphics[width=\columnwidth]{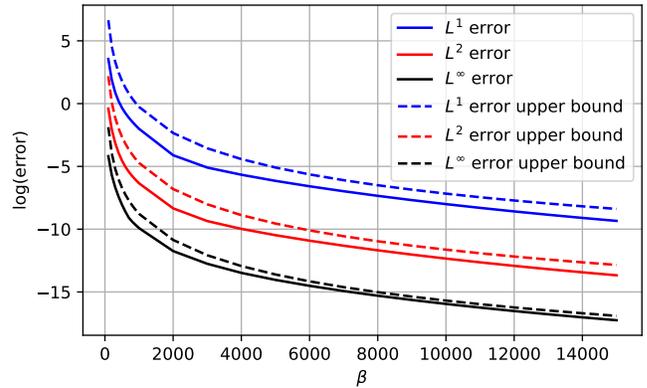}
        \caption{Error metrics along with their upper bounds \eqref{eq:nninterpl1errsb}-%\eqref{eq:nninterpl2errsb}, and 
            \eqref{eq:nninterplinferrsb} for the zeroth-order interpolation estimator \eqref{eq:nnest}.
        }
        \label{fig:nninterp}
    \end{figure}

    \begin{figure}[t]
        \centering
        \includegraphics[width=\columnwidth]{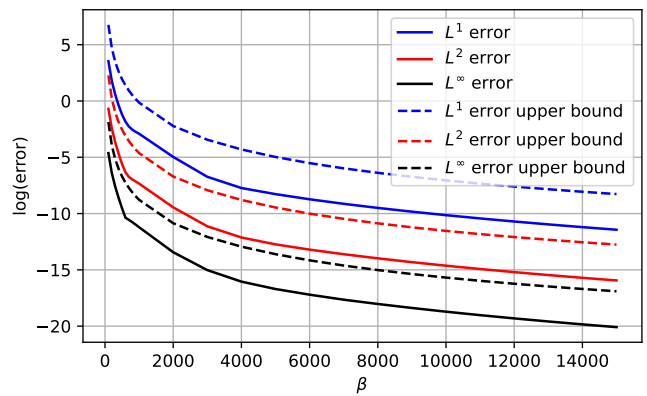}
        \caption{Error metrics along with their upper bounds \eqref{eq:lininterpl1errsb}-%, \eqref{eq:lininterpl2errsb}, and 
            \eqref{eq:lininterplinferrsb} for the first-order interpolation estimator \eqref{eq:linest}.
            %(\protect\input{\figpath{experiment_2002.txt}}).
        }
        \label{fig:linearinterp}
    \end{figure}

    \begin{figure}[t]
        \centering
        \includegraphics[width=\columnwidth]{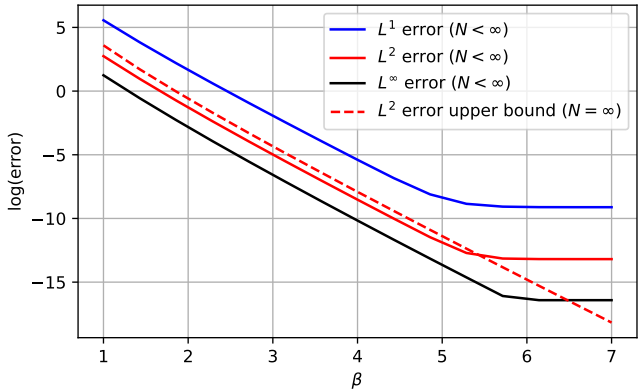}
        \caption{Error metrics along with the  upper bound for the $L^2$ error
            \eqref{eq:sincinterpavgerrmappars} for sinc interpolation.}
        \label{fig:sincinterp}
    \end{figure}

    \begin{figure}[t]
        \centering
        \includegraphics[width=\columnwidth]{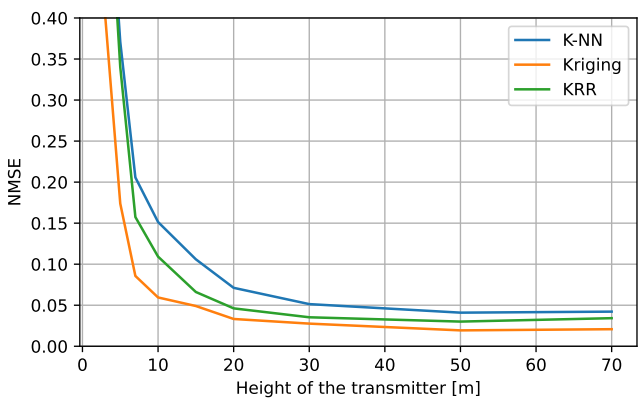}
        \caption{
            \begin{changes}
                Norm mean square error vs. transmitter heights.
            \end{changes}
        }
        \label{fig:nmsevsheight}
    \end{figure}

\end{bullets}

\subsection{Experiment with Ray-Tracing Data}
\label{sec:raytracing}
\begin{changes}

    \begin{bullets}
        \blt[Overview]This section presents an experiment where a power map is generated using ray-tracing software. The goal is to verify the claim that power maps are more difficult to estimate  the closer transmitters are to $\mapregion$ in a 2D scenario with a realistic channel model.

        \blt[Simulation parameters] To this end, a collection of power maps was generated, each one for a different height of the transmitters.
        \begin{bullets}
            \blt[Large map]The values of the map are obtained using a 3D model of  downtown  Ottawa on a rectangular grid with 1 m spacing  constructed on a horizontal
            region of size $47\times56$ m and height 2 m.
            \blt[transmitters]In all maps, 5 transmitters are deployed. The x,y-coordinates of these transmitters are the same across maps. Their z-coordinates are equal within one map but they  differ across maps.  The transmitters use isotropic antennas and operate at 2.4 GHz with  power $\txpow[\srcind]=\txpow =
                10~\text{dBm}~\forall \srcind$.

            \blt[Receivers]At each Monte Carlo iteration, a smaller map is generated by drawing a sub-region of size $32\times32$ m uniformly at random from the large map of the considered height. The locations of  $\spnum=100$ measurements collected by receivers with isotropic antennas are then drawn uniformly at random from the grid points that lie inside the sub-region but outside the buildings.

            \blt[Estimators] Three simple estimators are used:
            \begin{bullets}
                \blt[KNN] (i) The $K$-nearest neighbors estimator with $K=5$~\cite{romero2022cartography},
                \blt[Kriging] (ii) simple Kriging with $\sigma_s = 3, \delta _s=50$ m~\cite{shrestha2023empirical}, and
                \blt[KRR] (iii) kernel ridge regression (KRR) with a Gaussian kernel of width 10 m and a regularization parameter of 0.001~\cite{romero2017spectrummaps}.
            \end{bullets}

            \blt[metric]The performance metric is the  normalized mean square error (NMSE) defined as
            \begin{align}
                \text{NMSE}\define
                \frac{\expected\{\|\pmapvec  - \pmapestvec\|_2^2\}}{\expected\{\|\pmapvec\|_2^2\}},
            \end{align}
            where $\pmapvec$ and $\pmapestvec$ are respectively the vectors collecting the values of the true and estimated power maps at the grid points that lie outside buildings. The expectations are, as indicated, over choices of the sub-region and the measurement locations.
        \end{bullets}

        \blt[Results] Fig.~\ref{fig:nmsevsheight} plots the NMSE  of the three
        aforementioned estimators vs. the height of the transmitters. As
        expected, the NMSE decreases for all estimators as the transmitters
        become further away from the mapped region, which provides evidence in
        favor of the aforementioned claim.

        Although it was analytically shown
        that this always occurs in free-space and it was empirically observed
        that it holds in Fig.~\ref{fig:nmsevsheight}, it is important to note
        that this is not necessarily the case in all situations. For example, in
        a setup with $\srcnum=1$, if the transmitter is placed above the center
        of a building with a horizontal metal rooftop, the transmitted signal
        will not reach the ground  when the transmitter is right on the rooftop.
        This results in the map being identically 0 and, therefore,  the
        estimation error for the considered estimators will be 0. However, if the transmitter is a certain height above the building, the rooftop will not block the signal everywhere. Thus, the error in this case will  be strictly positive and, therefore, it will violate the claim. The conclusion is that this claim can be used as a guideline but it need not be accurate in all situations.

    \end{bullets}

\end{changes}

\section{Conclusions}
\label{sec:conclusions}
\begin{bullets}

    \blt[RME problem]This paper studied the problem of reconstructing a power
    map produced by a set of incoherent sources.
    \blt[Variability]The variability of these maps was characterized via upper and lower bounds. Remarkably, power maps are seen to be spatially low-pass.
    \blt[Interpolation]%
    \begin{bullets}%
        \blt[0th and 1st]Three function reconstruction error metrics were upper bounded for estimators based on zeroth-order, first-order, and sinc interpolation. A simple numerical experiment demonstrates that the bounds are tight and accurately predict the decrease rate with respect to the distance of the sources to the mapped region.
        \blt[proximity coef]This justifies the introduction of the \emph{proximity coefficient}, which is proportionally related to most of the reconstruction bounds and indicates that the difficulty of the RME problem increases with the transmitted power and decreases with the distance from the sources to the mapped region.
        \blt[sinc] The analysis suggests that the sinc interpolator results in a much smaller reconstruction error than zeroth- and first-order interpolators. However, the finite length of the sampling interval in practice implies that the error of the sinc interpolator will be significantly large unless the sources are very close to the mapped region.
        \blt[ray-tracing]\begin{changes}An experiment with ray-tracing  data reveals that the difficulty of the RME problem also tends to increase with the proximity of the sources in non-free space propagation environments.\end{changes}

        \blt[limitations]Being the first theoretical analysis in this context, this work suffers from several limitations. As a result, future work may address the estimation of radio maps in higher dimensions and account for noise, correlation among the transmitters, and propagation effects such as reflection, refraction, absorption, and diffraction. Bounds for more sophisticated estimators would also be of interest. It is thus the hope of the authors that this paper opens the door to a fertile research topic in this context.

    \end{bullets}

    % \blt[Future directions]
    % \begin{journalonly}
    %     \begin{bullets}
    %         \blt[further channel models]

    %         \blt[multipath]

    %         \blt[noisy measurements]

    %         \blt[sinc interp in finite-length intervals] Error
    %         analysis. Countermeasures.

    %         \blt[parametric] Error of parametric methods.  \ra assume known
    %         locations and use e.g. LS

    %         \blt[db scale]

    %     \end{bullets}
    % \end{journalonly}
\end{bullets}

\printmybibliography

%\clearpage
\appendices
\section{Proof of Theorem~\ref{prop:approxdiff}}
\label{sec:approxdiff}

\begin{bullets}

    % \blt[extension] Start by considering the extension of $\mapregion[\dimregion]$ to $\rfield^3$ given by
    % \begin{align}
    % \mapregion = \{[\locx;\locy;\locz]\in\rfield^3~|~[\locx]\in \mapregion[\dimregion]~\text{and}~\locy=\locz=0\}
    % \end{align}
    % if $\dimregion=1$ and 
    % \begin{align}
    % \mapregion = \{[\locx;\locy;\locz]\in\rfield^3~|~[\locx;\locy]\in \mapregion[\dimregion]~\text{and}~\locz=0\}
    % \end{align}
    % if $\dimregion=2$.

    \blt[subset] If conditions \ref{eq:condsubset} and
    \ref{eq:condrdists} hold, it is clear that $\fspmapset[\dimregion]$
    contains the set
    \begin{align}
        \label{eq:rdistssubset}
        \fspmaprdset{\dimregion} \define \bigg\{  \pmap ~| & ~
        \pmap(\loc) = \sum_{\srcind=1}^{\srcnum}\frac{\srccoef[\srcind]}{
        \|\loc - \srcloc[\srcind]\|^2+\rdists  },              \\&~\srcloc[\srcind] \in \mapregion[\dimregion],~\srccoef[\srcind]\geq 0,~\srcnum\in \mathbb{N}
        \bigg\}.\nonumber
    \end{align}

    \blt[rkhs]

    \begin{mylemma}
        \label{prop:rkhs} For any $\rdists>0$, $\spanc~
            \fspmaprdset{\dimregion}$ is a reproducing-kernel Hilbert space
        (RKHS) with kernel
        \begin{align}
            \label{eq:kerneldef}
            \kernel(\loc,\locp)\define  \frac{1}{
                \|\loc - \locp\|^2+\rdists  }.
        \end{align}
        Besides, $\kernel$ is universal.
    \end{mylemma}

    \begin{IEEEproof}
        It was shown in~\cite{schoenberg1938metric} that
        functions of the form $\kernel(\loc,\locp)=\auxfun(\|\loc -
            \locp\|^2)$ are positive definite kernels if $\auxfun$ can be
        written as
        \begin{align}
            \label{eq:schoenbergcond}
            \auxfun(t)= \int_{\rfield_+} e^{-t\sigma}d\measure(\sigma)
        \end{align}
        for some finite Borel measure $\measure$. Noting that, in
        \eqref{eq:kerneldef}, $\auxfun(t) = 1/(t+\rdists)$ and selecting
        $\measure$ such that
        \begin{align}
            \label{eq:measuredef}
            \measure(\auxset)= \int_{\auxset} e^{-\rdists\sigma}d\sigma
        \end{align}
        for all Borel sets $\auxset\subset \rfield_+$, it is easy to see
        that \eqref{eq:schoenbergcond} holds for $\kernel$ in
        \eqref{eq:kerneldef}. Therefore, $\kernel$ in
        \eqref{eq:kerneldef} is a positive definite kernel.

        Noting that
        \begin{align}
            \label{eq:spanfspmapset}
            \spanv\fspmaprdset{\dimregion} = \bigg\{\pmap~| & ~
            \pmap(\loc) = \sum_{\srcind=1}^{\srcnum}\frac{\srccoef[\srcind]}{
                \|\loc - \srcloc[\srcind]\|^2+\rdists  }
            ,                                                   \\&~\srcloc[\srcind] \in \mapregion[\dimregion],~\srccoef[\srcind]\in
            \cfield,~\srcnum\in \mathbb{N}
            \bigg\}\nonumber
        \end{align}
        shows that $\spanc~\fspmaprdset{\dimregion}$ is an RKHS with kernel $\kernel$.

        Finally, observe that the Borel measure associated with $\kernel$
        in \eqref{eq:kerneldef} is not concentrated at zero; cf.
        \eqref{eq:measuredef}. Thus, it follows from \cite[Theorem
            17]{micchelli2006universal} that $\kernel$ is universal.

    \end{IEEEproof}

    \blt[approximate $\pmap$] Due to the universality of $\kernel$,
    for any $\tol>0$ and continuous $\pmap$, there exists $\auxpmap\in
        \spanc~\fspmaprdset{\dimregion}$ such that $\sup_{\loc \in
            \mapregion[\dimregion]}|\pmap(\loc)-\auxpmap(\loc)|<\tol/2$.
    Since $\auxpmap\in \spanc~\fspmaprdset{\dimregion}$, it follows that there
    exists $\auxpmapt\in \spanv~\fspmaprdset{\dimregion}$ such that
    $\sup_{\loc \in\mapregion[\dimregion]
        }|\auxpmap(\loc)-\auxpmapt(\loc)|<\tol/2$.
    Besides, since $\pmap$ takes real values, it follows that
    \begin{salign}
        \label{eq:infndist}
        % \sup_{\loc \in
        % \mapregion}|\pmap(\loc)-\reb{\auxpmapt(\loc)}|&\leq
        \sup_{\loc \in\mapregion[\dimregion]}&|\pmap(\loc)-\reb{\auxpmapt(\loc)}|\leq
        \sup_{\loc \in\mapregion[\dimregion]}|\pmap(\loc)-\auxpmapt(\loc)|\\
        &\leq
        \sup_{\loc \in\mapregion[\dimregion]}|\pmap(\loc)-\auxpmap(\loc)|
        +\sup_{\loc \in\mapregion[\dimregion]}
        |\auxpmap(\loc)-\auxpmapt(\loc)|\\
        &\leq \frac{\tol}{2}+\frac{\tol}{2}=\tol,
    \end{salign}
    where the second inequality follows from the triangle inequality and the properties of $\sup$.

    Thus, it remains only to write $\reb{\auxpmapt}$ as the
    difference between two functions in $\fspmapset[\dimregion]$. To this end,
    note that, since $\auxpmapt\in \spanv~\fspmaprdset{\dimregion}$, it follows
    (cf. \eqref{eq:spanfspmapset}) that there exist
    $ \srcloc[\srcind] \in \mapregion[\dimregion],~\srccoef[\srcind]\in
        \cfield$ and $\srcnum\in \mathbb{N}$ such that
    $\auxpmapt(\loc) = \sum_{\srcind=1}^{\srcnum}\srccoef[\srcind]
        \kernel(\loc,\srcloc[\srcind])$. If $\pmap_+$ and $\pmap_-$ are
    such that
    \begin{salign}
        \pmap_+(\loc)&=  \sum_{\srcind=1}^{\srcnum}\max(0,\reb{\srccoef[\srcind]})
        \kernel(\loc,\srcloc[\srcind])\\
        \pmap_-(\loc)&=  \sum_{\srcind=1}^{\srcnum}\max(0,-\reb{\srccoef[\srcind]})
        \kernel(\loc,\srcloc[\srcind]),
    \end{salign}
    then, it is easy to verify that $\pmap_+, \pmap_-\in
        \fspmaprdset{\dimregion}\subset \fspmapset[\dimregion]$ and
    $\pmap_+-\pmap_-= \reb{\auxpmapt}$. Substituting this last
    expression into \eqref{eq:infndist} yields
    \eqref{eq:approxdiffcond}, which concludes the proof.

\end{bullets}

\section{Proof of Lemma \ref{prop:boundedderivative}}
\label{proof:boundedderivative}
\begin{salign}
    |\pmap'(\locx)|&=2 \sum_{\srcind=1}^{\srcnum}\frac{\srccoef[\srcind]
        | \srclocx[\srcind] - \locx  | }{[
                (\locx - \srclocx[\srcind])^2 + \rdists[\srcind]   ]^2}\\
    &\leq 2 \sum_{\srcind=1}^{\srcnum}\sup_{\srclocx[\srcind], \locx}\frac{\srccoef[\srcind]
        | \srclocx[\srcind] - \locx  | }{[
                (\locx - \srclocx[\srcind])^2 + \rdists[\srcind] ]^2  }\\
    &= 2 \sum_{\srcind=1}^{\srcnum}\sup_{\auxvar}\frac{\srccoef[\srcind]
        \auxvar }{[
                \auxvar^2 + \rdists[\srcind] ]^2  }\jlac
    = \frac{3^{3/2}}{8} \sum_{\srcind=1}^{\srcnum}
    \frac{\srccoef[\srcind]
    }{ \rdist[\srcind]^3  }
\end{salign}

\section{Proof of Theorem~\ref{prop:varbounds}}
\label{sec:varbounds}

\begin{changes}
    Without loss of generality, assume that $\locx=0$ and $\locxdelta{}\geq 0$.
    Proving the upper bound in \eqref{eq:varbounds} can be equivalently  phrased as finding an upper bound for the value that a power map can take at $\locxdelta{}$ given that it takes the value  $\pmapval$ at 0, where $\pmapval\in \rfield_+$ is arbitrary. Formally, one needs to find an upper bound for
    \begin{align}
        \label{eq:varboundsupper1}
        \sup \{ \pmap(\locxdelta{}) ~|~ \pmap\in \fspmapset[1],~\pmap(0)=\pmapval\}.
    \end{align}
    Clearly, this supremum is upper bounded by
    \begin{align}
        \label{eq:varboundsupper}
        \sup \{ \pmap(\locxdelta{}) ~|~ \pmap\in \auxfspmapset[1],~\pmap(0)=\pmapval\}.
    \end{align}
    where $\auxfspmapset[1]$ is any set such that $\fspmapset[1]\subset \auxfspmapset[1]$. Due to \eqref{eq:minsrcregdist}, this condition is satisfied if one enlarges $\fspmapset[1]$ so that $\srcregion[1]=\rfield^1$ and
    $\mindist^2\in\rdistsset([\srclocx])~\forall\srclocx$.

    From \eqref{eq:friis1D}, the $\sup$ in \eqref{eq:varboundsupper} can be  expressed as the solution to
    \begin{salign}
        %\label{eq:friis1D}
        &\maximize_{
        \srcnum, \{(\srclocx[\srcind],\rdists[\srcind],\srccoef[\srcind])\}_{\srcind=1}^{\srcnum}
        }~~  \sum_{\srcind=1}^{\srcnum}\frac{\srccoef[\srcind]}{
            (\locxdelta{} - \srclocx[\srcind])^2 + \rdists[\srcind]   }      \\
        &\st~~        \sum_{\srcind=1}^{\srcnum}\frac{\srccoef[\srcind]}{
            \srclocx[\srcind]^2 + \rdists[\srcind]   }
        = \pmapval,\quad {\rdist[\srcind]} \geq
        \mindist~\forall\srcind.
    \end{salign}
    Now introduce auxiliary variables $\pmapval_1,\pmapval_2,\ldots,\pmapval_{\srcnum}$ and rewrite the problem as
    \begin{salign}[eq:varboundsallvarmax]
        %\label{eq:friis1D}
        &\maximize_{
        \srcnum, \{\pmapval_\srcind\}_{\srcind=1}^{\srcnum}, \{(\srclocx[\srcind],\rdists[\srcind],\srccoef[\srcind])\}_{\srcind=1}^{\srcnum}
        }~~  \sum_{\srcind=1}^{\srcnum}\frac{\srccoef[\srcind]}{
            (\locxdelta{} - \srclocx[\srcind])^2 + \rdists[\srcind]   }      \\
        &\st~~        \sum_{\srcind=1}^{\srcnum}
        \pmapval_\srcind= \pmapval,~
        \pmapval_\srcind = \frac{\srccoef[\srcind]}{
            \srclocx[\srcind]^2 + \rdists[\srcind]   },~ \rdist[\srcind] \geq
        \mindist~\forall\srcind.
    \end{salign}
    By optimizing first with respect to $\{(\srclocx[\srcind],\rdists[\srcind],\srccoef[\srcind])\}_{\srcind=1}^{\srcnum}$, it is easy to see that \eqref{eq:varboundsallvarmax} is equivalent to
    \begin{salign}[eq:varboundssubpvarmax]
        %\label{eq:friis1D}
        \maximize_{
        \srcnum, \{\pmapval_\srcind\}_{\srcind=1}^{\srcnum}
        }~~ & \sum_{\srcind=1}^{\srcnum} \pmapoptval(\pmapval_\srcind)      \\
        \st~~       & \sum_{\srcind=1}^{\srcnum}
        \pmapval_\srcind= \pmapval,
        \label{eq:varboundssubpvarmaxx}
    \end{salign}
    where $\pmapoptval(\pmapval_\srcind)$ is the optimal value of the problem
    \begin{salign}[eq:varboundsmainvarmax]
        \maximize_{
            \srclocx[\srcind],\rdists[\srcind],\srccoef[\srcind]
        }~~ & \frac{\srccoef[\srcind]}{
            (\locxdelta{} - \srclocx[\srcind])^2 + \rdists[\srcind]   }      \\
        \st~~
        & \pmapval_\srcind = \frac{\srccoef[\srcind]}{
            \srclocx[\srcind]^2 + \rdists[\srcind]   },~ \rdist[\srcind] \geq
        \mindist~\forall\srcind.
    \end{salign}
    To solve \eqref{eq:varboundsmainvarmax}, optimize first with respect to $\srccoef[\srcind]$ to obtain
    \begin{salign}[eq:varboundssubtovar]
        \maximize_{
            \srclocx[\srcind],\rdists[\srcind]
        }~~ & \frac{
            \pmapval_\srcind (\srclocx[\srcind]^2 + \rdists[\srcind]  )
        }{
            (\locxdelta{} - \srclocx[\srcind])^2 + \rdists[\srcind]   }      \\
        \st~~
        & \rdist[\srcind] \geq
        \mindist.
    \end{salign}
    Setting the derivative of the objective with respect to $\srclocx[\srcind]$ equal to zero yields
    \begin{align}
        \srclocx[\srcind] = \frac{\locxdelta{}\pm \sqrt{\locxdelta{}^2 + 4\rdists[\srcind]}}{2} .
    \end{align}
    The $+$ solution is a maximum and the $-$ solution is a minimum. Substituting the former  in \eqref{eq:varboundssubtovar} results in
    \begin{salign}%[eq:varboundssubtovar]                
        &\pmapoptval(\pmapval_\srcind)=
        % =\sup_{\rdist[\srcind] \geq
        %     \mindist
        % }~~  \frac{
        %     \pmapval_\srcind (\srclocx[\srcind]^2 + \rdists[\srcind]  )
        % }{
        %     (\locxdelta{} - \srclocx[\srcind])^2 + \rdists[\srcind]   }\\&=
        %
        \pmapval_\srcind\sup_{\rdist[\srcind] \geq
            \mindist
        }~~  \frac{
            \left(\locxdelta{}+ \sqrt{\locxdelta{}^2 + 4\rdists[\srcind]}\right)^2 + 4\rdists[\srcind]
        }{
            \left(\locxdelta{}- \sqrt{\locxdelta{}^2 + 4\rdists[\srcind]}\right)^2 + 4\rdists[\srcind]  }.
    \end{salign}
    Letting $\auxrdist =\sqrt{\locxdelta{}^2 + 4\rdists[\srcind]}/\locxdelta{}$ yields
    \begin{salign}
        \pmapoptval(\pmapval_\srcind)&=\pmapval_\srcind\sup_{
            \auxrdist \geq\boundfun(\locxdelta{})
        }~~  \frac{
            \left(\locxdelta{}+
            \auxrdist\locxdelta{}
            \right)^2 + \locxdelta{}^2(\auxrdist^2-1)
        }{
            \left(\locxdelta{}- \auxrdist\locxdelta{}\right)^2 + \locxdelta{}^2(\auxrdist^2-1)  }
        \\&=
        \pmapval_\srcind\sup_{
            \auxrdist \geq\boundfun(\locxdelta{})
        }~~  \frac{
            \auxrdist+1
        }{
            \auxrdist-1
        } = \pmapval_\srcind~  \frac{
            \boundfun(\locxdelta{})+1
        }{
            \boundfun(\locxdelta{})-1
        }.
        \label{eq:varboundsuppergs}
    \end{salign}
    Substituting \eqref{eq:varboundsuppergs} into
    \eqref{eq:varboundssubpvarmax} and using
    \eqref{eq:varboundssubpvarmaxx} yields a problem that does not
    depend on $\srcnum$ and
    $\{\pmapval_\srcind\}_{\srcind=1}^{\srcnum}$. Its optimal value is
    therefore the desired upper bound in \eqref{eq:varbounds}. The lower
    bound in \eqref{eq:varbounds} can be obtained by following a similar
    reasoning. To see that the bounds are tight, it suffices to note
    that \eqref{eq:varboundsupper1} equals \eqref{eq:varboundsupper} if
    $\fspmapset[1]=\auxfspmapset[1]$.

\end{changes}

\section{Proof of Theorem~\ref{prop:ftbounds}}
\label{sec:ftbounds}

\begin{bullets}
    \blt[FT] Let us start by considering the following result, which provides an explicit form for the Fourier transform of $\pmap$:
    \begin{mylemma}
        \label{prop:pmapft}
        It holds that:
        \begin{align}
            \pmapft(\wnx)=\pi\sum_{\srcind=1}^{\srcnum}\frac{\srccoef[\srcind]}{\rdist[\srcind]}e^{-j\wnx  \srclocx[\srcind]}
            e^{-\rdist[\srcind]|\wnx|}
        \end{align}

    \end{mylemma}

    \begin{IEEEproof}
        It is easy to show that, for any $\aconst>0$, it follows that
        \begin{align}
            \fourier\brackets{
                \frac{2\aconst}{ \locx^2+ \aconst^2 }
            } = 2\pi e^{-a|\wnx|},
        \end{align}
        where $\fourier$ denotes the Fourier transform.
        Therefore,
        \begin{salign}
            \pmapft(\wnx)& \define \fourier\brackets{\pmap(\locx)}\\
            &=\fourier\brackets{\sum_{\srcind=1}^{\srcnum}\frac{\srccoef[\srcind]}{
                    (\locx - \srclocx[\srcind])^2 + \rdists[\srcind]   }
            }\jlac
            \begin{intermed}
                =\sum_{\srcind=1}^{\srcnum}\fourier\brackets{\frac{\srccoef[\srcind]}{
                        (\locx - \srclocx[\srcind])^2 + \rdists[\srcind]   }
                }
            \end{intermed}\\
            &=\sum_{\srcind=1}^{\srcnum}e^{-j\wnx  \srclocx[\srcind]}\fourier\brackets{\frac{\srccoef[\srcind]}{
                    \locx^2 + \rdists[\srcind]   }
            }\jlac
            \begin{intermed}
                =\sum_{\srcind=1}^{\srcnum}\frac{\srccoef[\srcind]}{2\rdist[\srcind]}e^{-j\wnx  \srclocx[\srcind]}\fourier\brackets{\frac{2\rdist[\srcind]}{
                        \locx^2 + \rdists[\srcind]   }
                }
            \end{intermed}
            \jlac
            \begin{intermed}
                =\sum_{\srcind=1}^{\srcnum}\frac{\srccoef[\srcind]}{2\rdist[\srcind]}e^{-j\wnx  \srclocx[\srcind]}
                2\pi e^{-\rdist[\srcind]|\wnx|}
            \end{intermed}
            \\
            &=\pi\sum_{\srcind=1}^{\srcnum}\frac{\srccoef[\srcind]}{\rdist[\srcind]}e^{-j\wnx  \srclocx[\srcind]}
            e^{-\rdist[\srcind]|\wnx|}.
        \end{salign}
        % \begin{conferenceonly}
        %     The proof is omitted due to lack of space.
        % \end{conferenceonly}
    \end{IEEEproof}

    It follows from \Cref{prop:pmapft} that
    \begin{salign}
        |\pmapft(\wnx)|
        &=\left|\pi\sum_{\srcind=1}^{\srcnum}\frac{\srccoef[\srcind]}{\rdist[\srcind]}e^{-j\wnx  \srclocx[\srcind]}
        e^{-\rdist[\srcind]|\wnx|}\right|\\
        &\leq\pi\sum_{\srcind=1}^{\srcnum}\left|\frac{\srccoef[\srcind]}{\rdist[\srcind]}e^{-j\wnx  \srclocx[\srcind]}
        e^{-\rdist[\srcind]|\wnx|}\right|\\
        &=\pi\sum_{\srcind=1}^{\srcnum}\frac{\srccoef[\srcind]}{\rdist[\srcind]}
        e^{-\rdist[\srcind]|\wnx|}\jlac
        \leq\left[\frac{\pi}{\rdistmin}\sum_{\srcind=1}^{\srcnum}\srccoef[\srcind]
            \right]e^{-\rdistmin|\wnx|},
    \end{salign}
    which establishes \eqref{eq:pmapftbound}.

    \blt[high pass energy]The high-pass energy of $\pmap$ can be upper bounded as
    \begin{salign}
        \int_{\bw}^\infty &|\pmapft(\wnx)|^2 d\wnx
        =\int_{\bw}^\infty \left|\pi\sum_{\srcind=1}^{\srcnum}\frac{\srccoef[\srcind]}{\rdist[\srcind]}e^{-j\wnx  \srclocx[\srcind]}
        e^{-\rdist[\srcind]|\wnx|}\right|^2 d\wnx \\
        &\leq \pi^2\int_{\bw}^\infty \left[\sum_{\srcind=1}^{\srcnum}\left|\frac{\srccoef[\srcind]}{\rdist[\srcind]}e^{-j\wnx  \srclocx[\srcind]}
        e^{-\rdist[\srcind]|\wnx|}\right|\right]^2 d\wnx \\
        &= \pi^2\int_{\bw}^\infty \left[\sum_{\srcind=1}^{\srcnum}\frac{\srccoef[\srcind]}{\rdist[\srcind]}
            e^{-\rdist[\srcind]\wnx}\right]^2 d\wnx \jlac
        \begin{intermed}
            = \pi^2\int_{\bw}^\infty \sum_{\srcind=1}^{\srcnum}\frac{\srccoef[\srcind]}{\rdist[\srcind]}
            e^{-\rdist[\srcind]\wnx}
            \sum_{\srcindp=1}^{\srcnum}\frac{\srccoef[\srcindp]}{\rdist[\srcindp]}
            e^{-\rdist[\srcindp]\wnx}
            d\wnx
        \end{intermed}
        \\
        &= \pi^2\sum_{\srcind=1}^{\srcnum}\sum_{\srcindp=1}^{\srcnum}\frac{\srccoef[\srcind]}{\rdist[\srcind]}
        \frac{\srccoef[\srcindp]}{\rdist[\srcindp]}
        \int_{\bw}^\infty e^{-(\rdist[\srcind]+\rdist[\srcindp])\wnx}
        d\wnx \\
        &= \pi^2\sum_{\srcind=1}^{\srcnum}\sum_{\srcindp=1}^{\srcnum}\frac{\srccoef[\srcind]}{\rdist[\srcind]}
        \frac{\srccoef[\srcindp]}{\rdist[\srcindp]}
        \frac{ e^{-(\rdist[\srcind]+\rdist[\srcindp])\bw}}{\rdist[\srcind]+\rdist[\srcindp]}
        \\&=\pi^2\auxvec\transpose\auxmat\auxvec,
    \end{salign}
    where $\auxvec\define [{\srccoef[1]}/{\rdist[1]};\ldots;{\srccoef[\srcnum]}/{\rdist[\srcnum]}]$ and  the $\srcnum \times \srcnum$ matrix $\auxmat$ is such that its  $(\srcind,\srcindp)$-th entry is $
        \auxmatent_{\srcind,\srcindp}\define
        { e^{-(\rdist[\srcind]+\rdist[\srcindp])\bw}}/({\rdist[\srcind]+\rdist[\srcindp]})$. Since $\auxmat$ is symmetric, its eigenvalues are real. In particular, its largest eigenvalue $\eigmax(\auxmat)$ is real. Since all entries of $\auxmat$ are positive, it follows necessarily that $\eigmax(\auxmat)>0$ and
    \begin{journalonly}
        \begin{salign}
            \int_{\bw}^\infty |\pmapft(\wnx)|^2 d\wnx
            &\leq \pi^2 \|\auxvec\|^2 \eigmax(\auxmat).
        \end{salign}
    \end{journalonly}
    \begin{conferenceonly}
        $\int_{\bw}^\infty |\pmapft(\wnx)|^2 d\wnx
            \leq \pi^2 \|\auxvec\|^2 \eigmax(\auxmat)$.
    \end{conferenceonly}
    Furthermore, since all the entries of $\auxmat$ are positive, $ \eigmax(\auxmat)$ is a Perron-Frobenius eigenvalue and, therefore, satisfies~\cite[eq. (2)]{pillai2005perron} that $\eigmax(\auxmat)\leq \max_{\srcind}\sum_{\srcindp}\auxmatent_{\srcind,\srcindp}$. It follows that
    \begin{salign}
        \int_{\bw}^\infty |\pmapft(\wnx)|^2 d\wnx
        &
        \begin{intermed}
            \leq  \pi^2
            \left[\sum_{\srcind=1}^{\srcnum}\frac{\srccoef[\srcind]^2}{\rdists[\srcind]}\right]
            \max_{\srcind}\sum_{\srcindp=1}^{\srcnum}
            \frac{ e^{-(\rdist[\srcind]+\rdist[\srcindp])\bw}}{\rdist[\srcind]+\rdist[\srcindp]}
        \end{intermed}\jlac
        \begin{intermed}
            \leq  \pi^2\srcnum
            \left[\sum_{\srcind=1}^{\srcnum}\frac{\srccoef[\srcind]^2}{\rdists[\srcind]}\right]
            \frac{ e^{-2\rdistmin\bw}}{2\rdistmin}
        \end{intermed}
        \jlac
        \leq
        \frac{\pi^2\srcnum\sum_{\srcind=1}^{\srcnum}\srccoef[\srcind]^2}{2\rdistmin^3}  e^{-2\rdistmin\bw},
    \end{salign}
    which establishes \eqref{eq:hpenergyub}.

    \blt[total energy lower bound]Finally, to upper bound the total energy, note that
    \begin{salign}
        \int_{0}^\infty |\pmapft(\wnx)|^2 d\wnx &=
        \frac{1}{2}\int_{-\infty}^\infty |\pmapft(\wnx)|^2 d\wnx \\
        &=\pi\int_{-\infty}^\infty |\pmap(\locx)|^2 d\locx
        \jlac
        \begin{intermed}
            =
            \pi\int_{-\infty}^\infty \left|
            \sum_{\srcind=1}^{\srcnum}\frac{\srccoef[\srcind]}{    (\locx - \srclocx[\srcind])^2 + \rdists[\srcind]   }
            \right|^2 d\locx
        \end{intermed}
        \\
        &\geq\pi\int_{-\infty}^\infty
        \sum_{\srcind=1}^{\srcnum}\left|\frac{\srccoef[\srcind]}{    (\locx - \srclocx[\srcind])^2 + \rdists[\srcind]   }
        \right|^2 d\locx \\
        &=\pi
        \sum_{\srcind=1}^{\srcnum}\int_{-\infty}^\infty \left[\frac{\srccoef[\srcind]}{    \locx^2 + \rdists[\srcind]   }
            \right]^2 d\locx.
    \end{salign}
    It is straightforward to verify that, for any $\aconst\neq 0$, it holds that
    \begin{align}
        \int \frac{1}{(\locx^2 + \aconst^2)^2} d\locx = \frac{1}{2\aconst^2}\left[
            \frac{\locx}{\locx^2 + \aconst^2} + \frac{1}{\aconst} \arctan \left(\frac{\locx}{\aconst}\right)
            \right].
    \end{align}
    Hence,
    \begin{salign}
        \int_{0}^\infty &|\pmapft(\wnx)|^2 d\wnx\nonumber
        \\&\geq
        \begin{intermed}
            \pi
            \sum_{\srcind=1}^{\srcnum}\srccoef[\srcind]^2\int_{-\infty}^\infty \frac{1}{ (   \locx^2 + \rdists[\srcind])^2   }
            d\locx
        \end{intermed}
        \jlac
        \begin{intermed}
            =
        \end{intermed}
        \pi
        \sum_{\srcind=1}^{\srcnum}
        \frac{\srccoef[\srcind]^2}{2\rdist[\srcind]^2}\left[
            \frac{\locx}{\locx^2 + \rdist[\srcind]^2} + \frac{1}{\rdist[\srcind]} \arctan \left(\frac{\locx}{\rdist[\srcind]}\right)
            \right]_{-\infty}^\infty\\
        &=\pi
        \sum_{\srcind=1}^{\srcnum}
        \frac{\srccoef[\srcind]^2}{2\rdist[\srcind]^2}
        \frac{\pi}{\rdist[\srcind]}
        \jlac
        =\frac{\pi^2}{2}
        \sum_{\srcind=1}^{\srcnum}
        \frac{\srccoef[\srcind]^2}{\rdist[\srcind]^3},
    \end{salign}
    which proves \eqref{eq:energylowerbound}.

\end{bullets}

\section{Proof of Theorem~\ref{prop:nninterperr}}
\label{sec:nninterperr}

\begin{bullets}%
    \blt[]

    \blt[notation]Let
    \begin{align}
        \label{eq:deriubdef}
        \deriub\define
        \frac{3^{3/2}}{8} \sum_{\srcind=1}^{\srcnum}
        \frac{\srccoef[\srcind]
        }{ \rdist[\srcind]^3  }
    \end{align}
    be the upper bound on the derivative of $\pmap$ provided by \eqref{eq:boundedderivative}.

    \blt[bounds in the interval] To prove Theorem~\ref{prop:nninterperr}, it is convenient to first establish the following result:
    \begin{mylemma}
        \label{prop:boundslocxmspind}
        If $\locx\in[\locxsp{\spind},\locxsp{\spind+1}]$, then
        \begin{salign}[eq:boundslocxmspind]
            \max[ & -\deriub ({\locx-\locxsp{\spind}}),
                -\deriub({\locxsp{\spind+1}-\locx})+ \pmapdelta{\spind}
            ]                                           \\&
            \leq
            \pmap(\locx)  - \pmapsv{\spind}             \\&\leq
            \min[
                \deriub ({\locx-\locxsp{\spind}}),
                \deriub({\locxsp{\spind+1}-\locx})+\pmapdelta{\spind}
            ],
        \end{salign}
        where $\pmapdelta{\spind}\define\pmap(\locxsp{\spind+1})-\pmap(\locxsp{\spind})$.
    \end{mylemma}

    \begin{IEEEproof}
        %\begin{journalonly}
        Given that $\pmap$ is differentiable, it follows from the mean-value
        theorem~\cite[Th. 5.10]{rudin1953} that, for any $\locxl<\locxu$,
        \begin{align}
            \exists \locx \in (\locxl,\locxu)~|~\pmap'(\locx)=\frac{\pmap(\locxu)-\pmap(\locxl)}{\locxu-\locxl}.
        \end{align}
        From $|\pmap'(\locx)|\leq \deriub$, it follows  that, for all
        $\locxl<\locxu$,
        \begin{align}
            \label{eq:meanvarbound}
            \left|\frac{\pmap(\locxu)-\pmap(\locxl)}{\locxu-\locxl}\right|\leq \deriub.
        \end{align}
        Setting $\locxl=\locxsp{\spind}$ and $\locxu=\locx>
            \locxsp{\spind}$ yields
        $\left|\pmap(\locx)-\pmap(\locxsp{\spind})\right|\leq \deriub
            ({\locx-\locxsp{\spind}})$  or, equivalently,
        \begin{align}
            \label{eq:boundslocxmspind1}
            -\deriub ({\locx-\locxsp{\spind}}) \leq \pmap(\locx)-\pmap(\locxsp{\spind})\leq \deriub ({\locx-\locxsp{\spind}}).
        \end{align}
        On the other hand, setting  $\locxu=\locxsp{\spind+1}$ and
        $\locxl=\locx<\locxsp{\spind+1}$ results in
        $\left|{\pmap(\locxsp{\spind+1})-\pmap(\locx)}\right|\leq
            \deriub({\locxsp{\spind+1}-\locx})$, which can also be written as
        \begin{intermed}
            \begin{align}
                \nonumber
                -\deriub({\locxsp{\spind+1}-\locx}) & \leq {\pmap(\locxsp{\spind+1})-\pmap(\locx)} \\&\leq
                \deriub({\locxsp{\spind+1}-\locx})
            \end{align}
            or
            \begin{align}
                -\deriub({\locxsp{\spind+1}-\locx}) \leq \pmap(\locx)  - \pmap(\locxsp{\spind+1})\leq
                \deriub({\locxsp{\spind+1}-\locx}).
            \end{align}
            Adding
            $\pmapdelta{\spind}$ yields
        \end{intermed}
        \begin{salign}
            \label{eq:boundslocxmspind2}
            -\deriub({\locxsp{\spind+1}-\locx})+ \pmapdelta{\spind} & \leq \pmap(\locx)  - \pmap(\locxsp{\spind}) \\& \leq
            \deriub({\locxsp{\spind+1}-\locx})+\pmapdelta{\spind}.
        \end{salign}
        Combining \eqref{eq:boundslocxmspind1} and
        \eqref{eq:boundslocxmspind2} yields \eqref{eq:boundslocxmspind} for
        $\locx\in(\locxsp{\spind},\locxsp{\spind+1})$. The cases  $\locx
            =\locxsp{\spind}$ and $\locx=\locxsp{\spind+1}$ follow from
        continuity.
        %\end{journalonly}
        % \begin{conferenceonly}
        %     The proof, which is based on the mean-value
        %     theorem~\cite[Th. 5.10]{rudin1953}, is omitted here due to lack of space.
        % \end{conferenceonly}

    \end{IEEEproof}

    \blt[lemma for the max] To prove Theorem~\ref{prop:nninterperr}, it is also convenient to first establish the following result:
    \begin{mylemma}
        \label{prop:maxlemma}
        Let $0 \leq a \leq b$ and let $\Delta\in\rfield$. It holds that
        \begin{align}
            \label{eq:maxlemma}
            \max \big[ & \min(a, b - \Delta),
                \min(a, b + \Delta) \big]
            = a
        \end{align}

    \end{mylemma}

    \begin{IEEEproof}
        %\begin{journalonly}
        To prove \eqref{eq:maxlemma}, consider the
        following cases:

        (C1) $\Delta \leq - b + a$: In this case, it clearly holds that $b
            - \Delta \geq 2b - a \geq a$, which implies that $\min(a, b -
            \Delta) = a$. On the other hand, it also holds that $b + \Delta \leq
            a$, which in turn implies that $\min(a, b + \Delta) = b + \Delta$. Therefore, the left-hand side of \eqref{eq:maxlemma} becomes $\max \big[ a, b + \Delta \big]
            = a$.

        (C2) $- b + a \leq \Delta \leq b - a$: Since $b - \Delta \geq a $, it follows that $\min(a, b -
            \Delta) = a $. Furthermore, since $b + \Delta \geq a $, one has that $\min(a, b +
            \Delta) = a$.  Hence, the left-hand side of \eqref{eq:maxlemma} becomes
        $
            \max \big[ a, a \big]
            = a
        $.

        (C3) $\Delta \geq b - a$: Since $b - \Delta \leq a $, it follows that $\min(a, b -
            \Delta) = b - \Delta$. Since $b + \Delta \geq 2b - a \geq a $, it holds that  $\min(a, b + \Delta) = a$. Thus, the left-hand side of \eqref{eq:maxlemma} becomes
        $ \max \big[ b - \Delta, a \big]
            = a$.

        Noting that \eqref{eq:maxlemma} has been proved for all values of $\Delta$ concludes the proof.

        %\end{journalonly}
        % \begin{conferenceonly}
        %     The proof is omitted due to lack of space.
        % \end{conferenceonly}

    \end{IEEEproof}

    \blt[bounds for abs difference] For
    $\locx\in[\locxsp{\spind},\locxsp{\spind+1}]$, the nearest neighbor
    interpolator is given by
    \begin{align}
        \label{eq:nnestimatecases}
        \pmapest(\locxsp{\spind}
        ) & \define\begin{cases}
                       \pmapsv{\spind} \quad
                       \text{if } \locx \leq \locxmp{\spind}
                       \\
                       \pmapsv{\spind+1} \quad
                       \text{if } \locx \geq \locxmp{\spind},
                   \end{cases}
    \end{align}
    where $\locxmp{\spind}\define (\locxsp{\spind} + \locxsp{\spind+1})/2$.
    It follows from \eqref{eq:boundslocxmspind} and $a \leq x \leq b
        \implies |x| \leq \max(-a, b)$ that,  for
    $\locx\in[\locxsp{\spind},\locxmp{\spind}]$,

    \begin{salign}
        |&\pmap(\locx)  -  \pmapsv{\spind}|
        \leq
        \begin{intermed}
            \max \Bigg[
            \nonumber
        \end{intermed}
        \jlac
        \begin{intermed}
            - \max \big[ -\deriub ({\locx-\locxsp{\spind}}),
                -\deriub({\locxsp{\spind+1}-\locx})+ \pmapdelta{\spind}
                \big],
            \nonumber
        \end{intermed}
        \jlac
        \begin{intermed}
            \min \big[
                \deriub ({\locx-\locxsp{\spind}}),
                \deriub({\locxsp{\spind+1}-\locx})+\pmapdelta{\spind}
                \big]
            \Bigg]
        \end{intermed}
        \jlac
        % \begin{intermed}
        %     =
        % \end{intermed}
        \max \Bigg[
            \min \big[ \deriub ({\locx-\locxsp{\spind}}),
                \nonumber
                %\\&
                \deriub({\locxsp{\spind+1}-\locx})- \pmapdelta{\spind}
                \big],
            \nonumber
            \\&
            \min \big[
                \deriub ({\locx-\locxsp{\spind}}),
                %\\&
                \deriub({\locxsp{\spind+1}-\locx})+\pmapdelta{\spind}
                \big]
            \Bigg].
    \end{salign}

    Now applying \Cref{prop:maxlemma} yields
    \begin{align}
        \label{eq:diffboundfirst}
        |\pmap(\locx)  -  \pmapsv{\spind}| \leq \deriub ({\locx - \locxsp{\spind}}), \forall \locxdelta{\spind}.
    \end{align}

    Similarly, for $\locx\in[\locxmp{\spind}, \locxsp{\spind+1}]$ , one obtains
    \begin{intermed}
        \begin{salign}
            |\pmap(\locx)  -  \pmapsv{\spind +1}|
            \leq
            & \max \Bigg[
                - \max \big[ -\deriub ({\locxsp{\spind+1} - \locx}),
                    \nonumber
                    \\&
                    -\deriub({\locx - \locxsp{\spind}}) - \pmapdelta{\spind}
                    \big],
                \nonumber
                \\&
                \min \big[
                    \deriub (\locxsp{\spind+ 1} - {\locx}),
                    \nonumber
                    \\&
                    \deriub({\locx - \locxsp{\spind}}) - \pmapdelta{\spind}
                    \big]
                \Bigg]
            \\&=
            \max \Bigg[
                \min \big[ \deriub ({\locxsp{\spind+1} - \locx}),
                    \nonumber
                    \\&
                    \deriub({\locx - \locxsp{\spind}}) + \pmapdelta{\spind}
                    \big],
                \nonumber
                \\&
                \min \big[
                    \deriub (\locxsp{\spind+ 1} - {\locx}),
                    \nonumber
                    \\&
                    \deriub({\locx - \locxsp{\spind}}) - \pmapdelta{\spind}
                    \big]
                \Bigg].
        \end{salign}

        Applying  \Cref{prop:maxlemma} results in
    \end{intermed}
    \begin{salign}
        \label{eq:diffboundsecond}
        |\pmap(\locx)  -  \pmapsv{\spind +1}| \leq \deriub ({\locxsp{\spind +1} - \locx}), \forall \locxdelta{\spind}.
    \end{salign}

    Thus, combining \eqref{eq:diffboundfirst} and \eqref{eq:diffboundsecond} produces the bound
    \begin{salign}
        \label{eq:diffboundnn}
        |\pmap(\locx)  - \pmapest(\locx)|&\leq
        \begin{cases}
            \deriub ({\locx - \locxsp{\spind}})    & ~\text{if } \locxsp{\spind}\leq \locx < \locxmp{\spind}       \\
            \deriub ({\locxsp{\spind +1} - \locx}) & ~\text{if } \locxmp{\spind}\leq \locx \leq \locxsp{\spind+1}.
        \end{cases}
    \end{salign}

    \blt[bounds for each error metric]
    \begin{extendedonly}
        \begin{bullets}%
            \blt[L1]The next step is to bound the $L^1$ error. Using the above expressions, it follows that
            \begin{salign}
                \int_{\locxsp{\spind}}^{\locxsp{\spind+1}}&|\pmap(\locx)  -
                \pmapest(\locx)| d\locx
                \nonumber
                \jlac
                \begin{intermed}
                    =\int_{\locxsp{\spind}}^{\locxmp{\spind}}|\pmap(\locx)  -
                    \pmapest(\locx)| d\locx +
                    \int_{\locxmp{\spind}}^{\locxsp{\spind+1}}|\pmap(\locx)  -
                    \pmapest(\locx)| d\locx
                \end{intermed}
                \\&
                \leq
                \int_{\locxsp{\spind}}^{\locxmp{\spind}}
                |\deriub (\locx - \locxsp{\spind})| d\locx +
                \int_{\locxmp{\spind}}^{\locxsp{\spind+1}}
                |\deriub (\locxsp{\spind+1} - \locx)| d\locx
                \jlac
                \begin{intermed}
                    =
                    \deriub \Bigg[\frac{(\locx - \locxsp{\spind})^2}{2} \Bigg]^{\locxmp{\spind}}_{\locxsp{\spind}} + \deriub \Bigg[-\frac{(\locxsp{\spind+1} - \locx)^2}{2} \Bigg]^{\locxsp{\spind+1}}_{\locxmp{\spind}}
                \end{intermed}
                \jlac
                \begin{intermed}=
                    \deriub \Bigg[\frac{(\locxmp{\spind} - \locxsp{\spind})^2}{2} - 0 \Bigg]+ \deriub \Bigg[-0 + \frac{(\locxsp{\spind + 1} - \locxmp{\spind})^2}{2} \Bigg]
                \end{intermed}
                \\&=
                \frac{\deriub}{2} \Bigg[(\locxmp{\spind} - \locxsp{\spind})^2 + (\locxsp{\spind + 1} - \locxmp{\spind})^2 \Bigg]
                \jlac
                \begin{intermed}=
                    \frac{\deriub}{2} \Bigg[\Big(\frac{\locxsp{\spind+1} + \locxsp{\spind}}{2} - \locxsp{\spind}\Big)^2 + \Big(\locxsp{\spind + 1} - \frac{\locxsp{\spind+1} + \locxsp{\spind}}{2}\Big)^2 \Bigg]
                \end{intermed}
                \jlac
                \begin{intermed}=
                    \frac{\deriub}{2} \Bigg[\Big(\frac{\locxsp{\spind+1} - \locxsp{\spind}}{2} \Big)^2 + \Big(\frac{\locxsp{\spind+1} - \locxsp{\spind}}{2}\Big)^2 \Bigg]
                \end{intermed}
                \jlac
                \begin{intermed}
                    =
                    \deriub \Big(\frac{\locxsp{\spind+1} - \locxsp{\spind}}{2} \Big)^2
                \end{intermed}
                \\&=\frac{\deriub \locxdelta{\spind}^2}{4}
            \end{salign}
            Combining this bound for the $\spnum-1$ intervals yields
            \begin{align}
                %\label{eq:lininterpl1errexp}
                \|\pmap  -
                \pmapest\|_1 & \leq
                \sum_{\spind=1}^{\spnum-1} \int_{\locxsp{\spind}}^{\locxsp{\spind+1}}|\pmap(\locx)  -
                \pmapest(\locx)| d\locx \nonumber
                \\& \leq \frac{\deriub}{4} \sum_{\spind=1}^{\spnum-1} \locxdelta{\spind}^2,
            \end{align}
            which, combined with \eqref{eq:deriubdef}, proves \eqref{eq:nninterpl1errsb}.

            \blt[L2]For the $L^2$ error, one can write the following:
            \begin{salign}
                \int_{\locxsp{\spind}}^{\locxsp{\spind+1}}&|\pmap(\locx)  -
                \pmapest(\locx)|^2 d\locx
                \nonumber
                \jlac
                \begin{intermed}
                    =\int_{\locxsp{\spind}}^{\locxmp{\spind}}|\pmap(\locx)  -
                    \pmapest(\locx)|^2 d\locx +
                    \int_{\locxmp{\spind}}^{\locxsp{\spind+1}}|\pmap(\locx)  -
                    \pmapest(\locx)|^2 d\locx
                \end{intermed}
                \\&
                \leq
                \int_{\locxsp{\spind}}^{\locxmp{\spind}}
                \deriub^2 (\locx - \locxsp{\spind})^2 d\locx +
                \int_{\locxmp{\spind}}^{\locxsp{\spind+1}}
                \deriub^2 (\locxsp{\spind+1} - \locx)^2 d\locx
                \jlac
                \begin{intermed}
                    =
                    \deriub^2 \Bigg[\frac{(\locx - \locxsp{\spind})^3}{3} \Bigg]^{\locxmp{\spind}}_{\locxsp{\spind}} + \deriub^2 \Bigg[-\frac{(\locxsp{\spind+1} - \locx)^3}{3} \Bigg]^{\locxsp{\spind+1}}_{\locxmp{\spind}}
                \end{intermed}
                \\&
                \begin{intermed}=
                    \deriub^2 \Bigg[\frac{(\locxmp{\spind} - \locxsp{\spind})^3}{3} - 0 \Bigg]+ \deriub^2 \Bigg[-0 + \frac{(\locxsp{\spind + 1} - \locxmp{\spind})^3}{3} \Bigg]
                \end{intermed}
                \jlac=
                \frac{\deriub^2}{3} \Bigg[(\locxmp{\spind} - \locxsp{\spind})^3 + (\locxsp{\spind + 1} - \locxmp{\spind})^3 \Bigg]
                \jlac
                \begin{intermed}=
                    \frac{\deriub^2}{3} \Bigg[\Big(\frac{\locxsp{\spind+1} + \locxsp{\spind}}{2} - \locxsp{\spind}\Big)^3 + \Big(\locxsp{\spind + 1} - \frac{\locxsp{\spind+1} + \locxsp{\spind}}{2}\Big)^3 \Bigg]
                \end{intermed}
                \jlac
                \begin{intermed}=
                    \frac{\deriub^2}{3} \Bigg[\Big(\frac{\locxsp{\spind+1} - \locxsp{\spind}}{2} \Big)^3 + \Big(\frac{\locxsp{\spind+1} - \locxsp{\spind}}{2}\Big)^3 \Bigg]
                \end{intermed}
                \jlac
                \begin{intermed}
                    =
                    \frac{2}{3}\deriub^2 \Big(\frac{\locxsp{\spind+1} - \locxsp{\spind}}{2} \Big)^3
                \end{intermed}
                \jlac
                \begin{intermed}=
                    \frac{2}{3}\deriub^2 \Big(\frac{\locxdelta{\spind}}{2} \Big)^3
                \end{intermed}
                \jlac
                \begin{intermed}=
                    \frac{2}{3}\deriub^2 \frac{\locxdelta{\spind}^3}{8}
                \end{intermed}
                \\&=\frac{\deriub^2 \locxdelta{\spind}^3}{12}
            \end{salign}

            Combining this bound for the $\spnum-1$ intervals yields
            \begin{align}
                %\label{eq:lininterpl1errexp}
                \|\pmap  -
                \pmapest\|_{2}^{2} & \leq
                \int_{\locxsp{\spind}}^{\locxsp{\spnum}}|\pmap(\locx)  -
                \pmapest(\locx)|^2 d\locx \nonumber
                \\& \leq \frac{\deriub^2}{12} \sum_{\spind=1}^{\spnum-1} \locxdelta{\spind}^3,
            \end{align}
            which proves \eqref{eq:nninterpl2errsb} after substitution of  \eqref{eq:deriubdef}.

            \blt[Linf]
            Finally, for the $L^\infty$ error, it follows that
            \begin{salign}
                \|\pmap  &- \pmapest\|_\infty = \sup_{\locx\in[\locxsp{\spind},\locxsp{\spnum})}|\pmap(\locx)  -
                \pmapest(\locx)|
                \\& \leq
                \max_{\spind}\bigg[\max\bigg(
                    \sup_{\locx\in[\locxsp{\spind},\locxmp{\spind}]}\deriub (\locx - \locxsp{\spind}),
                    \nonumber
                    \\&
                    \sup_{\locx\in[\locxmp{\spind},\locxsp{\spind+1}]}\deriub (\locxsp{\spind +1}- \locx)
                    \bigg) \bigg]
                \\& =
                \max_{\spind}\bigg[\max\bigg(
                    \deriub (\locxmp{\spind} - \locxsp{\spind}),
                    \deriub (\locxsp{\spind +1}- \locxmp{\spind})
                    \bigg) \bigg]
                \jlac
                \begin{intermed}
                    =
                    \max_{\spind}\bigg[\max\bigg(
                        \deriub (\frac{\locxsp{\spind+1} + \locxsp{\spind}}{2} - \locxsp{\spind}),
                        \deriub (\locxsp{\spind +1}- \frac{\locxsp{\spind+1} + \locxsp{\spind}}{2})
                        \bigg) \bigg]
                \end{intermed}
                \jlac
                \begin{intermed}
                    =
                    \max_{\spind}\bigg[\max\bigg(
                        \deriub (\frac{\locxsp{\spind+1} - \locxsp{\spind}}{2}),
                        \deriub (\frac{\locxsp{\spind+1} - \locxsp{\spind}}{2})
                        \bigg) \bigg]
                \end{intermed}
                \jlac
                \begin{intermed}
                    =
                    \max_{\spind}\bigg[\max\bigg(
                        \deriub \frac{\locxdelta{\spind}}{2}, \deriub \frac{\locxdelta{\spind}}{2}
                        \bigg) \bigg]
                \end{intermed}
                \jlac
                \begin{intermed}
                    =
                    \max_{\spind}\bigg[\frac{m}{2} \locxdelta{\spind} \bigg]
                \end{intermed}
                \\&
                =
                \frac{m}{2} \max_{\spind} \locxdelta{\spind},
            \end{salign}
            which, together with \eqref{eq:deriubdef}, proves \eqref{eq:nninterplinferrsb}.

        \end{bullets}%
    \end{extendedonly}
    \begin{nonextendedonly}
        The rest of the proof involves integrating \eqref{eq:diffboundnn} to obtain the L$^1$ and L$^2$ errors and obtaining the suprema on each subinterval to obtain $L^\infty$. It is omitted due to lack of space.
    \end{nonextendedonly}

\end{bullets}%

\section{Proof of Theorem~\ref{prop:lininterperr}}
\label{sec:lininterperr}

Using \Cref{prop:boundslocxmspind}, it is possible to prove the following:
\begin{mylemma}
    The estimator  $\pmapest$ defined in \eqref{eq:linest} satisfies:
    \begin{salign}[eq:lininterpallerr]
        \label{eq:lininterpl1err}
        \|\pmap  -
        \pmapest\|_1 &\leq \frac{9}{32a} \deriub\sum_{\spind=1}^{\spnum-1} \locxdelta{\spind}^2
        \\
        \label{eq:lininterpl2err}
        \|\pmap  -
        \pmapest\|_2^2&\leq
        \frac{16\sqrt{2}-13}{96
        }\deriub^2
        \sum_{\spind=1}^{\spnum-1}\locxdelta{\spind}^3
        \\
        \label{eq:lininterplinferr}
        \|\pmap  -
        \pmapest\|_\infty
        &\leq
        \frac{
            \deriub
        }{
            2
        }\max_{\spind}\locxdelta{\spind}
    \end{salign}

\end{mylemma}

\begin{IEEEproof}
    \begin{bullets}
        \blt[diff bound]
        It follows from \eqref{eq:boundslocxmspind} that, for  $\locx\in[\locxsp{\spind},\locxsp{\spind+1}]$,
        \begin{align}
            \label{eq:boundsondiff}
            \lbound{\spind}(\locx) \leq
            \pmap(\locx)  - \pmapest(\locx)\leq
            \ubound{\spind}(\locx),
        \end{align}
        where
        \begin{salign}
            \lbound{\spind}(\locx)&\define -\min[\deriub ({\locx-\locxsp{\spind}}),
                \deriub({\locxsp{\spind+1}-\locx})- \pmapdelta{\spind}
            ]\nonumber \\&- \frac{
                \pmapdelta{\spind}
            }{
                \locxdelta{\spind}
            }
            (\locx - \locxsp{\spind})\\
            \ubound{\spind}(\locx)&\define
            \min[
                \deriub ({\locx-\locxsp{\spind}}),
                \deriub({\locxsp{\spind+1}-\locx})+\pmapdelta{\spind}
            ]\nonumber \\&-\frac{
                \pmapdelta{\spind}
            }{
                \locxdelta{\spind}
            }
            (\locx - \locxsp{\spind}).
        \end{salign}
        Letting $\locxmp{\spind}\define (\locxsp{\spind} +
            \locxsp{\spind+1})/2$, it is then straightforward to verify that
        $\ubound{\spind}(\locx)=\bound{\spind}(\locx;\pmapdelta{\spind})$ and
        $\lbound{\spind}(\locx)=-\bound{\spind}(\locx;-\pmapdelta{\spind})$,
        where

        \begin{align}
            \label{eq:boundsascases}
            \bound{\spind}(\locx;
            \pmapdelta{\spind}
            ) & \define\begin{cases}
                           \left(\deriub-\frac{
                               \pmapdelta{\spind}
                           }{
                               \locxdelta{\spind}
                           }\right)
                           (\locx - \locxsp{\spind})                                                           \\\quad\quad
                           \text{if } \locxsp{\spind}\leq \locx <  \locxmp{\spind}+\pmapdelta{\spind}/2\deriub \\
                           \left(\deriub +\frac{
                               \pmapdelta{\spind}
                           }{
                               \locxdelta{\spind}
                           }\right)( \locxsp{\spind+1}-\locx)                                                  \\\quad\quad \text{if }
                           \locxmp{\spind}+\pmapdelta{\spind}/2\deriub\leq \locx <\locxsp{\spind+1}.
                       \end{cases}
        \end{align}

        \begin{conferenceonly}
            Using the mean-value theorem to prove that $|\pmapdelta{\spind}|\leq \locxdelta{\spind} \deriub$, it can be readily shown that that both coefficients  $\deriub-
                \pmapdelta{\spind}/
                \locxdelta{\spind}
            $
            and
            $\deriub+
                \pmapdelta{\spind}/
                \locxdelta{\spind}
            $
            in \eqref{eq:boundsascases} are non-negative .
        \end{conferenceonly}
        \begin{journalonly}
            It can be readily shown that that both coefficients  $\deriub-
                \pmapdelta{\spind}/
                \locxdelta{\spind}
            $
            and
            $\deriub+
                \pmapdelta{\spind}/
                \locxdelta{\spind}
            $
            in \eqref{eq:boundsascases} are non-negative (just  substitute $\locxl=\locxsp{\spind}$ and
            $\locxu=\locxsp{\spind+1}$ in \eqref{eq:meanvarbound} to note  that
            $|\pmapdelta{\spind}|\leq \locxdelta{\spind} \deriub$).
        \end{journalonly}

        % See pdf notes 2023/03/15. 
        Since $a\leq x\leq b$ implies that $|x|\leq \max(-a,b)$, it follows
        from \eqref{eq:boundsondiff} that
        \begin{salign}[eq:abspmapptdif]
            |&\pmap(\locx)  - \pmapest(\locx)|\leq
            \max[-\lbound{\spind}(\locx),
                \ubound{\spind}(\locx)]\\
            &=\max[\bound{\spind}(\locx;
                -\pmapdelta{\spind}
                ),
                \bound{\spind}(\locx;
                \pmapdelta{\spind}
                )]\\
            &=\max[\bound{\spind}(\locx;
                -\pa{\spind}
                ),
                \bound{\spind}(\locx;
                \pa{\spind}
                )]\jlac
            \define \cbound{\spind}(\locx)
        \end{salign}
        for all $\locx\in[\locxsp{\spind},\locxsp{\spind+1}]$.

        Using \eqref{eq:boundsascases}, it is also easy to verify that
        $\bound{\spind}(\locx; \pmapdelta{\spind} ) =
            \bound{\spind}(2\locxmp{\spind}-\locx;- \pmapdelta{\spind} ) $. As a
        consequence, it is easy to see from \eqref{eq:abspmapptdif} that
        $\cbound{\spind}(2\locxmp{\spind}-\locx) =
            \cbound{\spind}(\locx)$. Thus,  \eqref{eq:abspmapptdif} can be
        alternatively expressed as
        \begin{salign}
            \label{eq:abspmapptdifaltx}
            |\pmap(\locx)  - \pmapest(\locx)|&\leq
            \begin{cases}
                \cbound{\spind}(\locx)                  & ~\text{if } \locxsp{\spind}\leq \locx < \locxmp{\spind}    \\
                \cbound{\spind}(2\locxmp{\spind}-\locx) & ~\text{if } \locxmp{\spind}\leq \locx < \locxsp{\spind+1}.
            \end{cases}
        \end{salign}

        Observe that $\locx\in [\locxmp{\spind},\locxsp{\spind+1}]$ if and only if
        $2\locxmp{\spind}-\locx\in [\locxsp{\spind},\locxmp{\spind}]$. Hence, it
        suffices to consider $\cbound{\spind}(\locx)$ in  $\locx\in
            [\locxsp{\spind},\locxmp{\spind}]$. In this interval, it is easy to
        see that $\cbound{\spind}(\locx)= \bound{\spind}(\locx;
            -\pa{\spind})$.

        \begin{extendedonly}
            \blt[l1 norm]The next step is to bound the $L^1$ error. Using the above expressions, it follows that
            \begin{salign}
                \int_{\locxsp{\spind}}^{\locxsp{\spind+1}}&|\pmap(\locx)  -
                \pmapest(\locx)| d\locx
                \nonumber
                \jlac
                \begin{intermed}
                    \int_{\locxsp{\spind}}^{\locxmp{\spind}}|\pmap(\locx)  -
                    \pmapest(\locx)| d\locx +
                    \int_{\locxmp{\spind}}^{\locxsp{\spind+1}}|\pmap(\locx)  -
                    \pmapest(\locx)| d\locx
                \end{intermed}
                \\&
                \leq
                \int_{\locxsp{\spind}}^{\locxmp{\spind}}
                \cbound{\spind}(\locx) d\locx +
                \int_{\locxmp{\spind}}^{\locxsp{\spind+1}}
                \cbound{\spind}(2\locxmp{\spind}-\locx) d\locx
                \\&=
                2\int_{\locxsp{\spind}}^{\locxmp{\spind}}
                \cbound{\spind}(\locx) d\locx
                \jlac=
                2\int_{\locxsp{\spind}}^{\locxmp{\spind}}
                \bound{\spind}(\locx;
                -\pa{\spind})
                d\locx
                \\&=
                2\int_{\locxsp{\spind}}^{
                    \locxmp{\spind}-\pa{\spind}/2\deriub
                }
                \bound{\spind}(\locx;
                -\pa{\spind})
                d\locx
                \\&+
                2\int_{
                    \locxmp{\spind}-\pa{\spind}/2\deriub
                }^{\locxmp{\spind}}
                \bound{\spind}(\locx;
                -\pa{\spind})
                d\locx
                \jlac
                \begin{intermed}
                    =
                    2 \left(\deriub+\frac{
                        \pa{\spind}
                    }{
                        \locxdelta{\spind}
                    }\right)
                    \int_{\locxsp{\spind}}^{
                        \locxmp{\spind}-\pa{\spind}/2\deriub
                    }
                    (\locx - \locxsp{\spind})
                    d\locx
                \end{intermed}
                \jlac
                \begin{intermed}
                    +
                    2
                    \left(\deriub -\frac{
                        \pa{\spind}
                    }{
                        \locxdelta{\spind}
                    }\right)
                    \int_{
                        \locxmp{\spind}-\pa{\spind}/2\deriub
                    }^{\locxmp{\spind}}
                    ( \locxsp{\spind+1}-\locx)
                    d\locx
                \end{intermed}
                \jlac
                \begin{intermed}
                    =
                    2 \left(\deriub+\frac{
                        \pa{\spind}
                    }{
                        \locxdelta{\spind}
                    }\right)
                    \left[
                        \frac{(\locx - \locxsp{\spind})^2}{2}
                        \right]_{\locxsp{\spind}}^{
                        \locxmp{\spind}-\pa{\spind}/2\deriub
                    }
                \end{intermed}
                \jlac
                \begin{intermed}
                    +
                    2
                    \left(\deriub -\frac{
                        \pa{\spind}
                    }{
                        \locxdelta{\spind}
                    }\right)
                    \left[
                        -\frac{ (\locxsp{\spind+1}-\locx)^2}{2}
                        \right]_{
                        \locxmp{\spind}-\pa{\spind}/2\deriub
                    }^{\locxmp{\spind}}
                \end{intermed}
                \jlac
                \begin{intermed}
                    =
                    \left(\deriub+\frac{
                        \pa{\spind}
                    }{
                        \locxdelta{\spind}
                    }\right)
                    \left[
                        (\locx - \locxsp{\spind})^2
                        \right]_{\locxsp{\spind}}^{
                        \locxmp{\spind}-\pa{\spind}/2\deriub
                    }
                \end{intermed}
                \jlac
                \begin{intermed}+
                    \left(\deriub -\frac{
                        \pa{\spind}
                    }{
                        \locxdelta{\spind}
                    }\right)
                    \left[
                        - (\locxsp{\spind+1}-\locx)^2
                        \right]_{
                        \locxmp{\spind}-\pa{\spind}/2\deriub
                    }^{\locxmp{\spind}}
                \end{intermed}
                \jlac
                \begin{intermed}=
                    \frac{\deriub \locxdelta{\spind} +
                        \pa{\spind}
                    }{
                        \locxdelta{\spind}
                    }
                    \left[
                        \frac{\deriub\locxdelta{\spind} - \pa{\spind}}{2\deriub}\right]^2
                \end{intermed}
                \jlac
                \begin{intermed}+
                    \frac{\deriub\locxdelta{\spind} -
                        \pa{\spind}
                    }{
                        \locxdelta{\spind}
                    }
                    \left[-\left(
                        \frac{\locxdelta{\spind}}{2}
                        \right)^2+
                        \left(\frac{\deriub\locxdelta{\spind} + \pa{\spind}}{2\deriub}\right)^2\right]
                \end{intermed}
                \jlac
                \begin{intermed}=
                    \frac{\deriub \locxdelta{\spind} +
                        \pa{\spind}
                    }{
                        4\deriub^2 \locxdelta{\spind}
                    }
                    \left[
                        \frac{\deriub\locxdelta{\spind} - \pa{\spind}}{1}\right]^2
                \end{intermed}
                \jlac
                \begin{intermed}+
                    \frac{\deriub\locxdelta{\spind} -
                        \pa{\spind}
                    }{
                        4\deriub^2 \locxdelta{\spind}
                    }
                    \left[
                        \frac{\deriub\locxdelta{\spind} + \pa{\spind}}{1}\right]^2
                    -\frac{\deriub\locxdelta{\spind} -
                        \pa{\spind}
                    }{
                        \locxdelta{\spind}
                    }
                    \left(
                    \frac{\locxdelta{\spind}}{2}
                    \right)^2
                \end{intermed}
                \jlac
                \begin{intermed}=
                    \frac{2\deriub \locxdelta{\spind} (\deriub \locxdelta{\spind} +
                        \pa{\spind})(\deriub \locxdelta{\spind} -
                        \pa{\spind})
                    }{
                        4\deriub^2 \locxdelta{\spind}
                    }-\frac{\deriub\locxdelta{\spind} -
                        \pa{\spind}
                    }{
                        4
                    }\locxdelta{\spind}
                \end{intermed}
                \jlac
                \begin{intermed}=
                    \frac{ (\deriub \locxdelta{\spind} +
                        \pa{\spind})(\deriub \locxdelta{\spind} -
                        \pa{\spind})
                    }{
                        2\deriub
                    }-\frac{\deriub\locxdelta{\spind} -
                        \pa{\spind}
                    }{
                        4
                    }\locxdelta{\spind}
                \end{intermed}
                \jlac
                \begin{intermed}=
                    \frac{ 2(\deriub \locxdelta{\spind} +
                        \pa{\spind})(\deriub \locxdelta{\spind} -
                        \pa{\spind})
                    }{
                        4\deriub
                    }-\frac{\deriub\locxdelta{\spind} -
                        \pa{\spind}
                    }{
                        4\deriub
                    }\locxdelta{\spind}\deriub
                \end{intermed}
                \jlac
                \begin{intermed}=
                    \frac{ 2(\deriub \locxdelta{\spind} +
                        \pa{\spind})
                        - \locxdelta{\spind}\deriub
                    }{
                        4\deriub
                    }
                    (\deriub \locxdelta{\spind} -
                    \pa{\spind})
                \end{intermed}
                \jlac
                \begin{intermed}=
                    \frac{ \deriub \locxdelta{\spind} +
                        2\pa{\spind}
                    }{
                        4\deriub
                    }
                    (\deriub \locxdelta{\spind} -
                    \pa{\spind})
                \end{intermed}
                \\&=
                \frac{ \deriub^2 \locxdelta{\spind}^2 +
                    \deriub \locxdelta{\spind}\pa{\spind} -2\pa{\spind}^2
                }{
                    4\deriub
                }.
            \end{salign} % checked with pyplot
            This is maximum when $\pa{\spind} = \deriub \locxdelta{\spind}/4$,
            which yields
            \begin{align}
                \int_{\locxsp{\spind}}^{\locxsp{\spind+1}}|\pmap(\locx)  -
                \pmapest(\locx)| d\locx & \leq \frac{9}{32} \deriub \locxdelta{\spind}^2.
            \end{align}
            Combining this bound for the $\spnum-1$ intervals yields
            \begin{align}
                %\label{eq:lininterpl1errexp}
                \int_{\locxsp{1}}^{\locxsp{\spnum}}|\pmap(\locx)  -
                \pmapest(\locx)| d\locx & \leq \frac{9}{32} \deriub\sum_{\spind=1}^{\spnum-1} \locxdelta{\spind}^2,
            \end{align}
            which proves \eqref{eq:lininterpl1err}.

            \blt[l2 norm]When it comes to the $L^2$ error, one can write the following:
            \begin{salign}
                \int_{\locxsp{\spind}}^{\locxsp{\spind+1}}&|\pmap(\locx)  -
                \pmapest(\locx)|^2 d\locx
                \nonumber
                \begin{intermed}
                    =
                    \int_{\locxsp{\spind}}^{\locxmp{\spind}}|\pmap(\locx)  -
                    \pmapest(\locx)|^2 d\locx +
                    \int_{\locxmp{\spind}}^{\locxsp{\spind+1}}|\pmap(\locx)  -
                    \pmapest(\locx)|^2 d\locx
                \end{intermed}
                \\&\leq
                \int_{\locxsp{\spind}}^{\locxmp{\spind}}
                \cbound{\spind}^2(\locx) d\locx +
                \int_{\locxmp{\spind}}^{\locxsp{\spind+1}}
                \cbound{\spind}^2(2\locxmp{\spind}-\locx) d\locx
                \\&=
                2\int_{\locxsp{\spind}}^{\locxmp{\spind}}
                \cbound{\spind}^2(\locx) d\locx
                \jlac=
                2\int_{\locxsp{\spind}}^{\locxmp{\spind}}
                \bound{\spind}^2(\locx;
                -\pa{\spind})
                d\locx
                \\&
                \begin{intermed}
                    =
                    2\int_{\locxsp{\spind}}^{
                        \locxmp{\spind}-\pa{\spind}/2\deriub
                    }
                    \bound{\spind}^2(\locx;
                    -\pa{\spind})
                    d\locx
                    +
                    2\int_{
                        \locxmp{\spind}-\pa{\spind}/2\deriub
                    }^{\locxmp{\spind}}
                    \bound{\spind}^2(\locx;
                    -\pa{\spind})
                    d\locx
                \end{intermed}
                \jlac
                \begin{intermed}=
                    2 \left(\deriub+\frac{
                        \pa{\spind}
                    }{
                        \locxdelta{\spind}
                    }\right)^2
                    \int_{\locxsp{\spind}}^{
                        \locxmp{\spind}-\pa{\spind}/2\deriub
                    }
                    (\locx - \locxsp{\spind})^2
                    d\locx
                \end{intermed}
                \jlac
                \begin{intermed}+
                    2
                    \left(\deriub -\frac{
                        \pa{\spind}
                    }{
                        \locxdelta{\spind}
                    }\right)^2
                    \int_{
                        \locxmp{\spind}-\pa{\spind}/2\deriub
                    }^{\locxmp{\spind}}
                    ( \locxsp{\spind+1}-\locx)^2
                    d\locx
                \end{intermed}
                \jlac
                \begin{intermed}=
                    2 \left(\deriub+\frac{
                        \pa{\spind}
                    }{
                        \locxdelta{\spind}
                    }\right)^2
                    \left[
                        \frac{(\locx - \locxsp{\spind})^3}{3}
                        \right]_{\locxsp{\spind}}^{
                        \locxmp{\spind}-\pa{\spind}/2\deriub
                    }
                \end{intermed}
                \jlac
                \begin{intermed}+
                    2
                    \left(\deriub -\frac{
                        \pa{\spind}
                    }{
                        \locxdelta{\spind}
                    }\right)^2
                    \left[
                        -\frac{ (\locxsp{\spind+1}-\locx)^3}{3}
                        \right]_{
                        \locxmp{\spind}-\pa{\spind}/2\deriub
                    }^{\locxmp{\spind}}
                \end{intermed}
                \jlac
                \begin{intermed}=
                    \frac{2}{3} \left(\deriub+\frac{
                        \pa{\spind}
                    }{
                        \locxdelta{\spind}
                    }\right)^2
                    \left[
                        (\locx - \locxsp{\spind})^3
                        \right]_{\locxsp{\spind}}^{
                        \locxmp{\spind}-\pa{\spind}/2\deriub
                    }
                \end{intermed}
                \jlac
                \begin{intermed}+
                    \frac{2}{3}\left(\deriub -\frac{
                        \pa{\spind}
                    }{
                        \locxdelta{\spind}
                    }\right)^2
                    \left[
                        - (\locxsp{\spind+1}-\locx)^3
                        \right]_{
                        \locxmp{\spind}-\pa{\spind}/2\deriub
                    }^{\locxmp{\spind}}
                \end{intermed}
                \jlac
                \begin{intermed}=
                    \frac{2}{3}\left(\frac{\deriub \locxdelta{\spind} +
                        \pa{\spind}
                    }{
                        \locxdelta{\spind}
                    }\right)^2
                    \left[
                        \frac{\deriub\locxdelta{\spind} - \pa{\spind}}{2\deriub}\right]^3
                \end{intermed}
                \jlac
                \begin{intermed}+
                    \frac{2}{3}\left(\frac{\deriub\locxdelta{\spind} -
                        \pa{\spind}
                    }{
                        \locxdelta{\spind}
                    }\right)^2
                    \left[-\left(
                        \frac{\locxdelta{\spind}}{2}
                        \right)^3+
                        \left(\frac{\deriub\locxdelta{\spind} + \pa{\spind}}{2\deriub}\right)^3\right]
                \end{intermed}
                \jlac
                \begin{intermed}=
                    %%%%%%%%%%
                    \frac{2}{3\cdot 8}\frac{\left(\deriub^2 \locxdelta{\spind}^2 -
                        \pa{\spind}^2\right)^2
                    }{
                        \deriub^3 \locxdelta{\spind}^2
                    }
                    \left[
                        \deriub\locxdelta{\spind} - \pa{\spind}\right]
                \end{intermed}
                \jlac
                \begin{intermed}+
                    \frac{2}{3\cdot 8}\frac{\left(\deriub^2\locxdelta{\spind}^2 -
                        \pa{\spind}^2\right)^2
                    }{
                        \deriub^3 \locxdelta{\spind}^2
                    }
                    \left[
                        \deriub\locxdelta{\spind} + \pa{\spind}\right]
                    -\frac{2}{3\cdot 8}\left(\deriub\locxdelta{\spind} -
                    \pa{\spind}\right)^2
                    \locxdelta{\spind}
                \end{intermed}
                \jlac
                \begin{intermed}=
                    %%%%%%%%%%
                    \frac{1}{6}\frac{\left(\deriub^2 \locxdelta{\spind}^2 -
                        \pa{\spind}^2\right)^2
                    }{
                        \deriub^3 \locxdelta{\spind}^2
                    }
                    \deriub\locxdelta{\spind}
                    -\frac{1}{12}\left(\deriub\locxdelta{\spind} -
                    \pa{\spind}\right)^2
                    \locxdelta{\spind}
                \end{intermed}
                \jlac
                \begin{intermed}=
                    %%%%%%%%%%
                    \frac{1}{6}\frac{\left(\deriub^2 \locxdelta{\spind}^2 -
                        \pa{\spind}^2\right)^2
                    }{
                        \deriub^2 \locxdelta{\spind}
                    }
                    -\frac{1}{12}\left(\deriub\locxdelta{\spind} -
                    \pa{\spind}\right)^2
                    \locxdelta{\spind}
                \end{intermed}
                \jlac
                \begin{intermed}=
                    %%%%%%%%%%
                    \frac{1}{12}\frac{2\left(\deriub^2 \locxdelta{\spind}^2 -
                        \pa{\spind}^2\right)^2
                        -\left(\deriub\locxdelta{\spind} -
                        \pa{\spind}\right)^2
                        \locxdelta{\spind}^2\deriub^2
                    }{
                        \deriub^2 \locxdelta{\spind}
                    }
                \end{intermed}
                \jlac
                \begin{intermed}=
                    %%%%%%%%%%
                    \frac{1}{12}\frac{
                        2\left(\deriub \locxdelta{\spind} +
                        \pa{\spind}\right)^2
                        \left(\deriub \locxdelta{\spind} -
                        \pa{\spind}\right)^2
                        -\left(\deriub\locxdelta{\spind} -
                        \pa{\spind}\right)^2
                        \locxdelta{\spind}^2\deriub^2
                    }{
                        \deriub^2 \locxdelta{\spind}
                    }
                \end{intermed}
                \jlac
                \begin{intermed}=
                    %%%%%%%%%%
                    \frac{\left(\deriub \locxdelta{\spind} -
                        \pa{\spind}\right)^2
                    }{12}\frac{2
                        \left(\deriub \locxdelta{\spind} +
                        \pa{\spind}\right)^2
                        -
                        \locxdelta{\spind}^2\deriub^2
                    }{
                        \deriub^2 \locxdelta{\spind}
                    }
                \end{intermed}
                \jlac
                \begin{intermed}=
                    %%%%%%%%%%
                    \frac{\left(\deriub \locxdelta{\spind} -
                        \pa{\spind}\right)^2
                    }{12}\frac{
                        2\deriub^2 \locxdelta{\spind}^2 +
                        4\deriub \locxdelta{\spind}
                        \pa{\spind}
                        +
                        2\pa{\spind}^2
                        -
                        \locxdelta{\spind}^2\deriub^2
                    }{
                        \deriub^2 \locxdelta{\spind}
                    }
                \end{intermed}
                \jlac
                \begin{intermed}=
                    %%%%%%%%%%
                    \frac{\left(\deriub \locxdelta{\spind} -
                        \pa{\spind}\right)^2
                    }{12}\frac{
                        \deriub^2 \locxdelta{\spind}^2 +
                        4\deriub \locxdelta{\spind}
                        \pa{\spind}
                        +
                        2\pa{\spind}^2
                    }{
                        \deriub^2 \locxdelta{\spind}
                    }
                \end{intermed}
                \jlac
                \begin{intermed}=
                    %%%%%%%%%%
                    \frac{\left(\deriub \locxdelta{\spind} -
                        \pa{\spind}\right)^2
                    }{12}\frac{
                        \deriub^2 \locxdelta{\spind}^2 +
                        4\deriub \locxdelta{\spind}
                        \pa{\spind}
                        +
                        2\pa{\spind}^2
                    }{
                        \deriub^2 \locxdelta{\spind}^2
                    }\locxdelta{\spind}
                \end{intermed}
                \jlac
                \begin{intermed}=
                    %%%%%%%%%%
                    \frac{\left(1 -\frac{
                            \pa{\spind}}{\deriub \locxdelta{\spind}}\right)^2
                    }{12}\frac{
                        \deriub^2 \locxdelta{\spind}^2 +
                        4\deriub \locxdelta{\spind}
                        \pa{\spind}
                        +
                        2\pa{\spind}^2
                    }{
                        1
                    }\locxdelta{\spind}
                \end{intermed}
                \jlac
                \begin{intermed}=
                    %%%%%%%%%%
                    \frac{\left(1 -\frac{
                            \pa{\spind}}{\deriub \locxdelta{\spind}}\right)^2
                    }{12}\frac{
                        1 +
                        4\frac{
                            \pa{\spind}}{\deriub \locxdelta{\spind} }
                        +
                        2\left(\frac{\pa{\spind}}{\deriub \locxdelta{\spind} }\right)^2
                    }{
                        1
                    }\locxdelta{\spind}^3 \deriub^2
                \end{intermed}
                \jlac
                =
                %%%%%%%%%%
                \frac{\locxdelta{\spind}^3 \deriub^2
                }{
                    12
                }\left(1 -\relvar{\spind}\right)^2
                (
                1 +
                4\relvar{\spind}
                +
                2\relvar{\spind}^2),
            \end{salign} % checked with sympy
            where $\relvar{\spind}\define{
                    \pa{\spind}}/({\deriub \locxdelta{\spind}})$.
            \begin{intermed}
                Setting the derivative equal to zero, one finds
                %%%%%%%%%%
                \begin{salign}
                    &4\relvar{\spind}^3-5\relvar{\spind}
                    +1 = 0\\
                    &(\relvar{\spind}-1)(4\relvar{\spind}^2+4\relvar{\spind}-1)=0
                    \\&
                    (\relvar{\spind}-1)\left(\relvar{\spind}-\frac{-1+\sqrt{2}}{2}\right)
                    \left(\relvar{\spind}-\frac{-1-\sqrt{2}}{2}\right)=0
                \end{salign}
            \end{intermed}
            The  maximum subject to $\relvar{\spind}\in[0,1]$ is attained when $
                \relvar{\spind}=\relvaropt{\spind}\define({-1+\sqrt{2}})/{2}
            $, which yields
            \begin{salign}
                \int_{\locxsp{\spind}}^{\locxsp{\spind+1}}|\pmap(\locx)  -
                \pmapest(\locx)|^2 d\locx&\leq
                \begin{intermed}
                    \frac{\locxdelta{\spind}^3 \deriub^2
                    }{
                        12
                    }\left(1 -\relvaropt{\spind}\right)^2
                    (
                    1 +
                    4\relvaropt{\spind}
                    +
                    2(\relvaropt{\spind})^2)
                \end{intermed}
                \jlac
                \begin{intermed}=
                    \frac{\locxdelta{\spind}^3 \deriub^2
                    }{
                        12
                    }\left(-\frac{13}{8} + 2\sqrt{2}\right)
                    =
                \end{intermed}
                \frac{16\sqrt{2}-13}{96
                }\locxdelta{\spind}^3 \deriub^2.
            \end{salign}
            Adding this error over the $\spnum-1$ intervals
            \begin{intermed} yields
                \begin{align}
                    %\label{eq:lininterpl2errexp}
                    \int_{\locxsp{1}}^{\locxsp{\spnum}}|\pmap(\locx)  -
                    \pmapest(\locx)|^2 d\locx & \leq
                    \frac{16\sqrt{2}-13}{96
                    }\deriub^2
                    \sum_{\spind=1}^{\spnum-1}\locxdelta{\spind}^3 ,
                \end{align}
                which
            \end{intermed}
            proves  \eqref{eq:lininterpl2err}.

            \blt[linf]Finally, for the $L^\infty$ error, note that
            \begin{salign}
                &\sup_{\locx\in[\locxsp{\spind},\locxsp{\spind+1})}|\pmap(\locx)  -
                \pmapest(\locx)|
                \\&
                =
                \max\bigg[
                \sup_{\locx\in[\locxsp{\spind},\locxmp{\spind})}|\pmap(\locx)  -
                \pmapest(\locx)|,
                \nonumber
                \\&
                \sup_{\locx\in[\locxmp{\spind},\locxsp{\spind+1})}|\pmap(\locx)  -
                    \pmapest(\locx)|
                    \bigg]\\&\leq
                \max[
                \sup_{\locx\in[\locxsp{\spind},\locxmp{\spind})}
                \cbound{\spind}(\locx),
                \sup_{\locx\in[\locxmp{\spind},\locxsp{\spind+1})}
                    \cbound{\spind}(2\locxmp{\spind}-\locx)
                ]
                \jlac
                \begin{intermed}
                    =
                    \max[
                    \sup_{\locx\in[\locxsp{\spind},\locxmp{\spind})}
                    \cbound{\spind}(\locx),
                    \sup_{\locx\in[\locxsp{\spind},\locxmp{\spind})}
                        \cbound{\spind}(\locx)]
                \end{intermed}
                \jlac
                \begin{intermed}=
                    \sup_{\locx\in[\locxsp{\spind},\locxmp{\spind})}
                    \cbound{\spind}(\locx)
                \end{intermed}
                \jlac
                \begin{intermed}=
                    \sup_{\locx\in[\locxsp{\spind},\locxmp{\spind})}
                    \bound{\spind}(\locx;
                    -\pa{\spind})
                \end{intermed}
                \jlac
                \begin{intermed}=
                    \bound{\spind}(
                    \locxmp{\spind}-\pa{\spind}/2\deriub
                    ;
                    -\pa{\spind})
                \end{intermed}
                \\&
                =
                \frac{
                    \deriub^2\locxdelta{\spind}^2 -
                    \pa{\spind}^2
                }{
                    2\deriub \locxdelta{\spind}
                }
                \jlac\leq
                \frac{
                    \deriub\locxdelta{\spind}
                }{
                    2
                }.
            \end{salign}

            Combining this bound for all intervals
            \begin{intermed}
                yields
                \begin{align}
                    %\label{eq:lininterplinferrexp}
                    \sup_{\locx\in[\locxsp{1},\locxsp{\spnum})}|\pmap(\locx)  -
                    \pmapest(\locx)|
                     & \leq
                    \frac{
                        \deriub
                    }{
                        2
                    }\max_{\spind}\locxdelta{\spind},
                \end{align}
                which
            \end{intermed}
            proves \eqref{eq:lininterplinferr}.

        \end{extendedonly}

        \begin{nonextendedonly}
            The rest of the proof involves integrating \eqref{eq:abspmapptdifaltx} to obtain the L$^1$ and L$^2$ errors and computing the suprema on each subinterval to obtain $L^\infty$. It is omitted due to lack of space.
        \end{nonextendedonly}

    \end{bullets}

\end{IEEEproof}

Finally, combining \eqref{eq:lininterpallerr} with
\eqref{eq:deriubdef} completes the proof.

\section{Proof of Theorem~\ref{prop:sincinterperr}}
\label{sec:sincinterperr}

\begin{bullets}
    \blt[error fixed offset]

    % Notes: 2022/03/11
    \begin{mylemma}
        \label{prop:sigerr}
        Let
        \begin{bullets}
            \blt $\sig[\sampind]\define\sig(\sampind\sampint)$ and
            \blt
            \begin{align}
                \sigest(\time)\define \sum_{\sampind=-\infty}^{\infty}\sig[\sampind]\sinc\left(\frac{\time -\sampind\sampint}{\sampint}\right).
            \end{align}
        \end{bullets}
        Then
        \begin{salign}
            \err \define& \int_{-\infty}^{\infty}|\sig(\time)-\sigest(\time)|^2d\time\\
            =&
            \frac{1}{2\pi}
            \sum_{\compind=-\infty, \compind\neq 0}^{\infty}\int_{-\infty}^{\infty}
            \left|\sigft_\compind\left(\freq \right)
            \right|^2d\freq
            \\&+
            \frac{1}{2\pi}\int_{-\infty}^{\infty}\left|
            \sum_{\compind=-\infty,\compind\neq0}^{\infty}\sigft_\compind\left(\freq \right)
            \right|^2d\freq,
        \end{salign}
        where
        \begin{align}
            \sigft_\compind(\freq) \define \begin{cases}
                                               \sigft(\freq + \compind \frac{2\pi}{\sampint}) & \text{ if }\freq\in [-\frac{\pi}{\sampint},\frac{\pi}{\sampint}) \\
                                               0                                              & \text{otherwise}.
                                           \end{cases}
        \end{align}

    \end{mylemma}

    \begin{IEEEproof}
        %\acom{we may want to prove convergence of these series and integrals                based on the assumption that $\sig$ has finite energy} 
        From Parseval's
        relation
        \begin{align}
            \label{eq:errparseval}
            \err = & \frac{1}{2\pi}\int_{-\infty}^{\infty}|\sigft(\freq)-\sigestft(\freq)|^2d\freq.
        \end{align}
        Note that
        \begin{salign}
            \sigest(\time)&
            \begin{intermed}
                =
                \sum_{\sampind=-\infty}^{\infty}\sig(\sampind\sampint)\sinc\left(\frac{\time -\sampind\sampint}{\sampint}\right)
            \end{intermed}
            \jlac
            = \sum_{\sampind=-\infty}^{\infty}\sig(\time)\delta(\time -\sampind\sampint) \ast \sinc\left(\frac{\time}{\sampint}\right).
        \end{salign}
        In the frequency domain:
        \begin{align}
            \sigestft(\freq)
            = & \frac{1}{2\pi}\left[ \sigft(\freq) \ast\frac{2\pi}{\sampint} \sum_{\compind=-\infty}^{\infty}\delta\left(\freq -\compind\frac{2\pi}{\sampint}\right)\right]
            \\&
            \cdot \sampint \cdot \rpulse_{-\pi/\sampint}^{\pi/\sampint}(\freq),
        \end{align}
        where
        \begin{align}
            \rpulse_a^b (\freq)\define\begin{cases}
                                          1 & \text{if }\freq\in[a,b) \\
                                          0 & \text{otherwise.}
                                      \end{cases}
        \end{align}
        Therefore,
        \begin{salign}
            \sigestft(\freq)
            &= \left[  \sum_{\compind=-\infty}^{\infty}\sigft\left(\freq -\compind\frac{2\pi}{\sampint}\right)\right] \cdot \rpulse_{-\pi/\sampint}^{\pi/\sampint}(\freq)\\
            \label{eq:sigestftassum}
            &=   \sum_{\compind=-\infty}^{\infty}\sigft_\compind\left(\freq \right).
        \end{salign}
        On the other hand, it is straightforward to see that
        \begin{align}
            \label{eq:sigftassum}
            \sigft(\freq) =  \sum_{\compind=-\infty}^{\infty}\sigft_\compind\left(\freq - \compind\frac{2\pi}{\sampint}\right).
        \end{align}
        Substituting \eqref{eq:sigestftassum} and \eqref{eq:sigftassum} into \eqref{eq:errparseval} yields
        \begin{salign}
            %\label{eq:errparseval}
            \err &
            \begin{intermed}
                =
                \frac{1}{2\pi}\int_{-\infty}^{\infty}\left|
                \sum_{\compind=-\infty}^{\infty}\sigft_\compind\left(\freq - \compind\frac{2\pi}{\sampint}\right)
                -
                \sum_{\compind=-\infty}^{\infty}\sigft_\compind\left(\freq \right)
                \right|^2d\freq
            \end{intermed}
            \jlac
            \begin{intermed}
                = \frac{1}{2\pi}\int_{-\infty}^{\infty}\left|
                \sum_{\compind=-\infty, \compind\neq 0}^{\infty}\sigft_\compind\left(\freq - \compind\frac{2\pi}{\sampint}\right)
                -
                \sum_{\compind=-\infty,\compind\neq0}^{\infty}\sigft_\compind\left(\freq \right)
                \right|^2d\freq
            \end{intermed}
            \jlac
            = \frac{1}{2\pi}\int_{-\infty}^{\infty}\left|
            \sum_{\compind=-\infty, \compind\neq 0}^{\infty}\sigft_\compind\left(\freq - \compind\frac{2\pi}{\sampint}\right)
            \right|^2d\freq \nonumber
            \\&
            +
            \frac{1}{2\pi}\int_{-\infty}^{\infty}\left|
            \sum_{\compind=-\infty,\compind\neq0}^{\infty}\sigft_\compind\left(\freq \right)
            \right|^2d\freq
            \\
            &
            \begin{intermed}
                = \frac{1}{2\pi}
                \sum_{\compind=-\infty, \compind\neq 0}^{\infty}\int_{-\infty}^{\infty}
                \left|\sigft_\compind\left(\freq - \compind\frac{2\pi}{\sampint}\right)
                \right|^2d\freq
                \nonumber
            \end{intermed}
            \jlac
            \begin{intermed}
                +
                \frac{1}{2\pi}\int_{-\infty}^{\infty}\left|
                \sum_{\compind=-\infty,\compind\neq0}^{\infty}\sigft_\compind\left(\freq \right)
                \right|^2d\freq
            \end{intermed}
            \jlac= \frac{1}{2\pi}
            \sum_{\compind=-\infty, \compind\neq 0}^{\infty}\int_{-\infty}^{\infty}
            \left|\sigft_\compind\left(\freq \right)
            \right|^2d\freq
            \\&+
            \frac{1}{2\pi}\int_{-\infty}^{\infty}\left|
            \sum_{\compind=-\infty,\compind\neq0}^{\infty}\sigft_\compind\left(\freq \right)
            \right|^2d\freq.
        \end{salign}

    \end{IEEEproof}

    \blt[avg error]For the following result, consider the shifted signal $\shiftsig(\time)\define \sig(\time-\sigshift)$ and its reconstruction
    \begin{align}
        \shiftsigest(\time)\define \sum_{\sampind=-\infty}^{\infty}\shiftsig(\sampind\sampint)\sinc\left(\frac{\time -\sampind\sampint}{\sampint}\right).
    \end{align}
    \begin{mylemma}
        \begin{bullets}
            \blt Let
            \begin{align}
                \label{eq:errphinu}
                \errp(\sigshift) \define & \int_{-\infty}^{\infty}|\shiftsig(\time)-\shiftsigest(\time)|^2d\time.
            \end{align}
            It holds that
            \begin{align}
                \label{eq:avgerrpp}
                \avgerrp \define\frac{1}{\sampint}\int_{0}^{\sampint}  \errp(\sigshift) d\sigshift = \frac{1}{\pi}\int_\highband|\sigft(\freq)|^2d\freq,
            \end{align}
            where $\highband\define (-\infty,-{\pi}/{\sampint}] \cup[\pi/\sampint,\infty)$.

        \end{bullets}
    \end{mylemma}

    \begin{IEEEproof}
        Let
        \begin{align}
            \shiftsigft_\compind(\freq) \define \begin{cases}
                                                    \shiftsigft(\freq + \compind \frac{2\pi}{\sampint}) & \text{if }\freq\in [-\frac{\pi}{\sampint},\frac{\pi}{\sampint}) \\
                                                    0                                                   & \text{otherwise}.
                                                \end{cases}
        \end{align}
        From \Cref{prop:sigerr}, it follows that
        \begin{align}
            \errp(\sigshift) = \errp_1(\sigshift) + \errp_2(\sigshift),
        \end{align}
        where
        \begin{salign}
            \label{eq:errorone}
            \errp_1(\sigshift) &\define \frac{1}{2\pi}
            \sum_{\compind=-\infty, \compind\neq 0}^{\infty}\int_{-\infty}^{\infty}
            \left|\shiftsigft_\compind\left(\freq \right)
            \right|^2d\freq\\
            \label{eq:errortwo}
            \errp_2(\sigshift)&\define
            \frac{1}{2\pi}\int_{-\infty}^{\infty}\left|
            \sum_{\compind=-\infty,\compind\neq0}^{\infty}\shiftsigft_\compind\left(\freq \right)
            \right|^2d\freq.
        \end{salign}
        Letting $\avgerrp_i \define
            (1/\sampint)\int_0^{\sampint}\errp_i(\sigshift)d\sigshift$, it follows
        that
        \begin{align}
            \label{eq:avgerr}
            \avgerrp = \avgerrp_1 + \avgerrp_2.
        \end{align}
        Noting that $\shiftsigft(\freq)= e^{-j\freq\sigshift}\sigft(\freq)$ yields
        \begin{salign}
            &\shiftsigft_\compind(\freq) \jlac
            \begin{intermed}
                = \begin{cases}
                    e^{-j(\freq + \compind \frac{2\pi}{\sampint})\sigshift}\sigft(\freq + \compind \frac{2\pi}{\sampint}) \alignchar \text{if }\freq\in [-\frac{\pi}{\sampint},\frac{\pi}{\sampint}) \jumpline
                    0                                                                                                     \alignchar
                    \text{otherwise}
                \end{cases}
                \nonumber
            \end{intermed}
            \jlac=e^{-j(\freq+\compind \frac{2\pi}{\sampint})\sigshift}
            \sigft_\compind(\freq).
        \end{salign}
        From \eqref{eq:errorone}, it is then easy to see that
        \begin{align}
            \label{eq:avgerr1}
            \errp_1(\sigshift) = \frac{1}{2\pi}\int_\highband|\sigft(\freq)|^2d\freq = \avgerrp_1.
        \end{align}
        %where the second equality follows from the fact that the right-hand             side does not depend on $\sigshift$.
        %
        On the other hand, from \eqref{eq:errortwo}, it follows that
        \begin{align}
             & \avgerrp_2
            \begin{intermed}
                = \frac{1}{2\pi\sampint}\int_0^{\sampint}
                \int_{-\infty}^{\infty}\left|
                \sum_{\compind=-\infty,\compind\neq0}^{\infty}\shiftsigft_\compind\left(\freq \right)
                \right|^2d\freq d\sigshift
            \end{intermed}
            \jlac
            \begin{intermed}
                = \frac{1}{2\pi\sampint}
                \int_{-\infty}^{\infty}\int_0^{\sampint} \left|
                \sum_{\compind=-\infty,\compind\neq0}^{\infty}
                e^{-j(\freq+\compind \frac{2\pi}{\sampint})\sigshift}
                \sigft_\compind(\freq)
                \right|^2d\sigshift d\freq
            \end{intermed}
            \\ \nonumber
             & = \frac{1}{2\pi\sampint}
            \int_{-\infty}^{\infty}\int_0^{\sampint} \left|
            \sum_{\compind=-\infty,\compind\neq0}^{\infty}
            e^{-j\compind \frac{2\pi}{\sampint}\sigshift}
            \sigft_\compind(\freq)
            \right|^2d\sigshift d\freq
            \jlac
            \begin{intermed}
                = \frac{1}{(2\pi)^2}
                \int_{-\infty}^{\infty}\int_0^{2\pi} \left|
                \sum_{\compind=-\infty,\compind\neq0}^{\infty}
                e^{-j\compind \sigshiftf}
                \sigft_\compind(\freq)
                \right|^2d\sigshiftf d\freq
            \end{intermed}
            \\
            \label{eq:delta2bfparseval}
             & = \frac{1}{(2\pi)^2}
            \int_{-\infty}^{\infty}\int_0^{2\pi} \left|
            \auxseqft(\sigshiftf,\freq)
            \right|^2d\sigshiftf d\freq,
        \end{align}
        where $
            \auxseqft(\sigshiftf,\freq) \define
            \sum_{\compind=-\infty,\compind\neq0}^{\infty}
            e^{-j\compind \sigshiftf}
            \sigft_\compind(\freq)
        $
        is the discrete-time Fourier  transform of
        \begin{align}
            \auxseq_\compind(\freq)\define\begin{cases}
                                              \sigft_\compind(\freq) & \text{if }\compind \neq 0 \\
                                              0                      & \text{otherwise}.
                                          \end{cases}
        \end{align}
        Applying Parseval's identity to \eqref{eq:delta2bfparseval}, it follows that
        \begin{salign}
            \avgerrp_2&
            = \frac{1}{2\pi}
            \int_{-\infty}^{\infty}
            \sum_{\compind=-\infty}^{\infty}
            |\auxseq_\compind(\freq)|^2
            d\freq\\
            &= \frac{1}{2\pi}
            \int_{-\infty}^{\infty}
            \sum_{\compind=-\infty,\compind\neq0}^{\infty}
            |\sigft_\compind(\freq)|^2
            d\freq\\
            \label{eq:avgerr2}
            &= \frac{1}{2\pi}
            \int_{\highband}
            |\sigft(\freq)|^2
            d\freq.
        \end{salign}
        Substituting \eqref{eq:avgerr1} and \eqref{eq:avgerr2} into \eqref{eq:avgerr} concludes the proof.

    \end{IEEEproof}

    \blt[final substitution]Finally, noting that $\shiftsigestcs{\sigshift}(\time)=
        \pmapest\sigshiftnot{-\sigshift}(\time-\sigshift)
    $ shows that $\errp(\sigshift)$ in \eqref{eq:errphinu} equals $\err(-\sigshift)$ in \eqref{eq:erroffsetsinc}. Since both functions are periodic with period $\sampint$, it follows that $\avgerrp$ in \eqref{eq:avgerrpp} equals $\avgerr$ in \eqref{eq:sincinterpavgerr}, which completes the proof.

    %Upon replacing $\shiftsigestcs{\sigshift}(\time)$ with  $\auxshiftsigestcs{-\sigshift}(\time)$, then    $\auxshiftsigestcs{\sigshift}(\time)$ with $\pmapest\sigshiftnot{\sigshift}(\time+\sigshift)$, and finally redefining $\errp(\sigshift)$ as $\errp(-\sigshift)$ yields \Cref{prop:sincinterperr}.

\end{bullets}

\section{Arbitrary Path-loss Exponent}
\label{sec:plexponent}

\begin{changes}

    \begin{bullets}
        \blt[no free space \ra no distance dependence]The results in this paper
        were obtained for free-space propagation, where the channel gain adheres
        to \eqref{eq:friisfull}. In more complex scenarios, the presence of
        obstacles introduces propagation phenomena such as reflection and
        diffraction. As a result, the channel gain  no longer depends on  the transmitter and receiver
        locations only through their distance. Still, one may be interested in predicting the channel gain
        of a given link based on its  distance.
        \blt[path loss exp]To this end, a common trick is
        to use \eqref{eq:friisfull} after replacing the square with a constant
        $\pathlossexp$ termed \emph{path-loss exponent} that is empirically
        adjusted. Although the resulting expression is  not physically accurate, one may wonder whether the results in this paper can be extended to an arbirary $\pathlossexp$.

        \blt[results]The answer is yes for the most part. For example,
        \begin{bullets}%        
            \blt[bouded derivatives]expressions \eqref{eq:friis1D} and \eqref{eq:boundedderivative} become
            \begin{salign}
                \label{eq:friis1Dexp}
                &\pmap(\loc)      = \pmap(\locx) = \sum_{\srcind=1}^{\srcnum}\frac{\srccoef[\srcind]}{
                    \left[(\locx - \srclocx[\srcind])^2 + \rdists[\srcind]\right]^{\pathlossexp/2} },
                \\
                \label{eq:boundedderivativeexp}
                &|\pmap'(\locx)|
                \begin{intermed}
                    =2 \sum_{\srcind=1}^{\srcnum}\frac{\srccoef[\srcind]
                        | \srclocx[\srcind] - \locx  | }{[
                                (\locx - \srclocx[\srcind])^2 + \rdists[\srcind]   ]^\pathlossexp}
                \end{intermed}                        \jumpline \alignchar
                \leq
                \begin{intermed}
                    2 \sum_{\srcind=1}^{\srcnum}\sup_{\srclocx[\srcind], \locx}\frac{\srccoef[\srcind]
                        | \srclocx[\srcind] - \locx  | }{[
                                (\locx - \srclocx[\srcind])^2 + \rdists[\srcind] ]^\pathlossexp  }
                \end{intermed}                        \jumpline \alignchar
                \begin{intermed}
                    = 2 \sum_{\srcind=1}^{\srcnum}\sup_{\auxvar}\frac{\srccoef[\srcind]
                        \auxvar }{[
                                \auxvar^2 + \rdists[\srcind] ]^\pathlossexp  }\jlac
                    =
                \end{intermed}
                \frac{(2\pathlossexp - 1)^{\pathlossexp-1/2}}{2^{\pathlossexp-1}\pathlossexp^{\pathlossexp}} \sum_{\srcind=1}^{\srcnum}
                \frac{\srccoef[\srcind]
                }{ \rdist[\srcind]^{2\pathlossexp-1}  }
            \end{salign}
            as a result of this generalization.
            \blt[reconstruction errors]Setting $\deriub$ equal to the right-hand side of  \eqref{eq:boundedderivativeexp} and following the same steps as in Appendices~\ref{sec:nninterperr} and \ref{sec:lininterperr}, one can readily generalize the error bounds for zeroth- and first-order interpolation.
            \blt[not generalizable results]On the other hand, the variability bounds in \Cref{prop:ftbounds} and the error bounds in Sec.~\ref{sec:sincinterp} are not easily generalizable to arbitrary path-loss exponents. This is because of the different nature of the proof techniques used therein.

            % readily extend the  This bound can also be used to derive the
            % reconstruction error bounds for the radio map estimators in the next
            % section for the case $\pathlossexp\neq 2$, i.e. let $\deriub =
            %     \frac{(2\pathlossexp -
            %         1)^{\pathlossexp-1/2}}{2^{\pathlossexp-1}\pathlossexp^{\pathlossexp}}
            %     \sum_{\srcind=1}^{\srcnum} \frac{\srccoef[\srcind] }{
            %         \rdist[\srcind]^{2\pathlossexp-1}  }$ for
        \end{bullets}

        \blt[conclusion] To sum up, some of the results in this paper can be extended to arbitrary path-loss exponents, but this is not fully meaningful as the model with $\pathlossexp\neq 2$ is not physically accurate. An analysis that accurately captures  actual propagation phenomena will be the subject of future publications.

    \end{bullets}%

\end{changes}

\end{document}